\documentclass[usenatbib,useAMS]{mnras}

\usepackage{graphicx}	
\usepackage{amsmath}	
\usepackage{amssymb}	
\usepackage{multicol}        
\usepackage{bm}
\usepackage{pdflscape}	
\usepackage{natbib}
\usepackage[varg]{txfonts}
\usepackage{hyperref}

\usepackage[T1]{fontenc}
\usepackage{ae,aecompl}

\title[A new model for heating of Solar North Polar Coronal Hole]{ A new model for heating of Solar North Polar Coronal Hole}

\author[E. Devlen et al.]{E. Devlen,$^{1,2}$\thanks{Contact e-mail: ebru.devlen@ege.edu.tr,} D. Zengin \c{C}amurdan,$^{1}$ M.  Yard{\i}mc{\i},$^{1}$ and E. R. Pek\"{u}nl\"{u},$^{1}$
\\
 $^{1}$University of Ege, Faculty of Science, Department of Astronomy \& Space Sciences, Bornova, 35100, Izmir, Turkey\\
 $^{2}$NORDITA, KTH Royal Institute of Technology and Stockholm University, Roslagstullsbacken 23, SE-106 91 Stockholm, Sweden \\}

\date{Last updated ??; in original form ??}

\pubyear{??}

\begin{document}
\label{firstpage}
\pagerange{\pageref{firstpage}--\pageref{lastpage}}
\maketitle

\begin{abstract}
This paper presents a new model of  North Polar Coronal Hole (NPCH) to study dissipation/propagation of MHD waves. We investigate the effects of the isotropic viscosity and  heat conduction on the propagation characteristics of the MHD waves in NPCH.
We first model NPCH by considering the differences in radial as well as in the direction perpendicular to the line of sight (\textit{los}) in  temperature, particle number density and non-thermal velocities between plumes and interplume lanes
for the specific case of \ion{O}{VI} ions. This model includes parallel and perpendicular (to the magnetic field) heat conduction and  viscous dissipation. Next, we derive the dispersion relations for the MHD waves in the case of absence and presence of parallel heat conduction. In the case of absence of parallel heat conduction, we find that MHD wave dissipation strongly depends on the  viscosity for modified acoustic and Alfven waves. The energy flux density of acoustic waves varies between $10^{4.7}$ and $10^7 \,erg\,cm^{-2}\,s^{-1}$ while the energy flux density of Alfven waves turned out to be between $ 10^6-10^{8.6} \,erg\,cm^{-2}\,s^{-1}$. But, solutions of the magnetoacustic waves show that the parallel heat conduction introduce anomalous dispersion to the NPCH plasma wherein the group velocity of waves exceeds the speed of light in vacuum. Our results suggests all these waves may provide 
significant source for the observed preferential accelerating and heating of \ion{O}{VI} ions, in turn coronal plasma heating and an extra accelerating agent for fast solar wind in NPCH.

\end{abstract}




\newpage

\section{Introduction}

Observations as well as theoretical studies on coronal heating and solar wind acceleration had long since drawn attention to the dissipation of MHD waves through various processes. Particularly, North Polar Coronal Hole (NPCH) have been observed by various satellites, for instance, Yohkoh, Wind, Soho, Hinode, TRACE, etc. The effective temperature of NPCH has been found as high as $1.0\times10^6-10^8$ K. Since the surface temperature of the Sun is much lower than the NPCH one, two questions are immediately required to be answered: 1) what is the continuous source of energy for NPCH, since the plasma with that much high temperature is bound to lose energy to the transition region below through heat conduction and optically thin emission? and 2) what kind of force causes solar wind acceleration? Observations revealed that NPCH may be regarded as collisional up to the radial distance 1.7 - 2.1 R range, where  $R=r/R_{\odot}$. Within this domain, MHD waves may donate their energies into random thermal motions of particles through microscopic mechanisms, i.e., viscosity, thermal conductivity and resistivity. 

\citet{bra65} pointed out that dissipative processes, we listed above, cause damping of MHD waves and thus heat the plasma. In the extensive review articles by \citet{ofm05} and \citet{cra09} it was shown that Alfven waves, fast and slow magnetosonic waves are the most likely candidates to heat the coronal plasma and accelerate the solar wind. Energy fluxes of acoustic waves are so low and the damping length scales are so short that they are excluded from the list of energy supplier candidates \citep{ath78,mei81}. \citet{suz02} claimed that acoustic waves with period $\tau \geq 60 \,s$ and energy flux density $2\times 10^5 \,erg\,cm^{-2}\,s^{-1}$ may heat the plasma to the temperatures of $T \geq 10^6$ K by dissipation of N-waves after the formation of the shocks. He also noted that other collective processes with larger length scales may heat the coronal plasma and accelerate the solar wind. 

Presence of MHD waves in NPCH is revealed by EUV observations carried out by SOHO, Hinode and TRACE satellites. \citet{ofm97}, \citet{Dmoo02}, \citet{bana09} and \citet{gup10} reported the presence of propagating and standing compressional waves in corona. Alfven waves are also observed  \citep{ Tom07,ban98, banb09, lan09, gup10,  bem12}. 

In the heliocentric distance range 1.4-2.6 $R_{\odot}$ the coronal heating and solar wind acceleration are shown by UVCS observations to be more effective \citep{fis95}. In earlier studies by \citet{hol86} and \citet{hol88}, the required energy flux density to heat the corona is evaluated as $4-5\times 10^5\,erg\,cm^{-2}\,s^{-1}$. When electron number density, magnetic field strength and the turbulent velocity amplitude are taking as $4.8\times 10^{-7}\,cm^{-3}$, $5 \,G$ and $\delta v^2 = 2\times (43.9 \,km\, s^{-1})^2$ respectively, it was shown that energy flux density of an MHD wave becomes $4.9\times 10^5 \,erg\,cm^{-2}\,s^{-1}$  at $r = 0.123\, R_{\odot}$  \citep{ban98}. \citet{Tom07} carried out multichannel polarimetric observations and detected Alfven waves with a phase speed $1-4 \times 10^6 \,m\, s ^{-1}$. They concluded that the observed Alfven waves were not energetic enough to heat the corona. However, \citet{mci11} managed to observe the Alfven waves and measured their amplitudes, phase speeds, the periods (of order of $100-500\,s$) and then calculated the energy flux densities of these waves which were found to be high enough to heat the NPCH and accelerate the solar wind.

\citet{ban11} reviewed the observational evidences of propagating MHD waves in coronal holes.  Periods of the MHD waves observed in coronal hole via remote sensing are listed in Table 1 of this paper. Periods of waves  detected in plumes and interplume lanes vary between $600 \,s$ and $1800 \,s$. Waves observed other regions of coronal holes (on-disk and off-limb) have a period range of $300 \,s$ to $\sim 5800 \,min$.

\citet{rud98} developed a model which is specifically applicable to the coronal hole. In this model, the authors considered the Alfven wave phase mixing in a magnetic field which consists of open field lines and has two dimensions wherein the vertical scale height is greater than that of the horizontal. In their model they considered viscosity as a free parameter since the kinematic viscosity coefficient in the solar corona has no certain value. Their model predicts that Alfven waves with a period falling in a $6-100\, s$ range are not damped near the plume boundary if the vertical height $H \approx 100\, Mm$ and $\nu \leq 10^5 \,m^2 \,s^{-1}$. On the other hand,  when the viscosity coefficient is raised to the value $\nu \leq 10^{10}\,m^2\, s^{-1}$ and the other parameters are kept unchanged,  Alfven waves are damped and transferred their energy to the magnetised plasma. 

Surface wave heating is also one of the possibilities for heating coronal hole \citep{nar98}. The authors evaluated that in a radial magnetic field configuration with a strength $10\, G$,  waves go through a strong dissipation within a radial distance range $2.0\, R_{\odot}$ (in a nonlinear regime) and $2.7\, R_{\odot}$ (in a linear regime). Their model predicts that in a magnetic field showing distortion in radial direction (``strong spreading'' in their jargon) surface waves are strongly dissipated by the coronal hole plasma.

The presence of linearly polarized spherical Alfven waves in coronal hole is also considered as a possible agent to heat the plasma and accelerate the solar wind \citep{nak00}. The authors investigated the weakly non-linear dynamics of the waves in a medium wherein only the shear viscosity plays an important role in dissipation. They remind the reader that there is no reliable information about the radial dependence of the viscosity coefficient in a coronal hole and assume constant viscosity.  Their conclusion is that in a plasma with normalized viscosity $< 10^{-5}$, linearly polarized Alfven waves with periods $\sim 300\, s$ are damped within a radial distance less than $10\, R_{\odot}$.

The upper parts of the chromosphere and the corona are believed to be characterized by strong resistivity, anisotropic viscosity and thermal conductivity \citep{rud00}. The authors considered slow surface waves propagating in the above-defined plasma threaded by equilibrium magnetic field. In the fluid description, they assumed that away from the dissipative region only the parallel (to the magnetic field) thermal conductivity and viscosity are important. However, at the layer where waves are dissipated they adopted the full Braginskii expressions for viscosity, electrical resistivity and the heat flux. In their conclusion, it is the resonant absorption that causes the surface wave damping.

Ultraviolet Coronograph Spectrometer white-light channel (UVCS/WLC) and Extreme Ultraviolet Imaging Telescope (EIT) observations revealed the presence of slow magnetoacoustic waves in polar plumes. Using these data \citep{ofm00} investigated the damping of waves propagating in a gravitationally stratified medium threaded by radially divergent magnetic field. Two-dimensional MHD model of the authors contains the non-linear effects of the solar wind and  viscosity. Dissipation length scale of the waves with amplitude $40\, km\,s^{-1}$ turns out to be $0.08\, R_{\odot}$. If the viscosity is taken into account for the waves with amplitude $15\, km\,s^{-1}$ and with a period $300\, s$, damping length scale becomes $0.14\, R_{\odot}$. Close to the Sun, dissipated  waves may contribute to the acceleration of the solar wind either by increasing the thermal pressure of the coronal plasma or by undergoing Landau damping and thus transferring momentum at the collisionless part of the hole which 
is further away from the Sun. 

\citet{Dev10}, in an MHD approximation, considered the wave propagation in NPCH wherein temperature and number density of \ion{O}{VI} ions show gradients both in the parallel and the perpendicular directions to the magnetic field. They also took into account the heat conduction in both directions to the magnetic field, but neglected viscosity. Even the fluid description of NPCH revealed the resonance absorption of the waves.

In this paper, we assume that the parallel and the perpendicular (to the magnetic field) heat conduction and viscosity affect the propagation characteristics of MHD waves in NPCH and these waves, in turn, interact with  \ion{O}{VI} ions and preferentially heat them. SOHO/UVCS data on \ion{Mg}{X} and \ion{O}{VI} ions showed that NPCH plasma becomes collisionless beyond the radial distance $1.75-2.1\, R_{\odot}$ \citep{Doy99}. Therefore, we considered NPCH plasma collisional up to radial distance $1.6-2.25\, R_{\odot}$ and used MHD approximation. This study investigates the effects of the isotropic viscosity and  heat conduction on the propagation characteristics of the MHD waves in NPCH.

This paper is organized as follows. In the next section we give details of our model of NPCH. In Section 3 we give basic MHD equations and linearized MHD equations. We obtain the dispersion relations of the driven waves in the case of absence and presence of parallel heat conduction, separately. We estimate the damping length scales and energy flux densities of these waves. Finally Section 4 concludes this paper with a brief summary and discussion of our major results.

\section{Model of NPCH}

SOHO/UVCS data revealed that NPCH was structured with so-called plumes and interplume lanes \citep{wil98}. In their Figure 1, \citet{wil98} drew attention to the fact that plumes are brighter, higher in density but cooler than interplume lanes. Temperature anisotropy for \ion{O}{VI} ions in NPCH were revealed by SOHO/UVCS data, i.e., $\textit{T}_{\perp}>\textit{T}_{\parallel}$, where subscripts refer to the perpendicular and the parallel directions to the magnetic field, respectively \citep{koh97,cra08}. Spectral line widths of \ion{O}{VI} ions taken from interplume lanes turned out to be wider than the ones taken from plumes. This, of course, indicates that perpendicular (to the magnetic field lines) heating in interplume lanes is more effective than that of plumes \citep{cra02, wil11}. Besides, minor ion species He$^{++}$, \ion{Si}{VIII}, \ion{Mg}{X}, etc. also confirmed the temperature anisotropy in NPCH.

Many authors considered NPCH as electrically quasi-neutral, i.e., $N_{e}\approx N_{p}\approx N$  \citep[e.g.][]{mar99,End01,voi02}. In Figure 2 of \citet{cra99} electron number densities in plumes and interplume lanes are plotted. This figure reveals that $N_{e}$ in plumes is about 10\% higher than that of interplume lanes. One can find electron number density plots with respect to heliocentric distance in \citet{wil11}'s study. The empirical fit by \citet{Doy99} seems to be the most consistent one with data points,

\begin{equation} 
N_{e}= \frac{1\times 10^8}{r^8}+\frac{2.5\times10^3}{r^4}+\frac{2.9\times10^5}{r^2} \quad cm^{-3},
\end{equation} where $r$ is the heliocentric distance. Following the authors who consider NPCH as quasi-neutral, we assume that $N_{e}=N_{p}$ and try to model NPCH in two dimensions: one is the heliocentric distance, i.e., $R=r/R_{\odot}$ and the other, $x$, is the direction perpendicular to both the magnetic field and the line of sight ($los$); $x$ is measured in arc seconds. Thus, proton number density may be expressed in $(R, x)$ space as below: 

\begin{figure}
\includegraphics[angle=0,scale=.50]{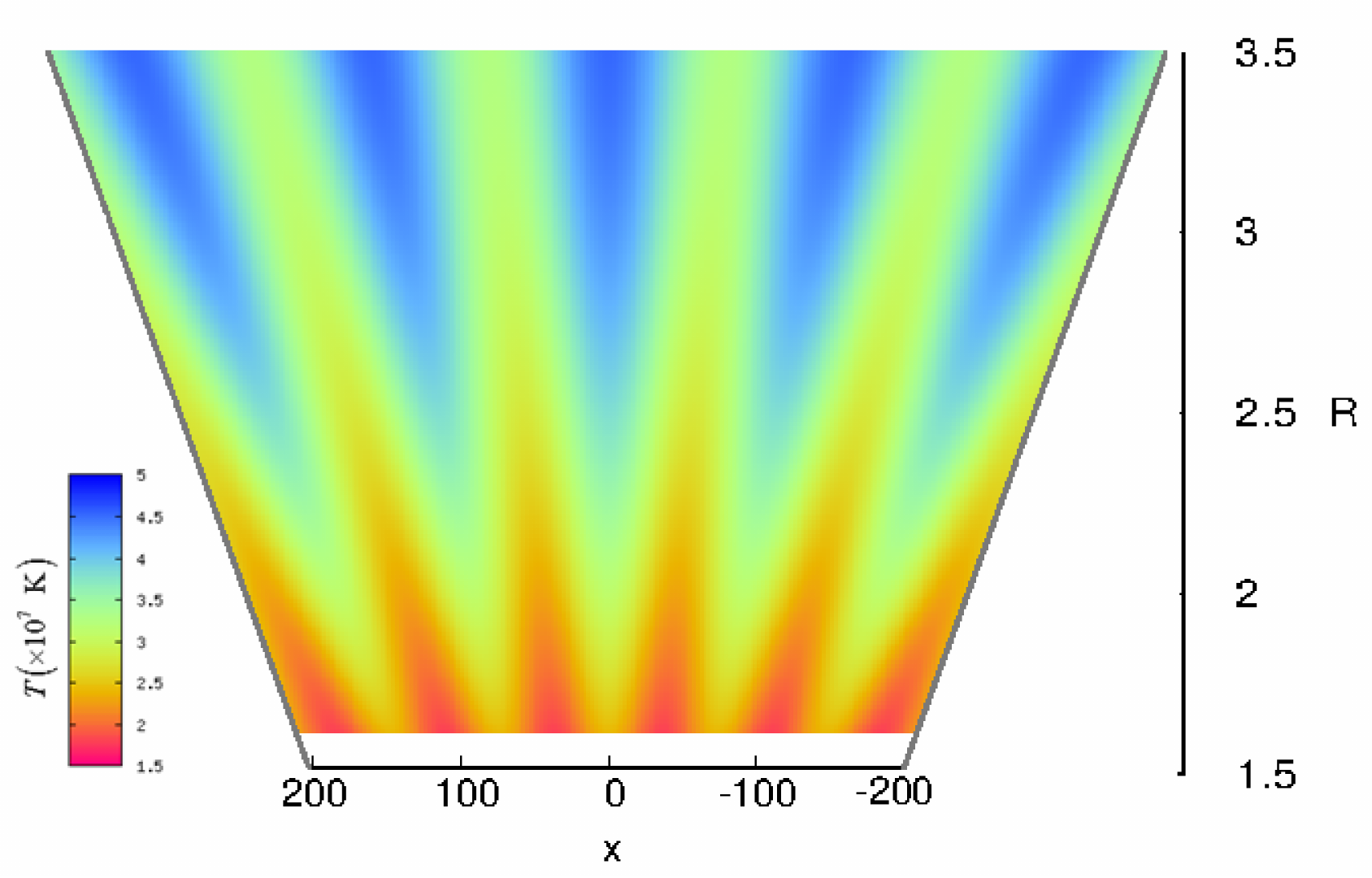}
\caption{Effective temperature variation of \ion{O}{VI} ions in NPCH. The width of the PL-IPL varies with $R$. The abscissa $x$ is in arc-seconds as it is also in Wilhelm et al. (1998).}
\end{figure}

\begin{equation}
N_p^{PL}(R,x)=N_p^{IPL}(R) (1+0.1 sin^2(2\pi x/92.16R)),
\end{equation}where superscripts IPL ad PL stand for interplume lanes and plumes, respectively.

Magnetic field in NPCH is assumed to be in the radial direction \citep{hol99},

\begin{equation}
B=1.5\left( f_{max} -1\right) R^{-3.5}+1.5R^{-2}  \quad  G,
\end{equation}where $f_{max} = 9$. Because of the absence of data as to how the magnetic filed varies in plumes and interplume lanes, and besides in $x$ direction, we assume that the equation (3) above is applied to both PL and IPL.


\ion{O}{VI} $\lambda 1032 $ $\AA$ line width is shown to vary both in the $R$ and the $x$ direction \citep{koh97}. \citet{wil98} formulated the total effective temperatures of minor ion species as the sum of ion temperature and the contribution from non-thermal processes, 

\begin{equation}
T_{\textit{eff}}=\frac{m_i}{2 k_B} v_{1/e}^2=T_i +\frac{m_i}{2 k_B}\xi^2,
\end{equation}where $v_{1/e}$ is the most probable speed of ion along the $los$ and $\xi$ is the most probable speed of an isotropic, Gaussian distributed, turbulent velocity field. \citet{Dol08} argued that the origin of non-thermal part of the line width is either Alfven wave dissipation or ion cyclotron resonance (ICR) after a turbulent cascade. It is almost impossible to separate these contributions to line widths. 

The effective temperature of the \ion{O}{VI} ions is given by \citet{Dev10}, 

\begin{equation}
T_{\textit{eff}}(R)=-7.941\times 10^7 R^2 + 4.9487 \times 10^8 R - 5.7625 \times 10^8.
\end{equation}

It was shown by \citet{ess99} that the value of $\xi$ for \ion{O}{VI} and \ion{Mg}{X} ions are quite close to each other and vary in the same manner with $R$. Taking this fact into account and by using \ion{Mg}{X} data for $\xi$, \citet{Dev10} derived a polynomial for the non-thermal part of the effective temperature of \ion{O}{VI} ions.

\begin{equation}
T_{\textit{eff}}^{\xi}=-4\times 10^6 R^2 + 3 \times 10^7 R -2\times 10^7.
\end{equation}

\citet{wil98} report that \ion{O}{VI} ion effective temperature in the interplume lanes is about 30\% higher than that of plumes (see Fig. 1). With this observational fact we may express the effective temperature in two dimensions as below:

\begin{equation}
T_{\textit{eff}}^{PL}(R,x)=T_{\textit{eff}}^{IPL}(R) (1-0.3 sin^2(2\pi x/92.16R)).
\end{equation}

The thermal conductivity and viscosity have anisotropic character due to the presence of magnetic field in NPCH. The parallel (to the magnetic field) thermal diffusivity for \ion{O}{VI} ions is given by \citep{bank66}

\begin{equation}
\kappa_{\parallel}=1.84\times10^{-8} T^{5/2} \,\,\,erg\,cm^{-1}s^{-1}K^{-1}.
\end{equation}For the plumes and interplume lanes under the typical coronal conditions $\kappa_{\parallel}=10^9-10^{10}$. In NPCH the ratio $\kappa_{\perp}/\kappa_{\parallel}$ is unknown, therefore we take it as a free parameter. For solar corona, \citet{van84} took it as $10^{-10}$. It may be argued that for solar coronal conditions $\kappa_{\perp}\ll\kappa_{\parallel}$. But it is not negligible when $\nabla_{\perp}^2 T\gg\nabla_{\parallel}^2 T$. \citet{van84} argued that even the ratio $\kappa_{\perp}/\kappa_{\parallel}$ assumes the value $10^{-14}$ it should be taken into consideration because it removes unphysical singularity in the case of $\kappa_{\perp}=0$ for thermal instability. 


The coefficient of kinematic viscosity in the isotropic case is given by 

\begin{equation}
 \rho\nu=2.21\times10^{-15} \frac{T^{5/2}}{ln \Lambda} \,\,\,g\,cm^{-1}s^{-1}
\end{equation}where $ln \Lambda$ is the Coulomb logarithm \citep{spi62}. \citet{rud98} argued that the viscosity coefficient for solar corona is quite uncertain. There is a same uncertainty on the coefficient of the conduction.
Bearing this fact in mind,  in this study we use both coefficients as free parameters remaining around theirs estimated values from Equation (8) and (9) for the model of NPCH. 

\section{Basic MHD Equations}

The basic equations for the investigation of wave equation and dissipation in a plasma are the continuity of mass, momentum, energy and magnetic induction, together with the magnetic flux  equations, in the forms given as \citep[e.g.][]{pri84}:

\begin{equation}
\frac{\partial{\rho}}{\partial{t}}+{\mathbf{v}\cdot \nabla\rho}+{\rho \nabla \cdot \mathbf{v}}=0,
\end{equation}

\begin{eqnarray}
\rho\frac{\partial\mathbf{v}}{\partial{t}} +\rho(\mathbf{v} \cdot \nabla)\mathbf{v}= \nonumber \\
-\nabla {p} + (\nabla \times\mathbf{B}) \times \frac{\mathbf{B}}{\mu_0} +\rho\nu \left[ \frac{4}{3} \nabla (\nabla \cdot \mathbf{v})- \nabla \times \nabla \times \mathbf{v} \right],
\end{eqnarray}

\begin{equation}
\frac{Dp}{Dt}-\gamma{\frac{p}{\rho}}\frac{D\rho}{Dt}=-(\gamma-1) \nabla  \cdot \mathbf{q},
\end{equation}

 \begin{equation}
\frac{\partial{\mathbf{B}}}{\partial{t}}= \nabla \times (\mathbf{v} \times\mathbf{B}),
\end{equation}

 \begin{equation}
\nabla \cdot \mathbf{B} =0,
\end{equation}where $\rho$ is the mass density, $\mathbf{v}$ is the fluid velocity, $p$ is the scalar pressure, $\mathbf{B}$ is the magnetic field, $\nu$ is the coefficient of kinematic viscosity (assumed uniform),  $\mathbf{q}=-\kappa_{\parallel}\nabla_{\parallel}T-\kappa_{\perp}\nabla_{\perp}T$ is the heat flux, where $\nabla_{\parallel}=\widehat{\mathbf{B}}(\widehat{\mathbf{B}}\cdot\nabla)$, $\nabla_{\perp}=\nabla-\nabla_{\parallel}$ and ${\widehat{\mathbf{B}}=\mathbf{B}/B}$ which is unit vector along the magnetic field, and the $\gamma$ is adiabatic index (its value is taken as $5/3$ for coronal plasma) and $D/Dt=\partial/\partial t+\mathbf{v}\cdot\nabla$ is a Lagrangian derivative.

We apply a standart Wentzel-Kramers-Brillouin (WKB) perturbation analysis on the equilibrium state. In this analysis, all the variables in the MHD equations are denoted by sums of equilibrium (denoted with a ``0'' subscript) and a small perturbed quantity (denoted with a ``1'' subscript), i.e. $\rho=\rho_{0}+\rho_{1}, \mathbf{v}=\mathbf{v}_{0}+\mathbf{v}_{1}$ etc.

The quasi-linear equations derived from equations (10)-(14) are

\begin{equation}
\frac{\partial{\rho_1}}{\partial{t}}+{(\mathbf{v}_1\cdot \nabla)\rho_0}+{\rho_0 (\nabla \cdot \mathbf{v}_1)}=0,
\end{equation}

\begin{eqnarray}
&&\rho_0\frac{\partial{\mathbf{v}_1}}{\partial{t}} +\rho_0(\mathbf{v}_1 \cdot \nabla)\mathbf{v}_1= -\nabla {p}_1 + (\nabla \times\mathbf{B}_1) \times \frac{\mathbf{B}_0}{\mu_0}  \nonumber\\
&&+\rho\nu \left[ \frac{4}{3} \nabla (\nabla \cdot \mathbf{v}_1)- \nabla \times (\nabla \times \mathbf{v}_1) \right],
\end{eqnarray}

\begin{eqnarray}
\frac{\partial{p}_1}{\partial{t}}+(\mathbf{v}_1 \cdot \nabla) p_0 - {c_s}^2 \left[ \frac{\partial{\rho}_1}{\partial{t}} + (\mathbf{v}_1 \cdot \nabla)  p_0 \right] =-0.6\nabla \cdot \mathbf{q}_{1},
\end{eqnarray}

 \begin{equation}
\frac{\partial{\mathbf{B}_1}}{\partial{t}}= \nabla \times (\mathbf{v}_1 \times\mathbf{B}_0),
\end{equation}

 \begin{equation}
\nabla \cdot \mathbf{B}_1 =0,
\end{equation}where $c_s$ is the sound speed. In the linear approximation, perturbed values are usually assumed to be negligible in comparison with the equilibrium values. In sec. 3.3  we will investigate the effects of the isotropic viscosity and  heat conduction on the propagation characteristics of the MHD waves in NPCH in linear approximation.  But, in the framework of quasi-linear MHD, the lowest order non-linear  term in rms velocity  $\sqrt{\langle\delta v^2 \rangle}$  is retained, i.e. we must  not neglect the $\mathbf{v}^2_{1}$ term. Because the observed perturbed velocity $\mathbf{v}_1$ of  \ion{Mg}{X}  falls within the range 95 $km\,s^{-1}$ and 140 $km\,s^{-1}$ in $1.6-2.25 \,R/R_{\odot}$ \citep{ess99}.  \citet{ess99} showed that the non-thermal velocities $(\delta v)$ of \ion{O}{VI} and \ion{Mg}{X} are quite close to each other and vary in the same manner with $R$. Hence we kept the $\mathbf{v}^2_{1}$ term in our first approach. But we neglect $\mathbf{B}^2_{1}$ term. Although we try very hard to find the magnitude of 
the perturbed magnetic field, i.e. $\mathbf{B}_{1}$ in the literature, but in vain. Therefore, we assume that the waves perturbing the magnetic field and giving it a radius of curvature so huge that $\nabla \times \mathbf{B}_{1}\simeq  \mathbf{B}_{1}/R_c$ becomes almost zero, where $R_c$ is the radius of curvature imposed by the propagation of waves!

The set of equations (15)-(19) is reduced to a single equation by differentiating equation (16) with respect to time and substituting for  ($\partial{\rho}_1 / \partial{t} $), ($\partial{p}_1 / \partial{t} $), ($\partial{\mathbf{B}}_1 / \partial{t} $) from equations (15), (17) and (18) respectively. After time differentiation, equation (16) reads as

\begin{eqnarray}
&& \rho_0\frac{\partial^2 \mathbf{v}_1}{\partial t^2}+ \rho_0\left( \frac{\partial \mathbf{v}_1}{\partial t} \cdot \nabla \right) \mathbf{v}_1+\rho_0 (\mathbf{v}_1 \cdot \nabla)\frac{\partial \mathbf{v}_1}{\partial t} \nonumber\\
&& =\nabla\lbrace (\mathbf{v}_1 \cdot \nabla)p_0+{c_s}^2[\rho_0(\nabla \cdot  \mathbf{v}_1)]+0.6 \nabla \cdot \mathbf{q}_{1}\rbrace \nonumber\\
 && +\lbrace \nabla \times [\nabla \times (\mathbf{v}_1 \times \mathbf{B}_0)]\rbrace \times \frac{\mathbf{B}_0}{\mu_0} \nonumber\\
&& +\rho \nu \left[ \frac{4}{3} \nabla \left( \nabla \cdot \frac{\partial{\mathbf{v}_1}} {\partial{t}}\right) -\nabla \cdot \nabla \frac{\partial{\mathbf{v}_1}} {\partial{t}} \right].
\end{eqnarray}

Remembering that the perturbation values vary as $\exp{[i(\mathbf{k}\cdot\mathbf{r}- \omega t)]}$  we make the below substitutions: $\partial / \partial t$ $\rightarrow$ $-i\omega$  and $\nabla$ $\rightarrow$ $ik$  and  we put $X=(\mathbf{v}_1 \cdot \mathbf{k})$, $Y=(\mathbf{v}_1 \cdot \mathbf{\widehat{B}})$. There is no harm in repeating that if  $\mathbf{v}_1 \cdot \mathbf{k}$ and $\mathbf{v}_1 \cdot \mathbf{\widehat{B}}$  does not vanish, two coupled (in $X$ and $Y$) equations may be constructed and they may be eliminated so that the dispersion relation is found  \citep{pri84}. Then we obtain

\begin{eqnarray}
&&\left( 2\omega X -\omega^2 +{v_A}^2 k^2 - i\omega\nu k^2 \right) \mathbf{v}_1 -{v_A}^2 k X \mathbf{\widehat{B}} \nonumber\\ &&+ \left(
\begin{array}{c}
   1.6X c_s^2 +0.6 \frac{1}{\rho_0} \left[ -H_\parallel k - \frac{H_\perp}{\xi}X \right]
\\ + {v_A}^2 X-kY {v_A}^2 - \nu 0.3i\omega X
\end{array}
\right)\mathbf{k}=0,
\end{eqnarray}where $H_\parallel =\kappa_{\|}^i \nabla_{\|} T_i $ and  $H_\bot =\kappa_{\bot}^i \nabla_{\bot} T_i $. We make dot product the equation (21) with $\mathbf{k}$, $\mathbf{\widehat{B}}$ , and $\mathbf{v}_1$, respectively and reach the coupled equations in $X$ and $Y$:

\begin{eqnarray}
&&2\omega X^2 + \left(
\begin{array}{c}
 -\omega ^2 + {v_A}^2 k^2 - \nu 1.3i \omega k^2
\\ +1.6 {c_s}^2 k^2- 0.6 \frac{1}{\rho_0} \frac{H_{\bot}} {\xi} k^2
\end{array}
\right)X -k^3 {v_A}^2 Y \nonumber\\
&& = 0.6 \frac{1}{\rho_0}  H_{\|} k^3,
\end{eqnarray}

 \begin{eqnarray}
&&2\omega XY+ \left( -\omega ^2 - i\omega\nu k^2\right)Y \nonumber\\
&&+ \left( 1.6 {c_s}^2 k - 0.6 \frac{1}{\rho_0} \frac{H_{\bot}} {\xi} k- \nu 0.3 i \omega k \right)X \nonumber\\
&&=0.6 H _{\parallel} k^2 \frac{1}{\rho_0},
\end{eqnarray}

\begin{eqnarray}
&&X^2 \left( 1.6 {c_s}^2 + {v_A}^2 - \nu 0.3i \omega -0.6 \frac{1}{\rho_0} \frac{H_{\bot}} {\xi} \right) \nonumber
\\ &&-2k v_A^2 XY + \left( 2\omega \xi ^2 - 0.6 \frac{1} {\rho_0} H_{\|} k \right)X \nonumber\\
&&=-\left( -\omega^2 +v_A^2 k^2 -\nu i \omega k^2   \right) \xi^2.
\end{eqnarray}From equations  (22) and (23), we find the equation (25) depending only on $X$:


\begin{equation}
4\omega^2 X^3 +X^2 2\omega\left[ Z1+Z2 \right] +X\left[Z1 Z2-2\omega Dk\right]-Dk Z2=0,
\end{equation}where ${Z1=\left( B+Ak\right)}$, ${Z2=\left( B+{v_A}^2k^2\right)}$, ${D=0.6\frac{{1}}{\rho_0} H_{\|} k^2}$, ${A=\left( 1.6{c_s}^2 k - 0.6 \frac{1}{\rho_0} \frac{H_{\bot}} {\xi} k-  0.3\nu i \omega k \right)}$ and ${B=\left( -\omega^2- i\omega\nu k^2 \right)}$.

In NPCH the effective temperature increases by 30\% in $x$-direction where  from the midpoint of a plume to the midpoint of neighbouring interplume lane. Distance in x-direction between from the midpoint of a plume and the midpoint of neighbouring interplume lane at $R$= 1.7 is about 27 885 $km$. The effective temperature change between these two points is about 30\%. Measuring from $R$= 1.7, the increase of the effective temperature by 30\% in $R$-direction takes place in 374 120 $km$. Thus, the temperature gradient  height scale is about $0.54\,R_{\odot}$. After $R=1.7$, the temperature change in $R-$direction will occur at $R=2.24$. We mentioned that $\kappa_{\perp}$ isn't negligible when $\nabla_{\perp}^2 T\gg\nabla_{\parallel}^2 T$ \citep{van84} in Section 2. If these height scales are used, one can easily see that $\nabla_{\perp}^2 T\gg\nabla_{\parallel}^2 T$ is satisfied. Therefore, especially to see the effect of $\kappa_{\perp}$ on the evolution of wave propagation in the $x-$direction, we may assume that temperature gradient does not depend on $R$ in the interval $R=1.7-2.24$ (i.e., $\nabla_{\parallel} T=0$) and thus  $H_{\parallel}=0$ and $D=0$. After that we will search 
solutions of equation (25) in two steps; by assuming $D$ is negligible, i.e., $H_{\parallel}=0$, in subsection 3.1, and by retaining it in equation (25) in subsection 3.2.

\subsection{In the absence of parallel heat conduction}

If $D$ = 0 then equation (25) becomes

\begin{equation}
4\omega^2 X^2 +2\omega X \left( Z1+Z2 \right) +\left(Z1Z2 \right)=0.
\end{equation}Equation (26) have two roots: ${X_1=-Z1/2\omega}$ and ${X_2=-Z2/2\omega}$. By substituting $X_1$ and $X_2$  into Equation (22) or (23),  we obtain a pair of ($X$ ,$Y$ ):

\begin{displaymath}
 \left(X_1,Y_1\right) = \left(-\frac{\left( B+Ak\right)}{2\omega}, -\frac{1}{2\omega} \frac{\left( B+Ak\right)}{k}\right) ,
\end{displaymath}

\begin{displaymath}
\left(X_2,Y_2\right) = \left(-\frac{\left( B+{v_A}^2 k^2\right)}{2\omega}, -\frac{1}{2\omega} \frac{A\left( B+{v_A}^2 k^2\right)}{{v_A}^2 k^2}\right).
\end{displaymath}

\begin{figure*}
\begin{center}
\includegraphics[angle=0,scale=.39]{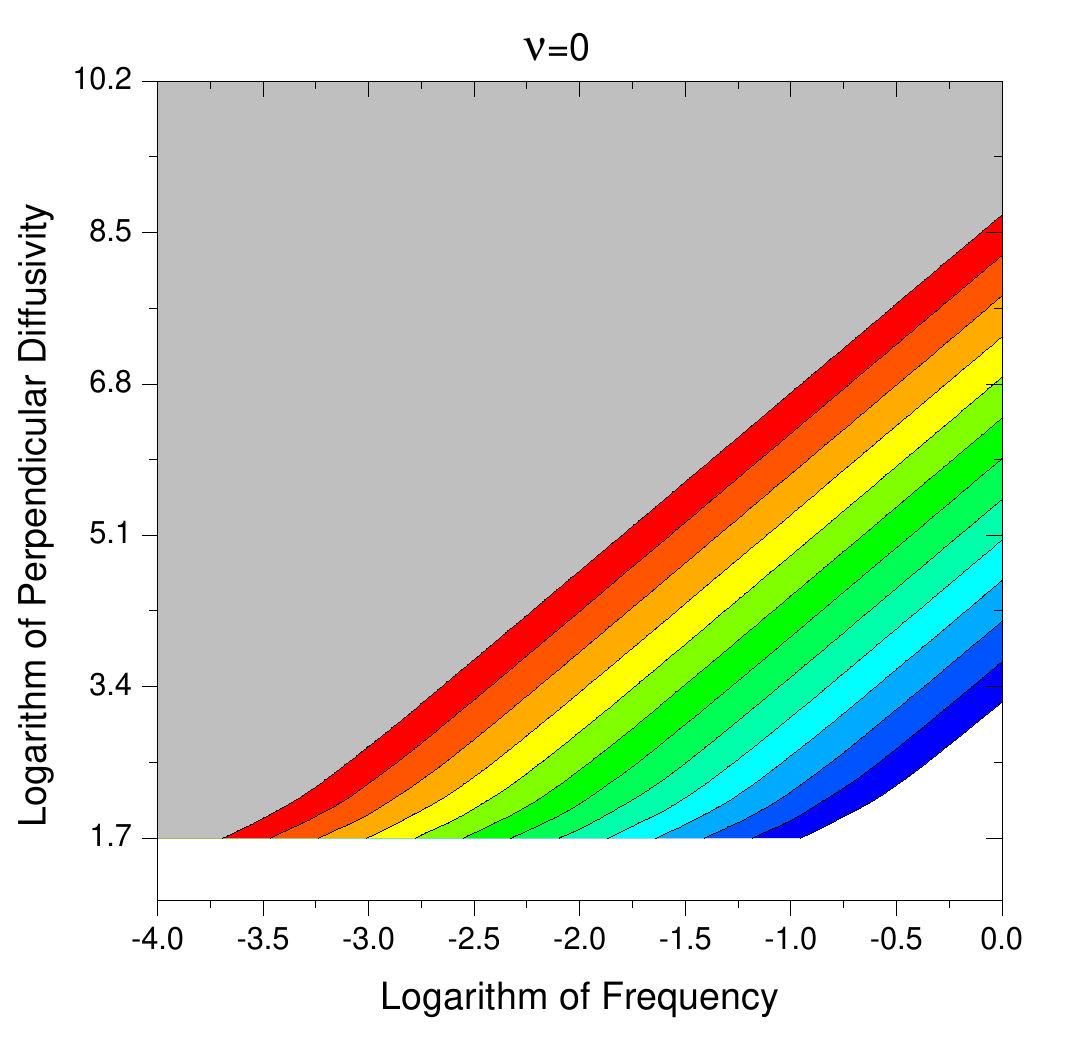}
\includegraphics[angle=0,scale=.39]{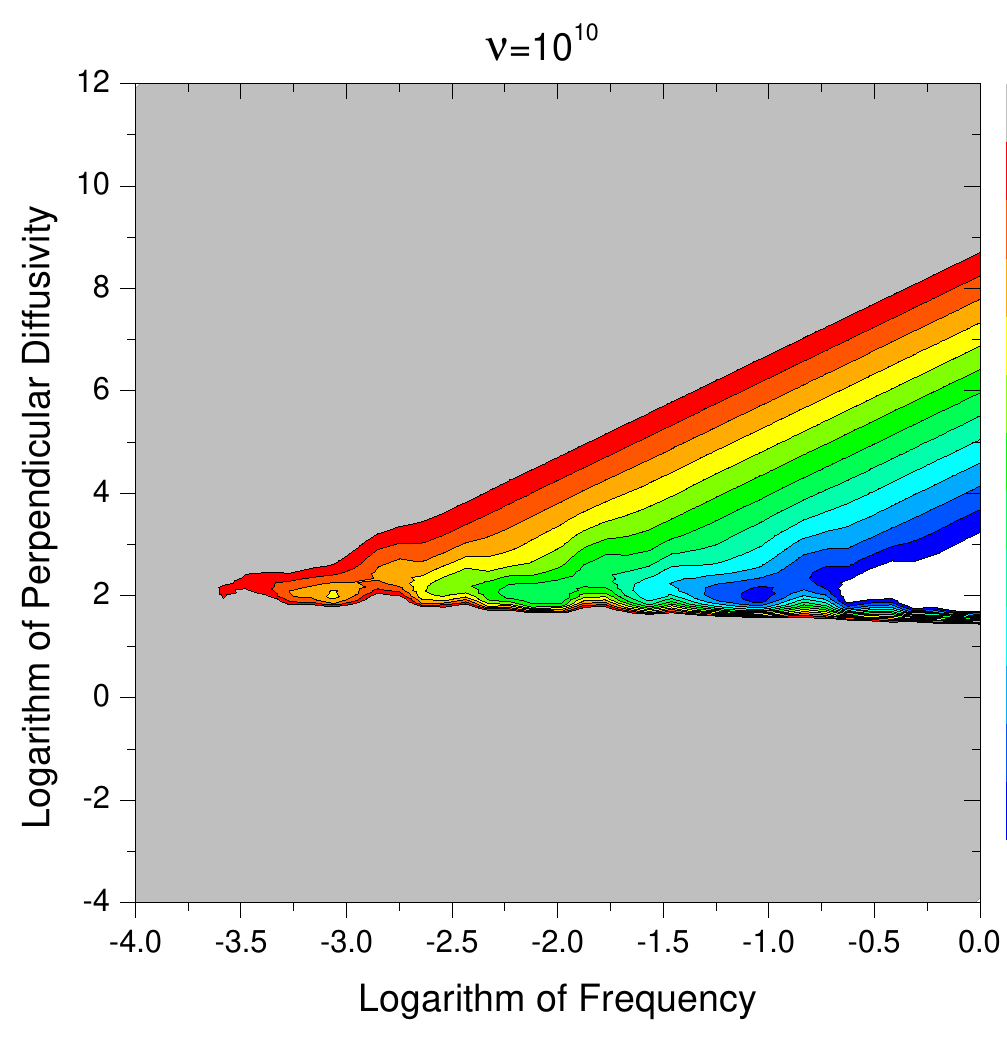}
\includegraphics[angle=0,scale=.39]{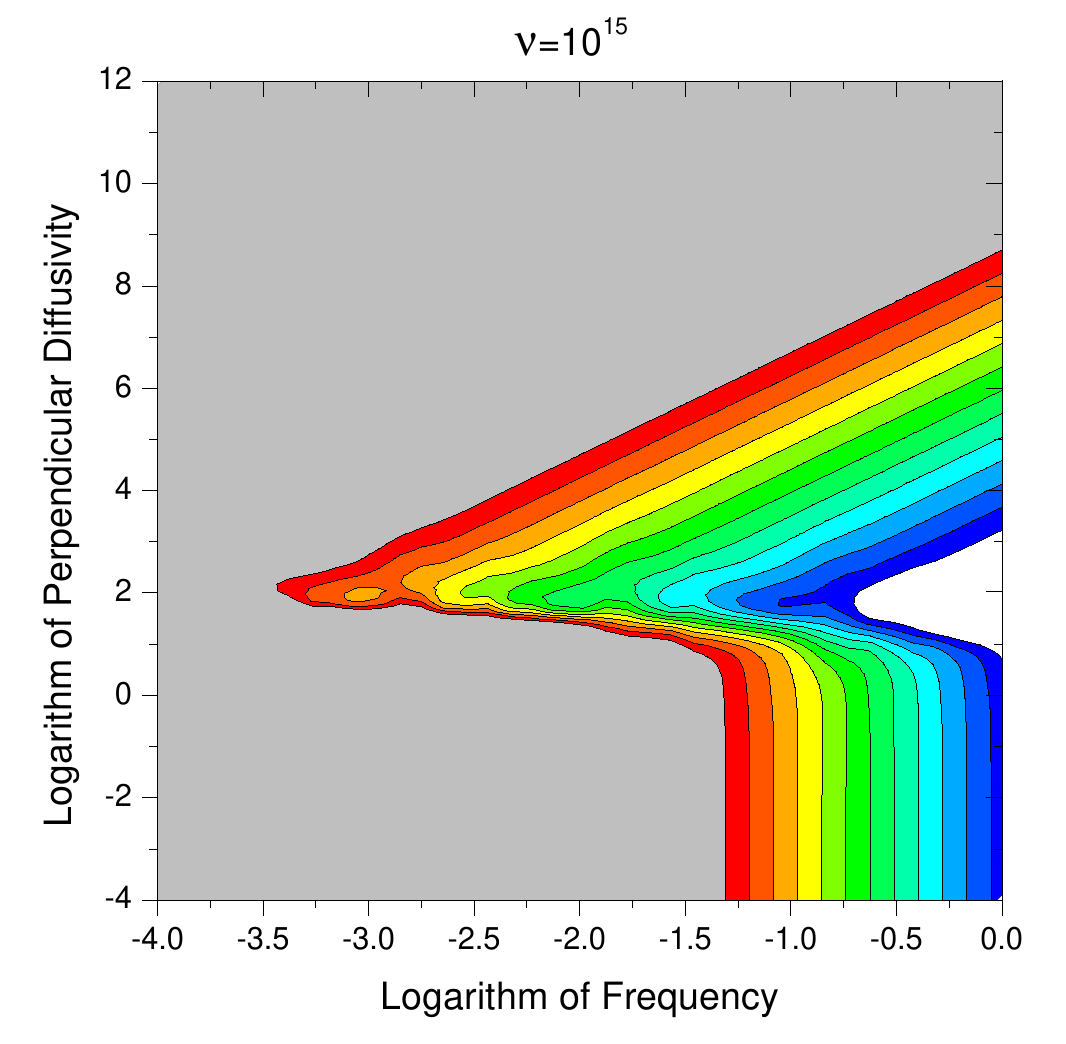}
\includegraphics[angle=0,scale=.39]{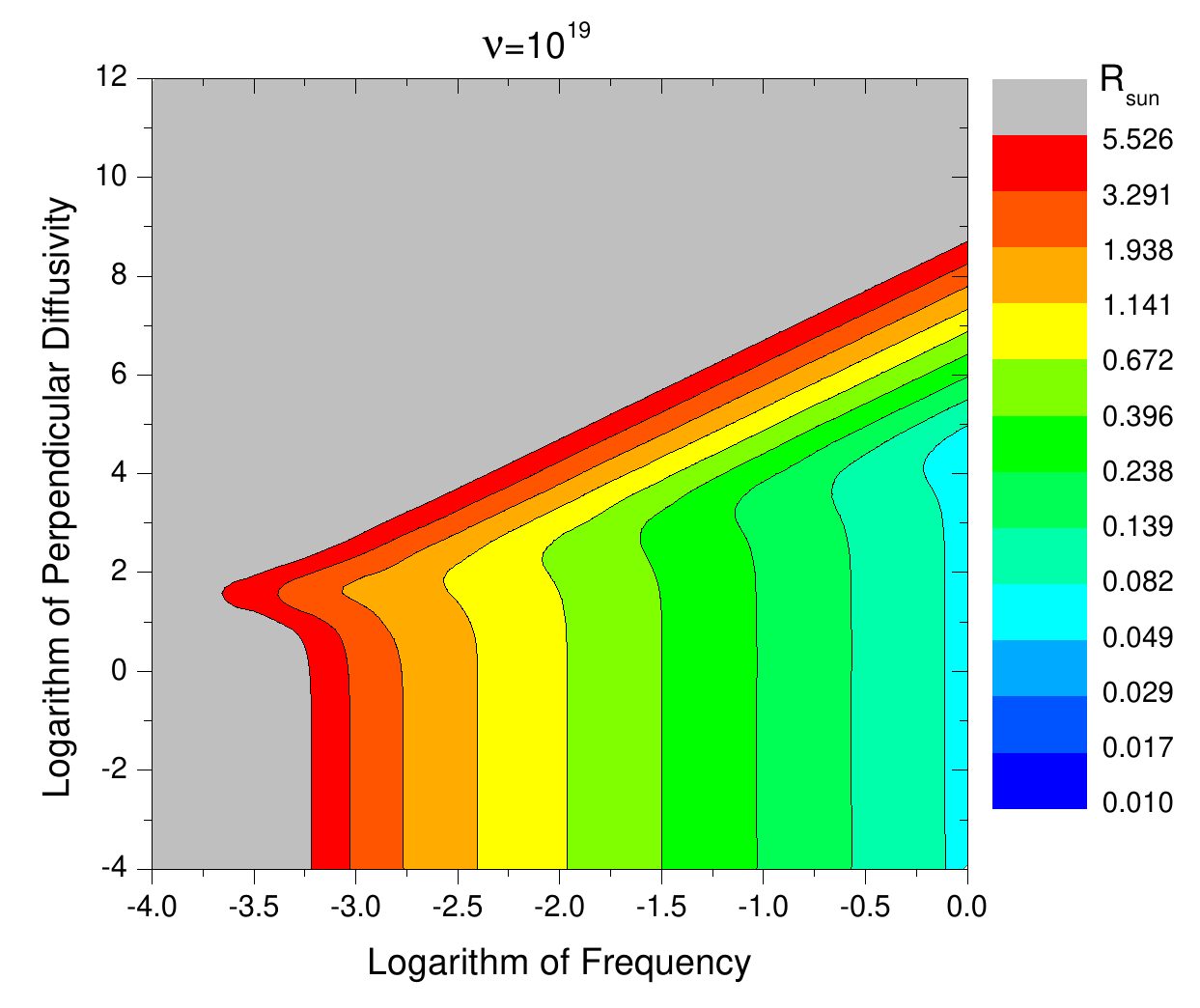}
\caption{Contours of the damping length scales in terms of solar radius plotted versus logarithms of perpendicular diffusivity and frequency at different viscosity values for the acoustic waves propagating in plume.}
\end{center}
\end{figure*}

\begin{figure*}
\begin{center}
\includegraphics[angle=0,scale=.6]{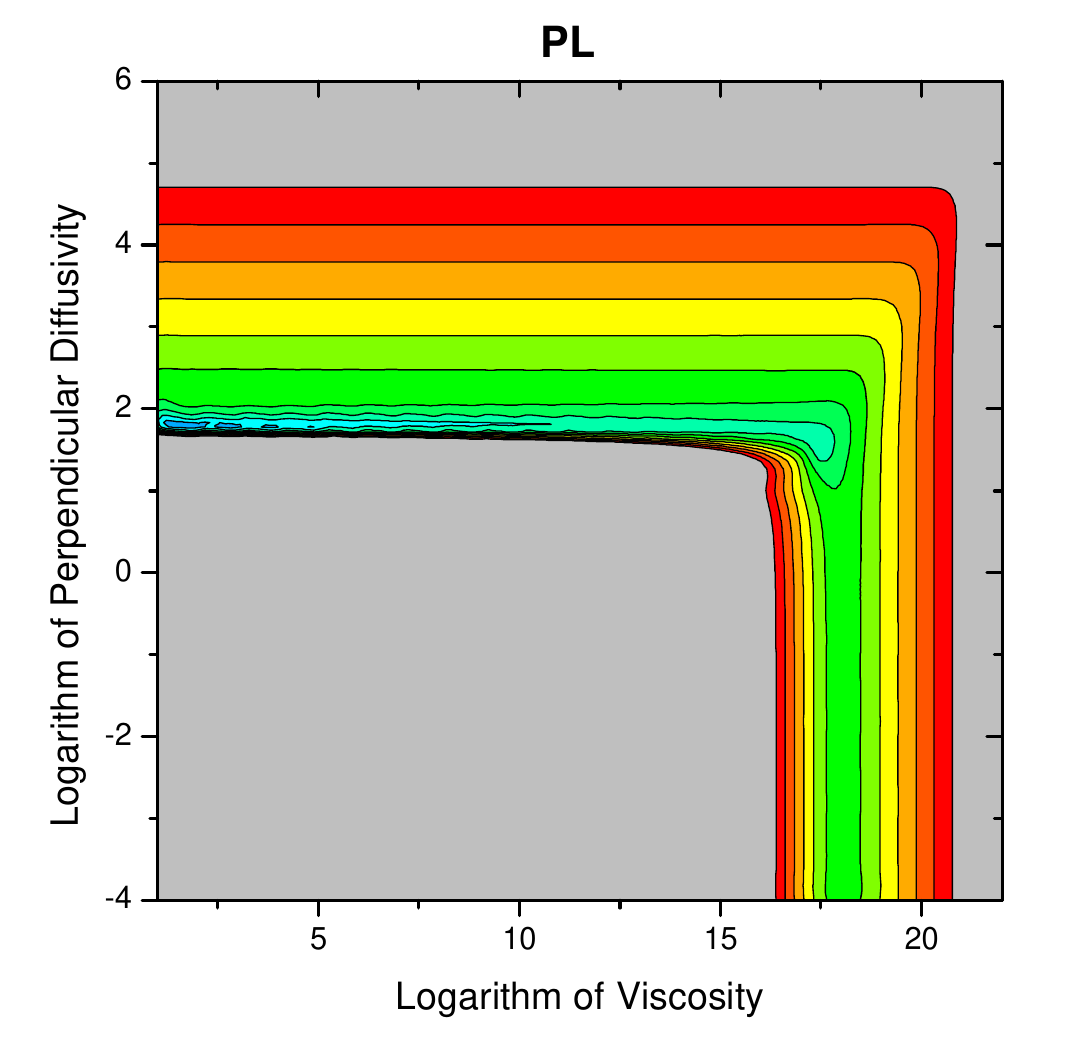}
\includegraphics[angle=0,scale=.6]{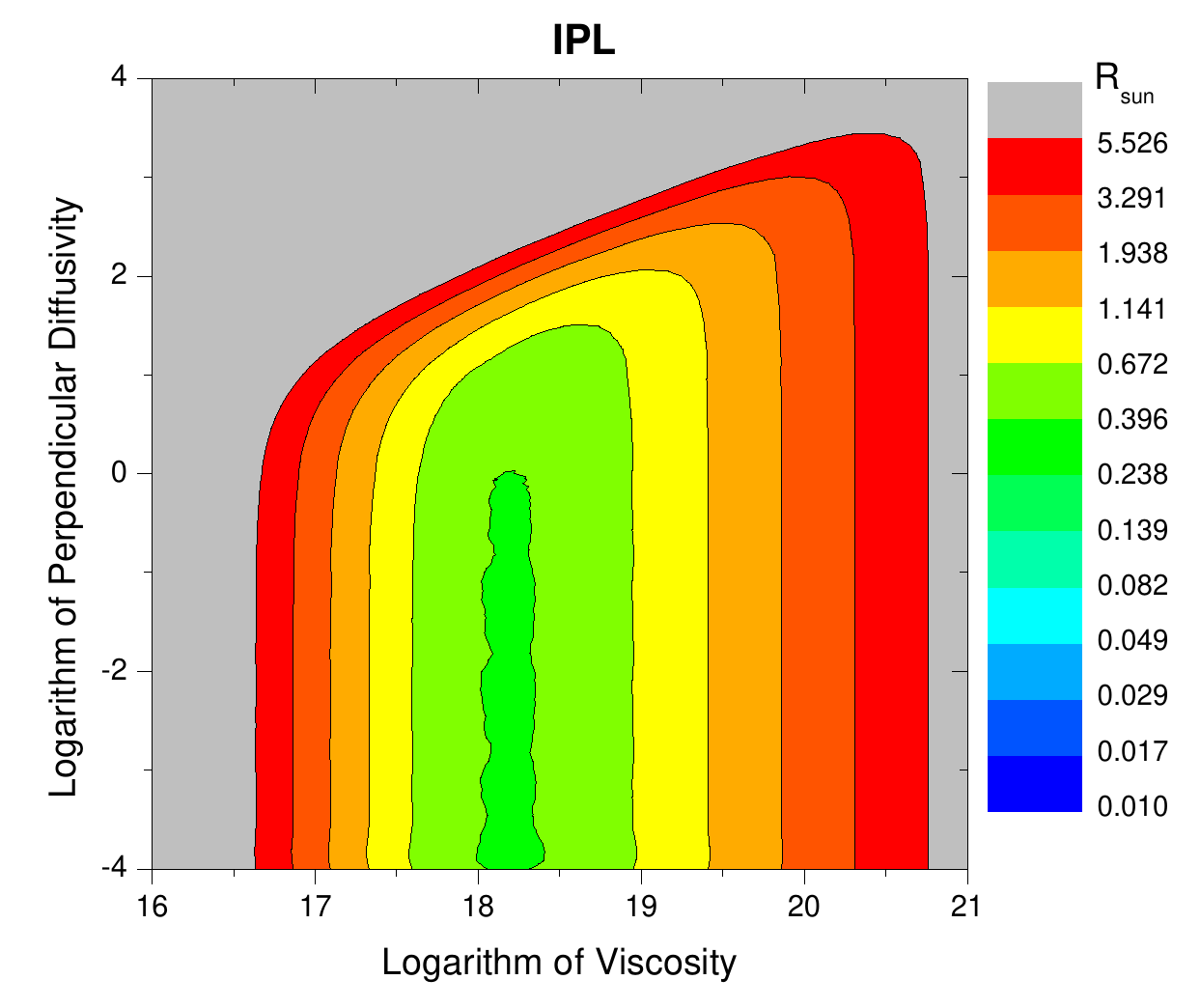}
\includegraphics[angle=0,scale=.6]{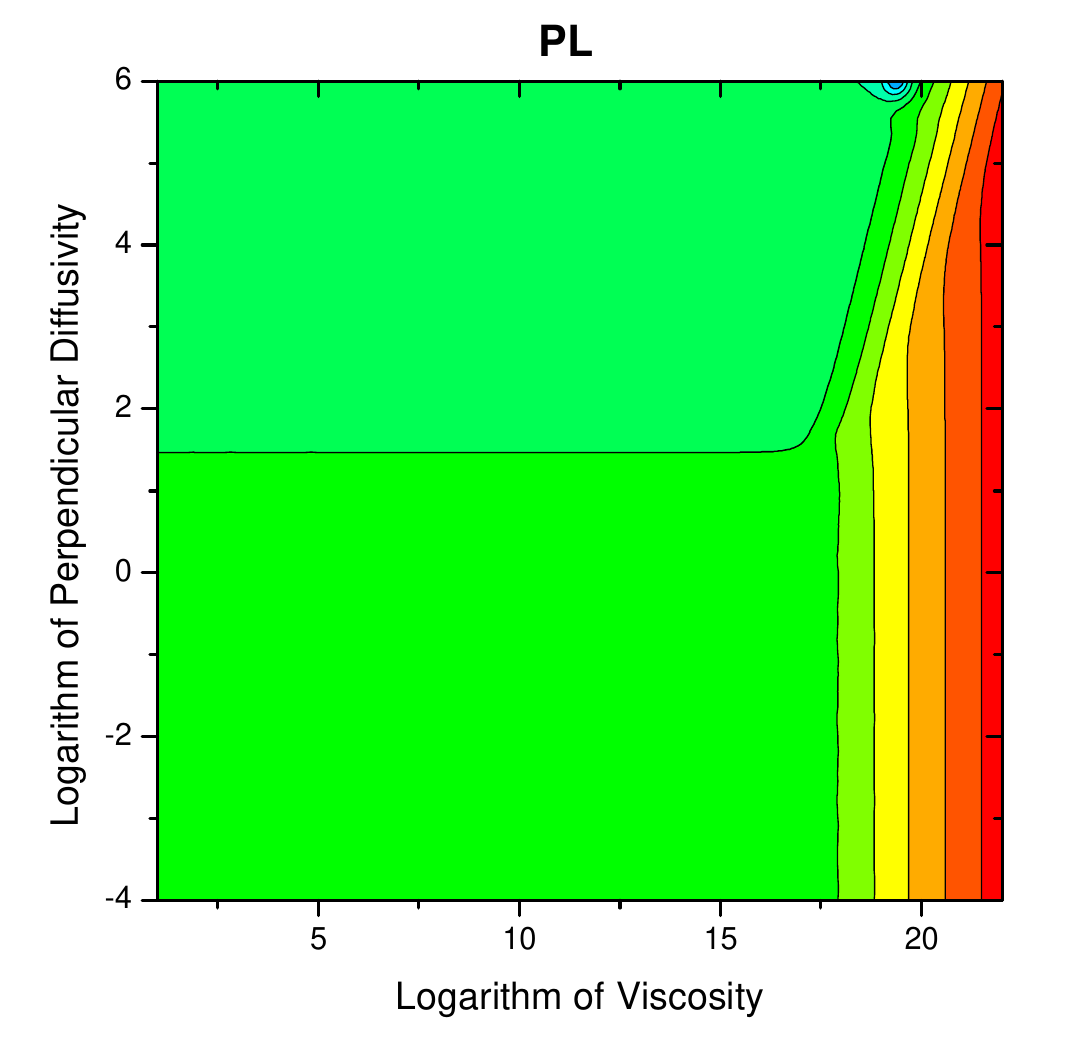}
\includegraphics[angle=0,scale=.6]{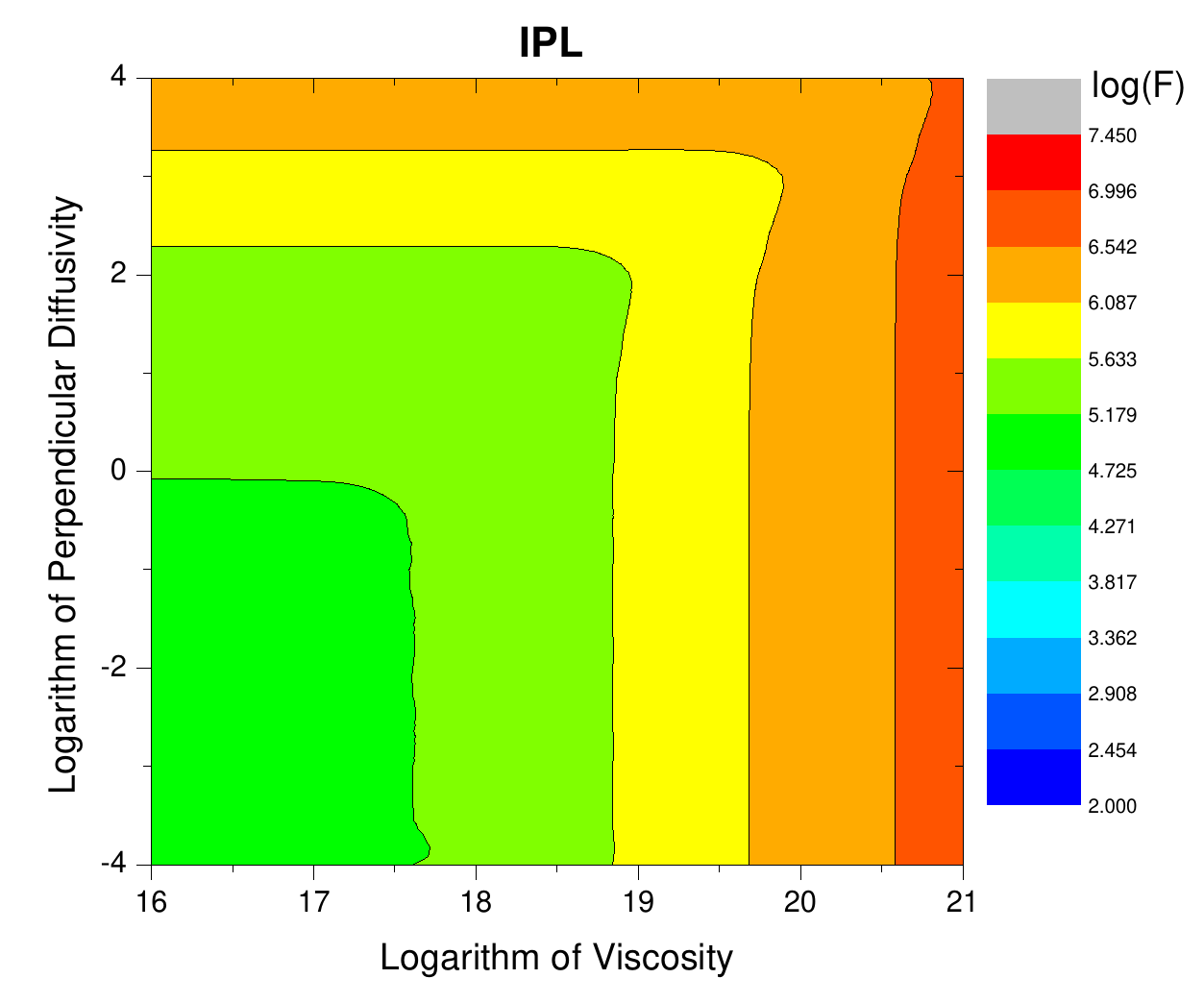}
\caption{Contours of the damping length scales (upper graphics) and the logarithm of energy flux densities (lower graphics) plotted versus logarithms of diffusivity and viscosity for the acoustic waves with frequency $0.01 \,rad/s$ propagating in plume(PL) and interplume lane (IPL).}
\end{center}
\end{figure*}

\begin{figure*}
\begin{center}
\includegraphics[angle=0,scale=.6]{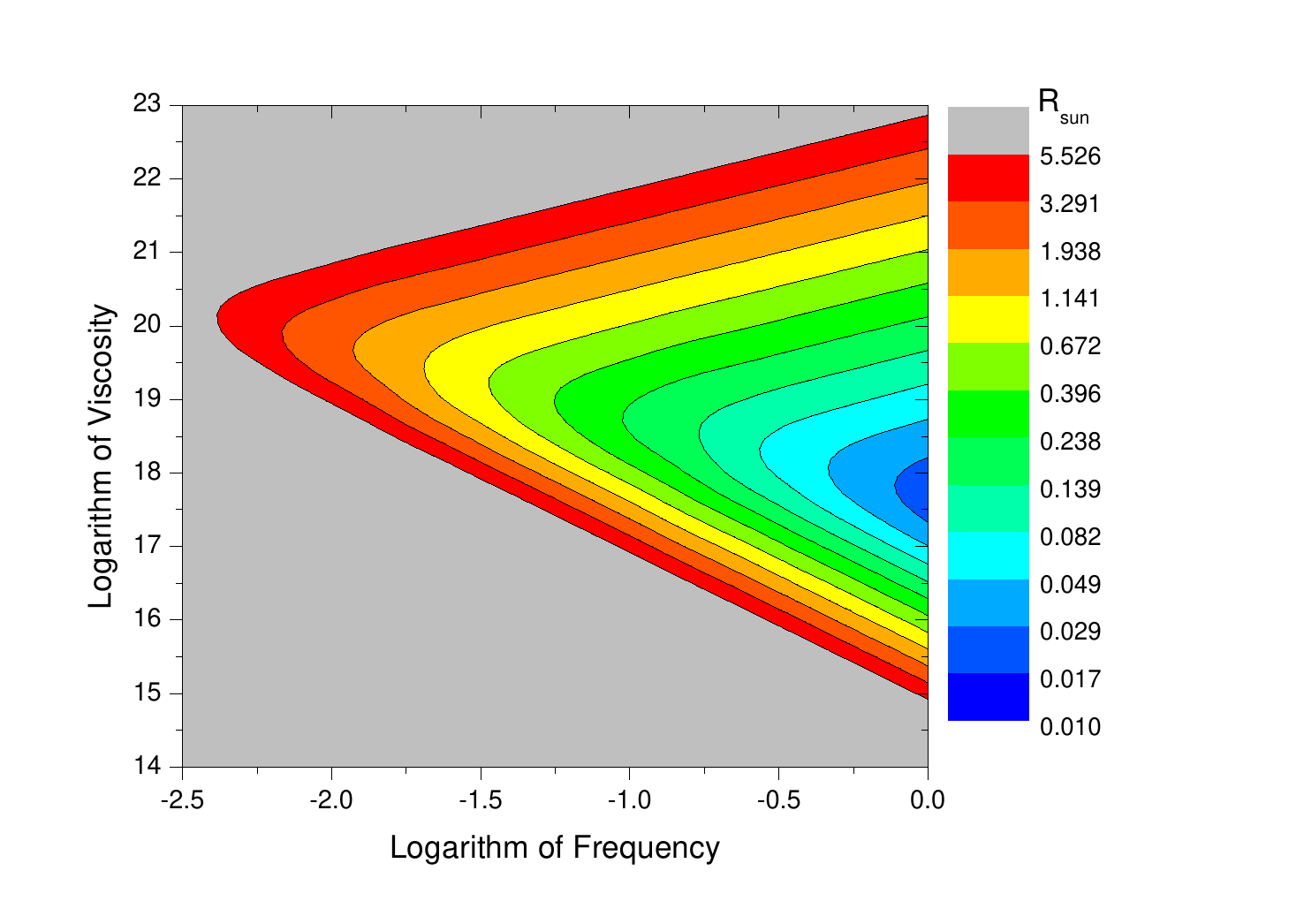}
\includegraphics[angle=0,scale=.6]{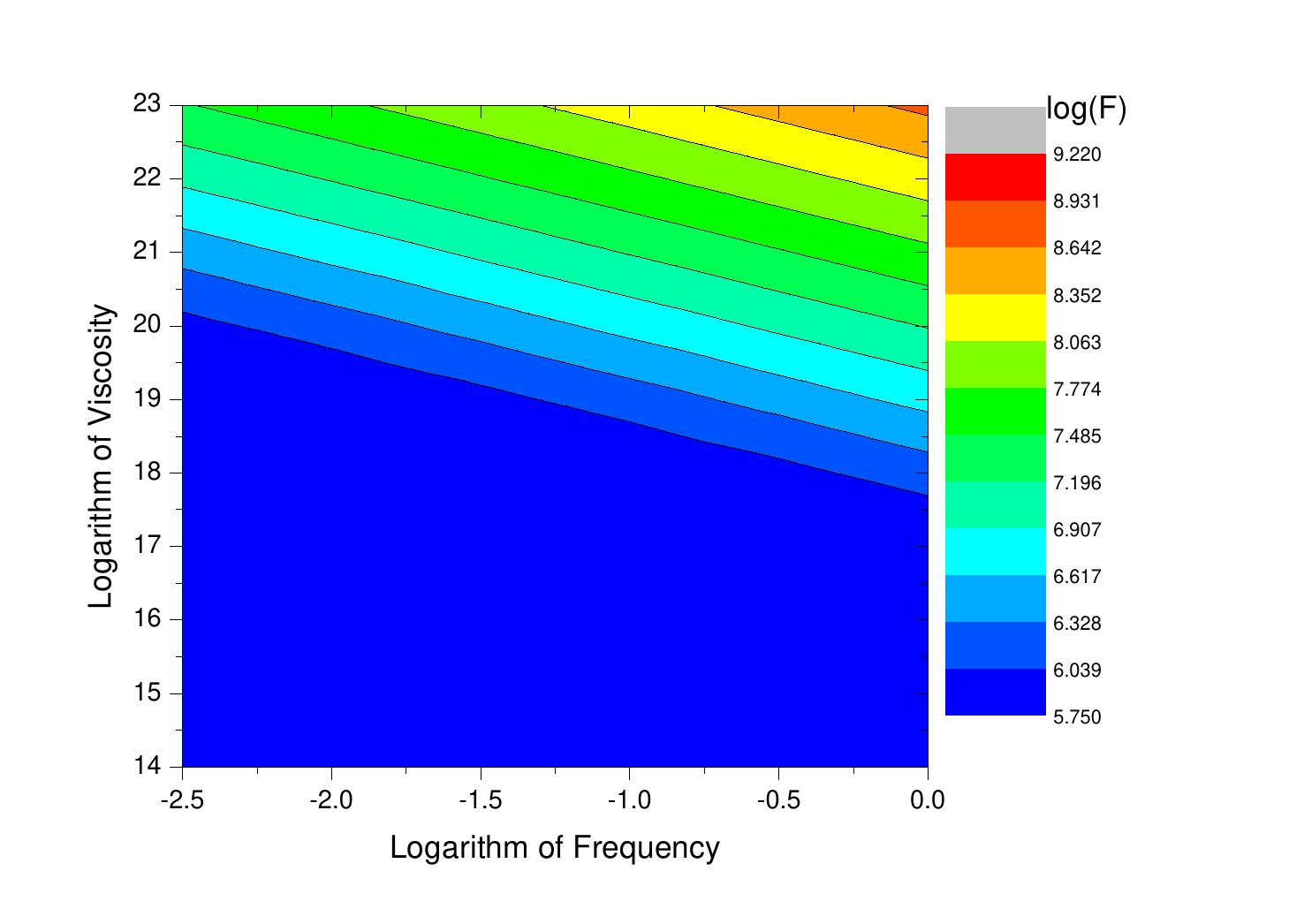}
\caption{Contours of the damping length scales in terms of solar radius (left) and logarithms of the energy flux densities (right) plotted versus logarithm of viscosity and  frequency for the Alfven waves propagating in plume.}
\end{center}
\end{figure*}

\subsubsection{Acoustic Waves in NPCH}

First, let us substitute ($X_1$,$Y_1$) into equation (24) and try to find the dispersion relation of the perturbation:

\begin{equation}
k^4 \left(M-i\omega\nu  \right)^2 - k^2 2\omega^2 \left(M-i\omega\nu +2 \xi^2 \right)+\omega^4=0,
\end{equation}where $M=\left( 1.6{c_s}^2 - 0.6 \frac{1}{\rho} \frac{H_{\bot}} {\xi} - 0.3 i \omega \nu\right).$

This mode may be labelled as ``acoustic wave modified by viscosity and perpendicular heat conduction'', nevertheless, for the sake of brevity we call it acoustic wave. Figure 2 shows the variation of the damping length scales plotted versus logarithms of perpendicular diffusivity and frequency for acoustic wave propagating in the plume ($x=40$) at the fiducial radius $R=1.7$ and different viscosity values. 
Due to the fact that the coefficients of heat conduction and viscosiy in the solar corona are not well-established, we adopted these coefficients as free parameters. Thus, we determined the necessary values of these parameters by considering the damping length scales which would cause the solar wind acceleration and heating corona. The damping length scale ($1/k_i$) of  this wave derived from equation (27) is given in terms of solar radius. In all the graphics showing the damping length scales between $0.01 R_{\odot}-5.5 R_{\odot}$  we used false colours. Distances greater than $ 5.5 R_{\odot}$  are indicated by grey colour and the ones shorter than $0.01 R_{\odot}$  by white colour.  We chose wave angular frequency range of $10^{-4} \,rad/s$ to $1 \,rad/s$ which corresponds to period interval $6 \,s>P>1048 \,min$.   
 
As shown the Figure 2, for the values of $\kappa_{\perp}$  greater than $\sim100 \,\,erg\,cm^{-1}s^{-1}K^{-1}$, it is seen that the value of the perpendicular diffusivity coefficient increases linearly with frequency for all the values of viscosity and all the damping length scales. For the values of $\kappa_{\perp}$  smaller than 100, the viscosity should have values higher than $10^{15} \,cm^{2}\,s^{-1}$ if the higher frequency waves are to be damped at the required distances. When the viscosity is about $10^{19}$ even the low frequency waves get damped at distances wherein coronal heating and solar wind acceleration take place.

Mechanical energy flux density of waves is given by \citet{pri84}:

\begin{equation}
F=\rho\langle\delta v^2\rangle \frac{\partial \omega}{\partial  k}
\end{equation}where  $\partial \omega /\partial k$  is the group velocity of the wave and $\langle\delta v^2\rangle=2\xi^2$  is the non-thermal velocity which is related to the small scale unresolved turbulent motions in Equation (4). 

Figure 3 shows the variation of the damping length scales of the waves with angular frequency $\omega= 0.01 \,\,rad/s$ propagating in plumes and interplume lanes (PL and IPL; upper graphics) and the variation of the energy flux density of the same waves with viscosity and perpendicular diffusivity (lower graphics). $\omega= 0.01 $ angular frequency waves correspond to $1.6 \,mHz$ frequency waves with a period of $\sim10.5 \,minute$. Within plumes, when the perpendicular diffusivity has a value between $\sim10^{1.8}-10^{4.6}\,\,erg\,cm^{-1}s^{-1}K^{-1}$, damping length scales are independent of viscosity. When the viscosity has higher values within the range $\sim10^{16}-10^{21} \,cm^{2}\,s^{-1}$, damping occurs independent of perpendicular diffusivity. Within IPLs damping of the waves occurs for the values of viscosity between $\sim10^{16.7}$ and $10^{20.7} \,cm^{2}\,s^{-1}$, otherwise the damping occurs at much greater distances.

When the damping length scales within PLs and IPLs are between $0.3 R_{\odot}$ and $5.5 R_{\odot}$, the energy flux density of the acoustic wave begins $ 10^{4.7}\,erg \,cm^{-2}\,s^{-1}$ and increases with increasing values of the diffusivity and viscosity(lower graphics in the Fig. 3). 

In the literature, the energies of acoustic waves are found to be inadequate for heating the corona (see \citet{ath78,mei81}). 
\citet{wit88} pointed out that most of the heat supply to the corona should be within $1-2 \,R_\odot$. Besides, in order to produce fast solar wind, a little more energy flux density ($1-2\times10^5\,erg \,cm^{-2}\,s^{-1}$) should be deposited at the sound point (a few solar radii). \citet{str12} studied the velocity maps of the solar wind and concluded that both the slow and the fast solar wind reveals the existence of an extra accelerating agent beyond $2.3\,R_\odot$.

Our study shows that when the variations of heat, density and temperature perpendicular to the magnetic field are taken into, the sound wave fluxes reach levels required for heating of the corona and  accelerating the solar wind. The energy flux density of acoustic waves varies between $ 10^{4.7}$ and $10^{7}\,erg \,cm^{-2}\,s^{-1}$. The damping length scale of these waves varies between $0.23\,R_{\odot}$ and $5.5\,R_{\odot}$. In even the negligible value range of perpendicular diffusivity (i.e. $10^{-4}-10^{4.8} \,\,erg\,cm^{-1}s^{-1}K^{-1}$) wave damping occurs as clearly seen in graphics. We may claim that these  waves may be responsible for perpendicular heating of \ion{O}{VI} ions and thus contribute to the heating of the coronal base in NPCH. Besides, our solution for the acoustic waves show that these waves may supply the observed extra acceleration for the fast solar wind because of its larger damping length scale ($2.2-5.5\,R_{\odot}$).This result is consistent with observations \citep{cra02,wil11}.

\begin{table*}
\caption{Variation of damping length scale and the energy flux density of Alfven and  acoustic waves with $\omega=0.1, 0.01$ and $0.001 \,rad/s $ propagating in plume ($x=40^{\arcsec}$) along the radial direction $R$. Viscosity and perpendicular diffusivity coefficients  are taken as $10^{20} \,cm^{2}$ s$^{-1}$ and $3 \,\,erg\,cm^{-1}s^{-1}K^{-1}$, respectively.} 
\centering
\begin{tabular}{l l c c c c }
\hline\hline

& \multicolumn{3}{c}{\textbf{Alfven}}  & \multicolumn{2}{c}{\textbf{Acoustic}} \\

\cline{1-6}
\cline{1-6}

$\omega$ ($rad/s$) & $R$ & $1/k_{i}$ ($R_{\sun}$) & $F$ ($erg\,cm^{-2}\, s^{-1}$) & $1/k_{i}$ ($R_{\sun}$) & $F$ ($erg\,cm^{-2}\, s^{-1}$)\\
\hline

    &1.7 & 0.6501 & $5.36\times 10^6$  & 0.7274 & $6.18\times 10^6$ \\

    &1.9 & 0.6531 & $3.63\times 10^6$  & 0.7266 & $4.19\times 10^6$ \\

0.1 &2.0 & 0.6531 & $3.03\times 10^6$  & 0.7263 & $3.51\times 10^6$ \\

    &2.24& 0.6531 & $2.11\times 10^6$  & 0.7256 & $2.41\times 10^6$ \\

    &2.5 & 0.6531 & $1.60\times 10^6$  & 0.7250 & $1.85\times 10^6$ \\

\cline{1-6}

    &1.7 & 2.4026 & $1.59\times 10^6$  & 2.2704 & $1.95\times 10^6$ \\

    &1.9 & 2.4729 & $1.08\times 10^6$  & 2.2734 & $1.31\times 10^6$ \\

0.01&2.0 & 2.4988 & $9.05\times 10^5$  & 2.2772 & $1.09\times 10^6$ \\

    &2.24& 2.5475 & $6.31\times 10^5$  & 2.2948 & $7.37\times 10^5$ \\

    &2.5 & 2.5566 & $4.74\times 10^5$  & 2.3218 & $5.53\times 10^5$ \\

\cline{1-6}

     &1.7 & 43.4073 & $6.80\times 10^5$  & 7.0291 & $5.96\times 10^5$ \\

     &1.9 & 51.8694 & $4.84\times 10^5$  & 7.5070 & $3.58\times 10^5$ \\

0.001&2.0 & 55.6892 & $4.13\times 10^5$  & 7.9150 & $2.76\times 10^5$ \\

     &2.24& 61.6644 & $2.96\times 10^5$  & 9.7629 & $1.65\times 10^5$ \\

     &2.5 & 62.4688 & $2.23\times 10^5$  & 13.0612 & $1.32\times 10^5$ \\

\hline\hline

\end{tabular}
\end{table*}

\subsubsection{Alfven waves in NPCH}

If we substitute ($X_2$,$Y_2$) into equation (24) we get the dispersion relation for Alfven waves:

\begin{equation}
\left(w^2+i\omega\nu k^2-{v_A}^2 k^2\right)=0.
\end{equation}

In Figure 4, it is shown how frequency and viscosity cause variation  of the damping length scale (left) in terms of solar radius and the energy flux density (right) of Alfven waves along the plume at the fiducial radius $R = 1.7$. There is a turning point of viscosity for each contours of the damping length scales. Up to the value of turning point firstly viscosity increases linearly with decreasing frequency and later increases with increasing frequency. For example the viscosity value of turning point is about $10^{20} \,cm^{2}\,s^{-1}$ for the damping length scales between $\sim2 R_{\odot}-5.5 R_{\odot}$. Wave flux is about $10^6-10^{7.1}\,erg\,cm^{-2}\,s^{-1}$ for the required damping length scales. When the frequency is set to a constant value, it is seen that wave flux increases with increasing viscosity. 

The damping length scales and the energy flux densities of this waves are similar for both the plumes and interplume lanes. 

The total Alfven wave energy flux required to heat the quiet corona is $3\times10^5\,erg\,cm^{-2}\,s^{-1}$ \citep{wit77}. We showed above that the energy flux density of Alfven waves falls within a range about $10^6 - 10^{8.5} \,erg\,cm^{-2}\,s^{-1}$. This amount is believed to replace the energy loss in NPCH by optically thin emission and by heat conduction to the transition region. The damping length scale and energy flux density are $2.16\,R_{\odot}$  and $1.1\times 10^{6} \,erg\,cm^{-2}\,s^{-1}$  for wave with frequency $\omega=0.01 \,rad/s$ at around Spitzer value of viscosity coeffcient ($\sim 5\times10^{19} \,cm^{2}\,s^{-1}$), respectively. Alfven waves may contribute to the heating of upper corona in NPCH and supply the observed extra acceleration for the fast solar wind. 

In the second part of the investigation, we searched the wave propagation characteristics of the acoustic and Alfven waves along the radial direction $R$. Taking $x=40^{\arcsec}$  as the distance  of the midpoint of plume we found the damping length scales and energy flux densities  along the radial direction. Table 1 shows these values of acoustic and Alfven waves having frequencies $\omega = 0.1, 0.01$ and $ 0.001 \,rad/s$ at some radial distances assuming viscosity has a value of $10^{20} \,cm^{2}\,s^{-1}$. Fluxes decrease with radial distance and this decrease is steeper at higher frequencies. For smaller viscosity values (e.g. of $10^{17} \,cm^{2}\,s^{-1}$) a similar variation occurs at frequencies $\omega = 0.1, 0.01$ and $ 0.001 \,rad/s$, but fluxes are smaller and variation along the radial distance is smoother for $\omega = 0.1 \,rad/s$.

\subsection{In the presence of parallel heat conduction}

If we take $D\neq0$ then equation (25) yields three $X$ values and thus three ($X$,$Y$) pairs as given below

\begin{displaymath}
\left( X_1,Y_1\right)=\left(\frac{G}{4\omega}, \frac{1}{2\omega} \frac{4D\omega-AG}{2B+G}\right),
\end{displaymath}

\begin{displaymath}
\left( X_2,Y_2\right)=\left(-\frac{\left( B+{v_A}^2 k^2 \right)}{2\omega}, -\frac{2\omega D+A\left(B+{v_A}^2 k^2 \right)} {2\omega{v_A}^2 k^2} \right),
\end{displaymath}

\begin{displaymath}
\left(X_3,Y_3\right)=\left(\frac{F}{4\omega}, \frac{1}{2\omega} \frac{4D\omega-AF}{2B+F}\right),
\end{displaymath}where $F=-Z1+\sqrt{Z1^2 +8\omega Dk}$, $G=-Z1-\sqrt{Z1^2 +8\omega Dk}$.

The square root in the  F and G terms is rewritten as $\sqrt{1+8\omega Dk/Z^2_1}$. Since the value of the second term under square root is less than one, we apply the Taylor series expansion, i.e., if the below condition is fulfilled

\begin{displaymath}
8\omega 0.6 \frac{1} {\rho_0} H_{\parallel} k^3 < \left[ -\omega^2 +\left( 1.6 {c_s}^2 -0.6 \frac{1}{\rho_0} \frac{H_{\perp}}{\xi}-\nu 1.3 i\omega\right) k^2\right] ^2.
\end{displaymath}This condition is fulfilled in our model.

\begin{figure*}
\begin{center}
\includegraphics[angle=0,scale=.49]{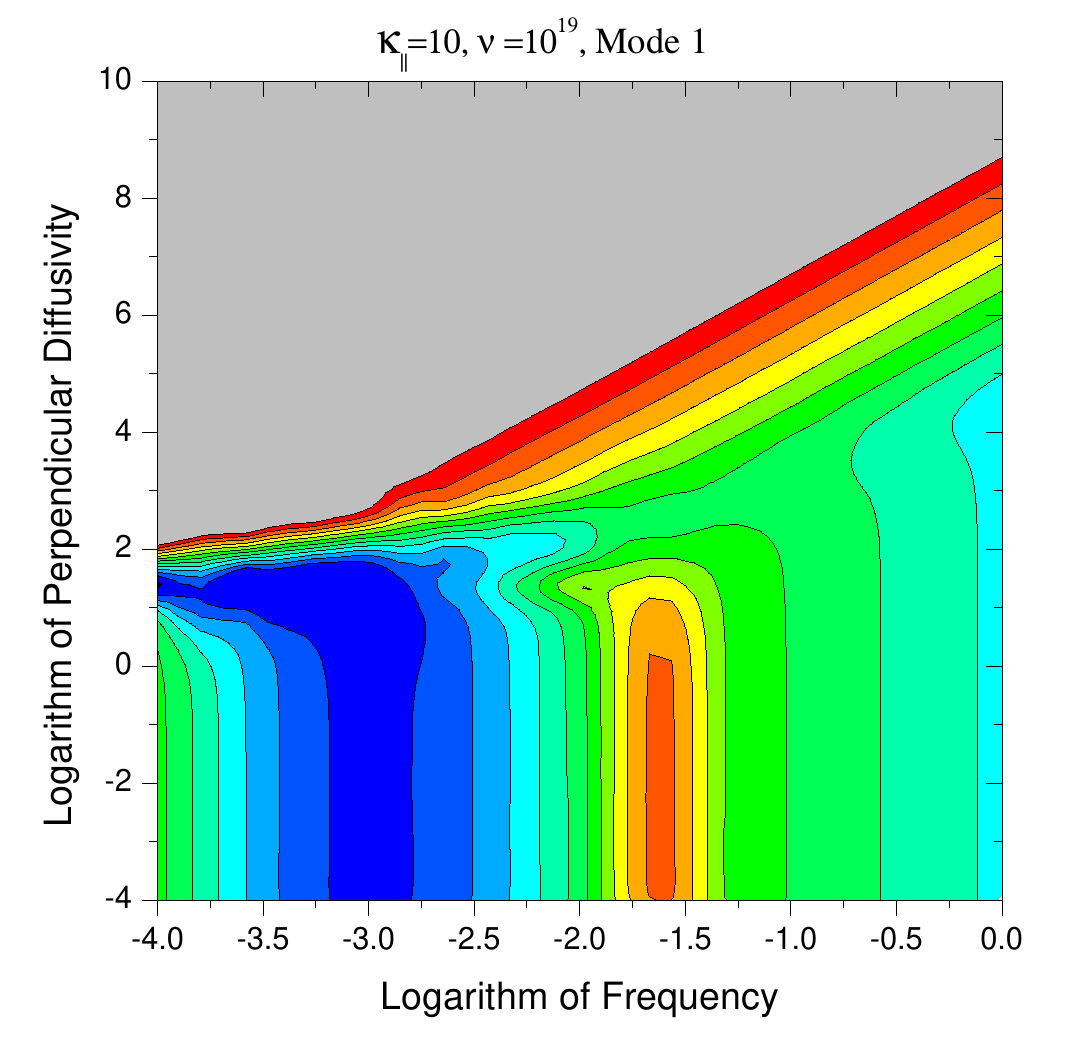}
\includegraphics[angle=0,scale=.49]{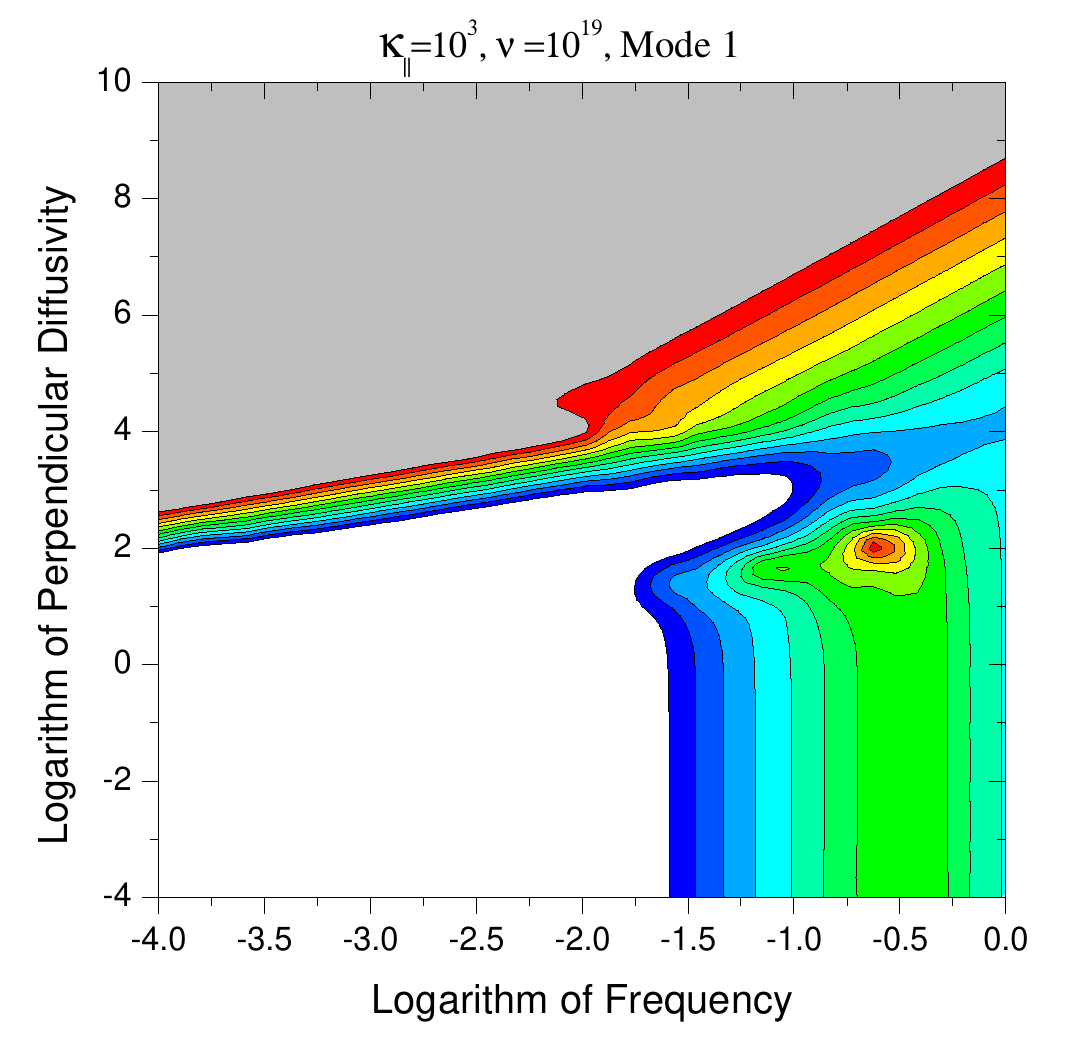}
\includegraphics[angle=0,scale=.54]{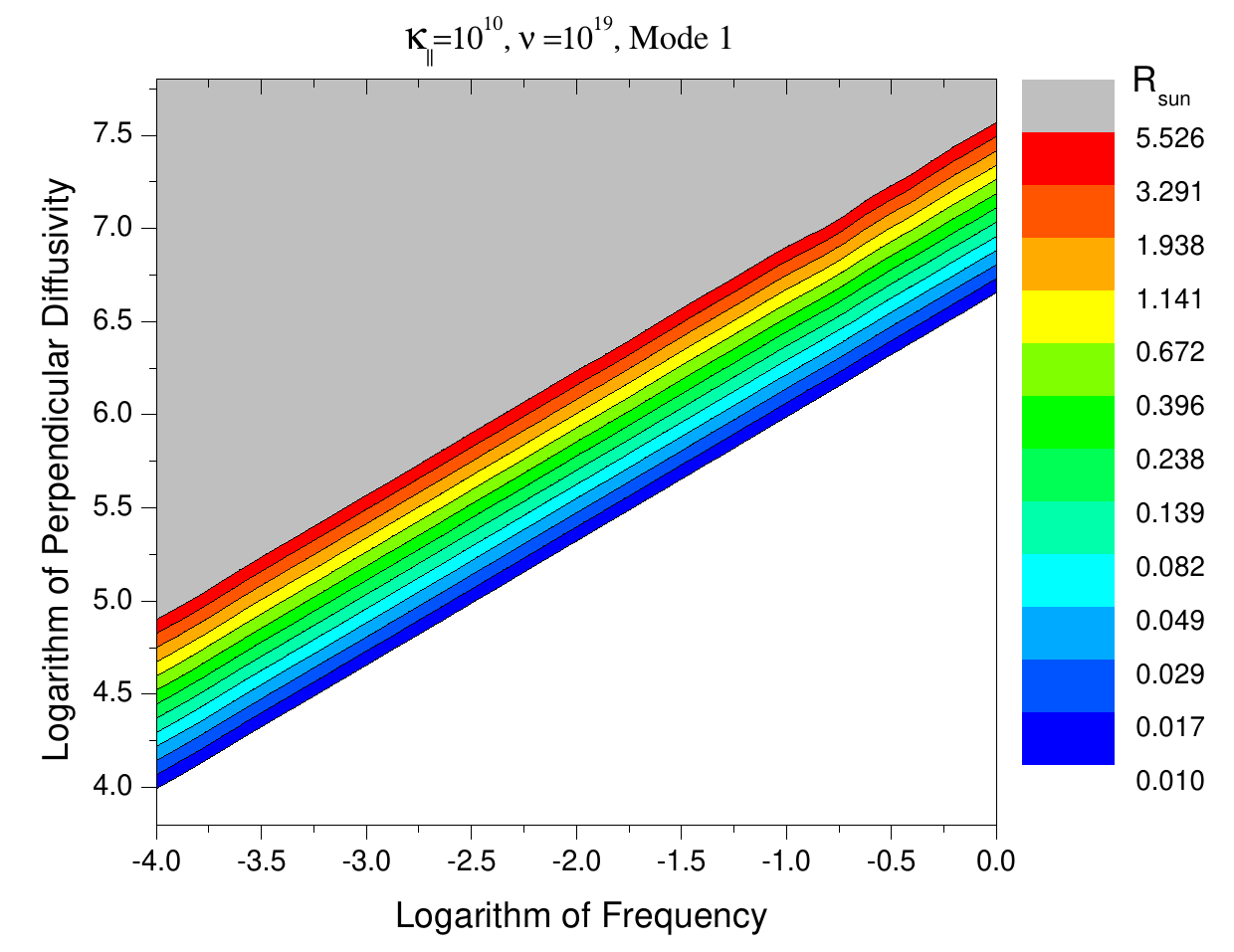}
\includegraphics[angle=0,scale=.49]{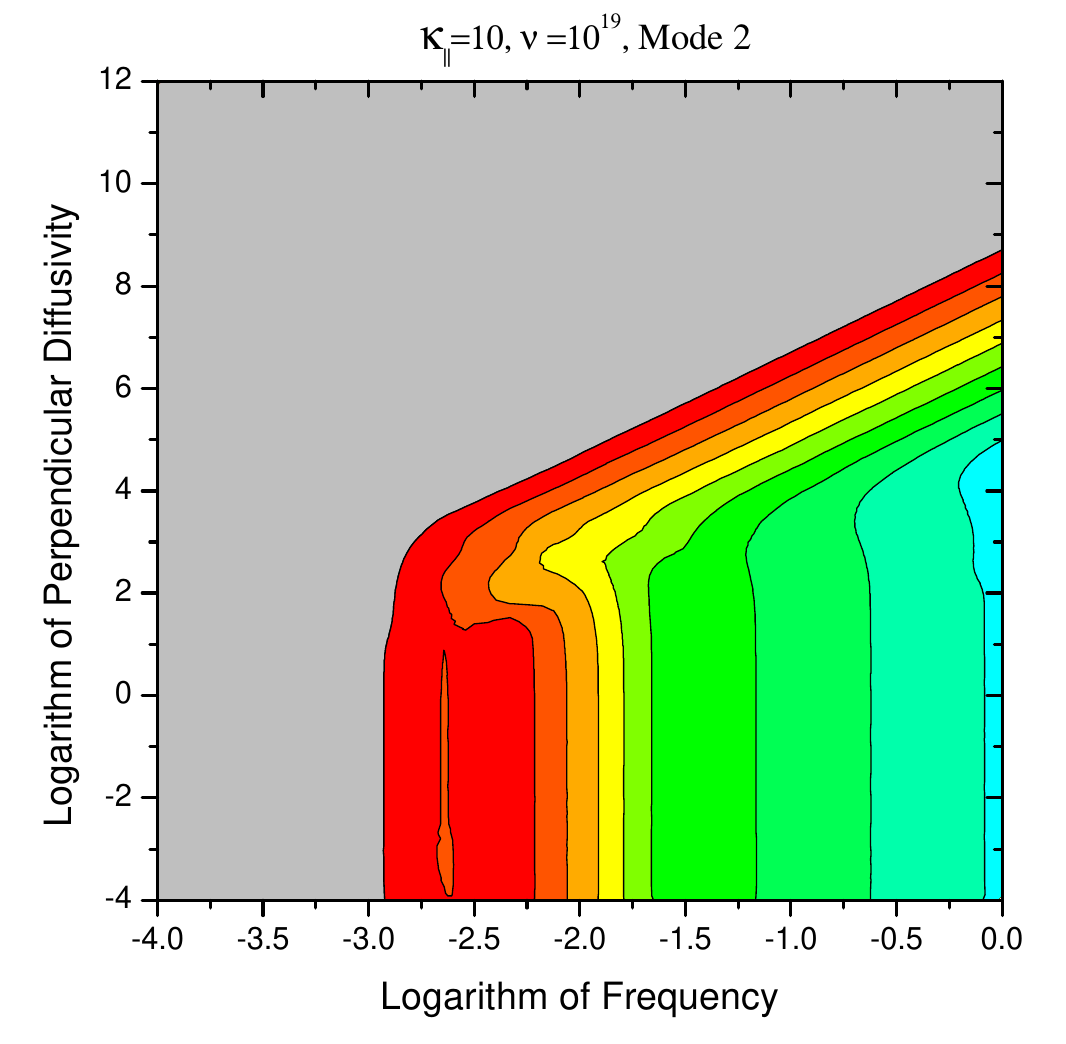}
\includegraphics[angle=0,scale=.49]{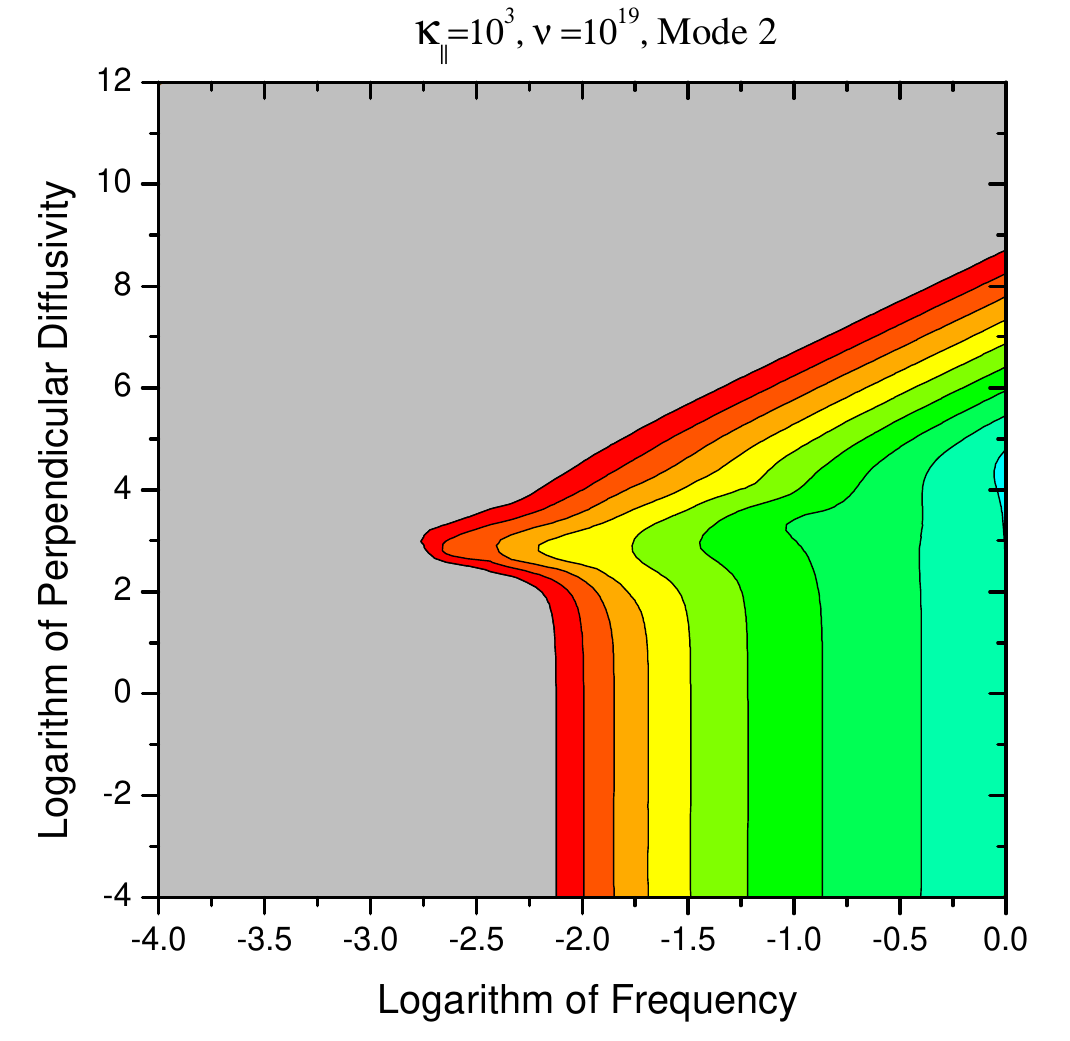}
\includegraphics[angle=0,scale=.54]{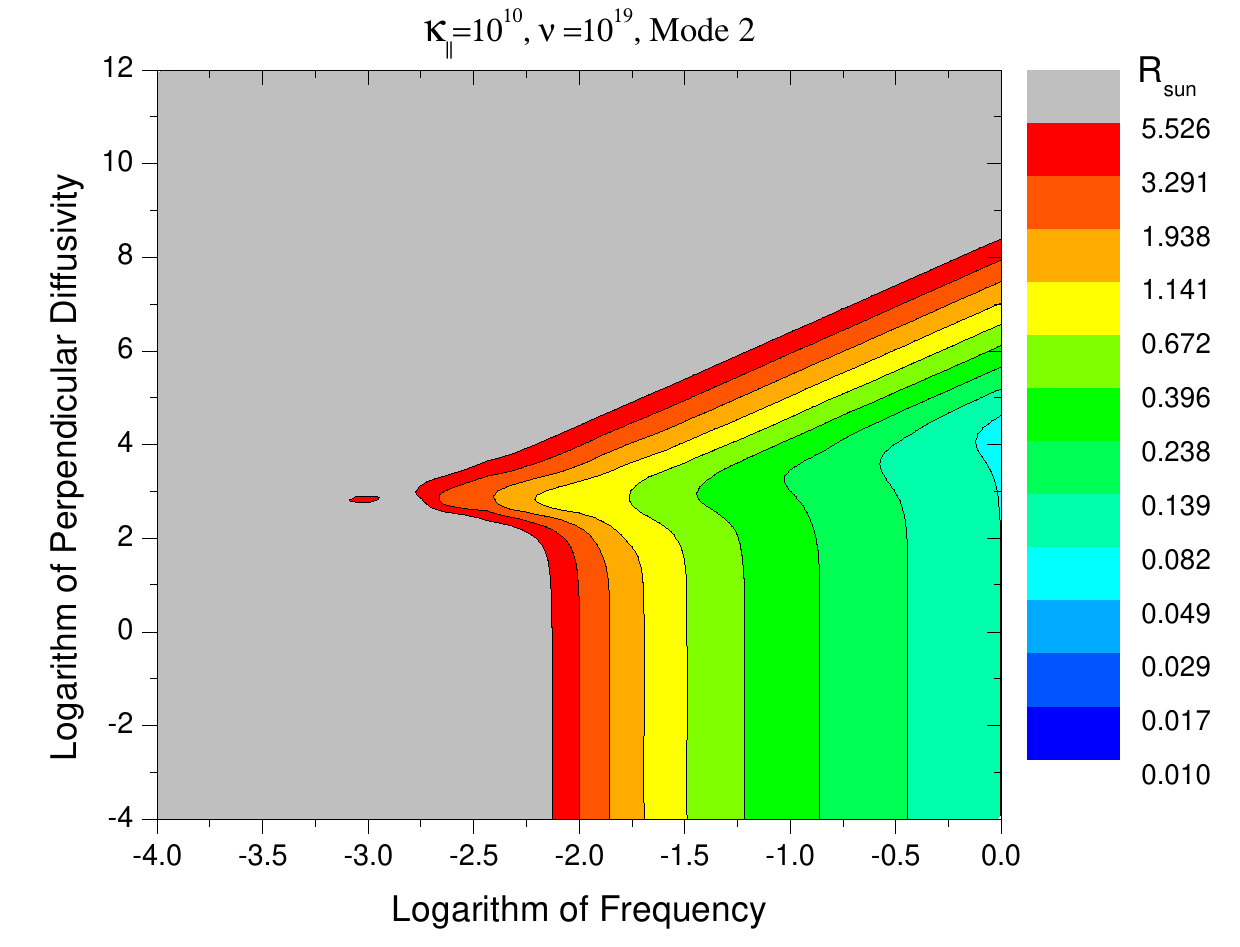}
\caption{The variation of damping length scales of Mode 1 (upper graphics) and Mode 2 (lower graphics) versus frequency and perpendicular diffusivity. Viscosity and three parallel diffusivity $\kappa_{\parallel}$  values are assumed to be $10^{19} \,cm^{2}\,s^{-1} $ and $10, 10^3, 10^{10} \,\,erg\,cm^{-1}s^{-1}K^{-1}$, respectively.}
\end{center}
\end{figure*}

\begin{figure*}
\begin{center}
\includegraphics[angle=0,scale=.48]{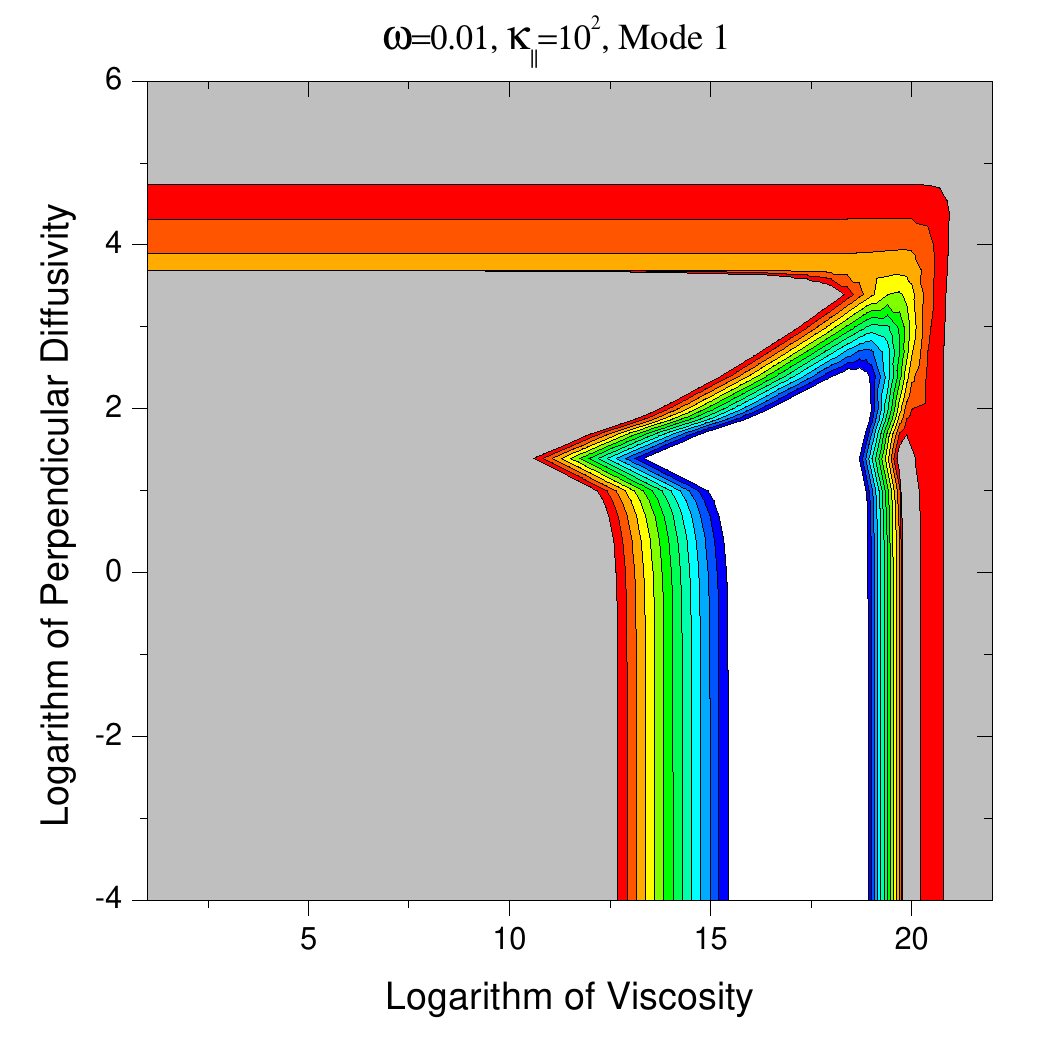}
\includegraphics[angle=0,scale=.53]{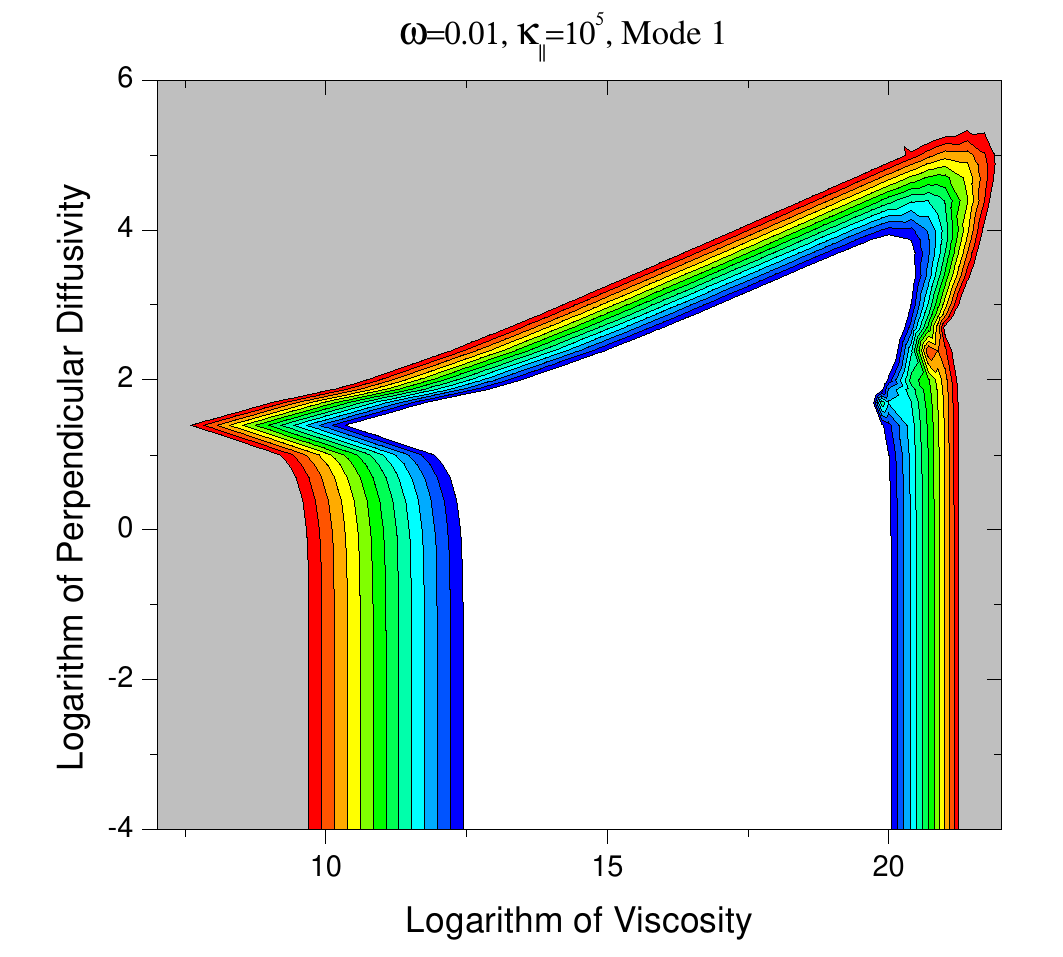}
\includegraphics[angle=0,scale=.52]{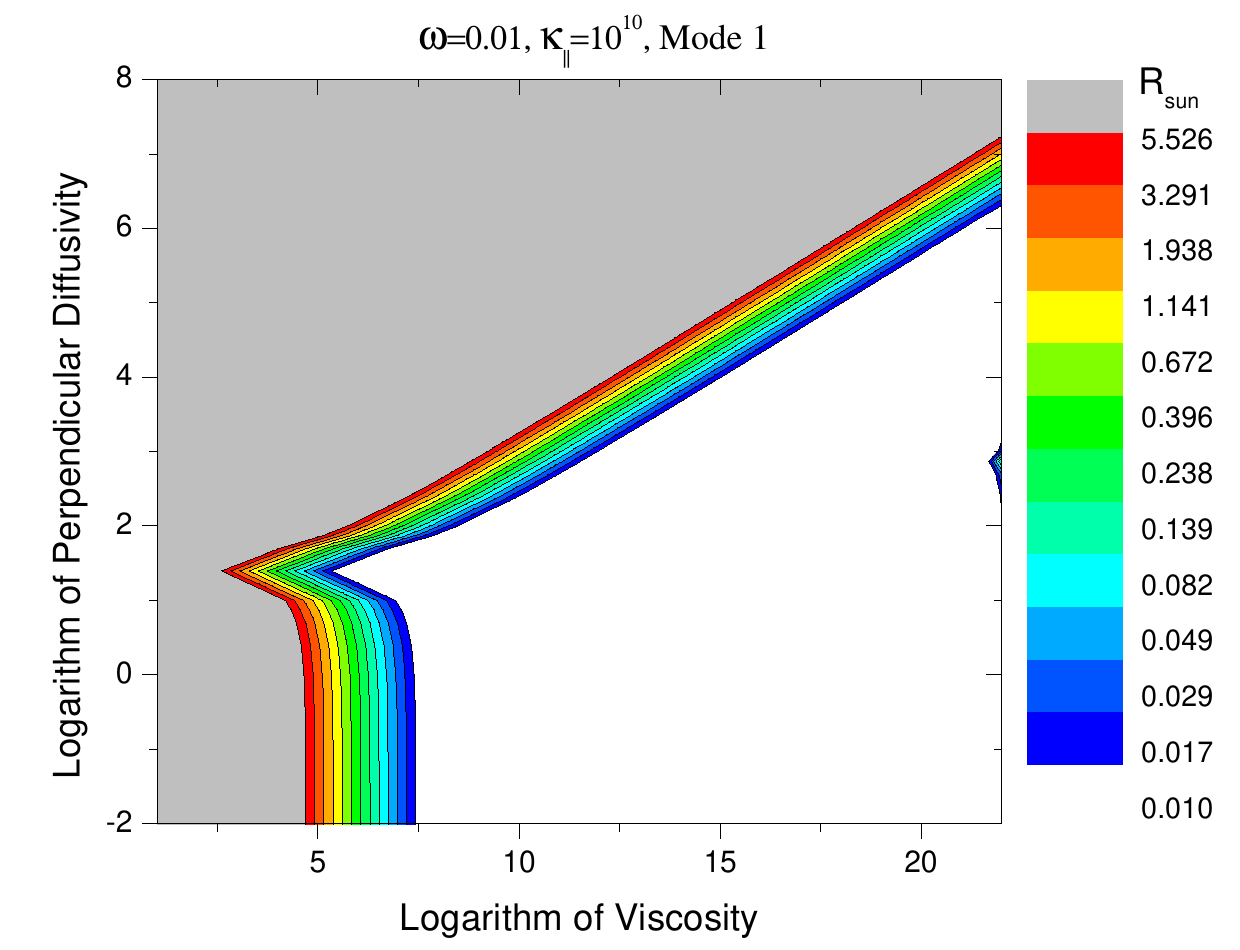}
\includegraphics[angle=0,scale=.49]{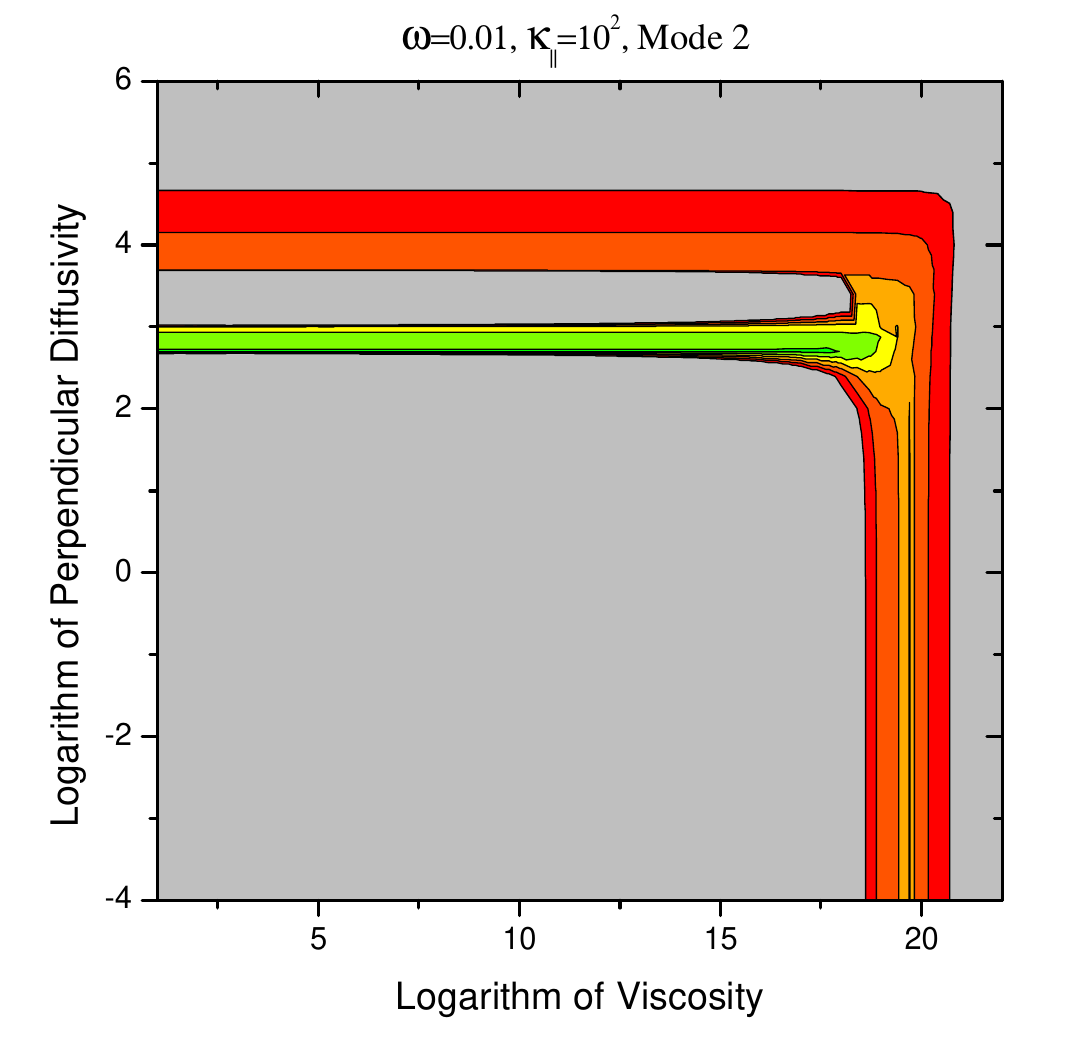}
\includegraphics[angle=0,scale=.49]{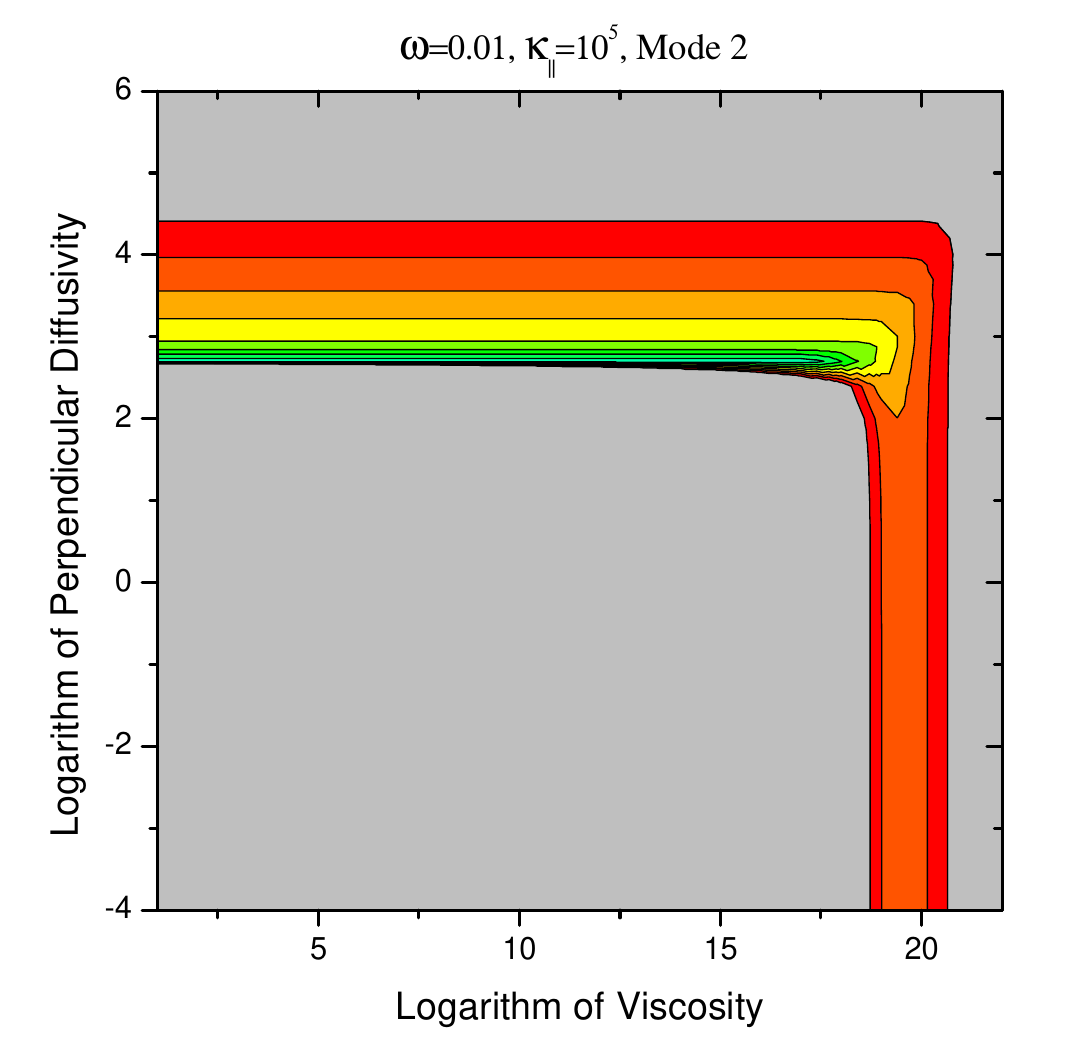}
\includegraphics[angle=0,scale=.5]{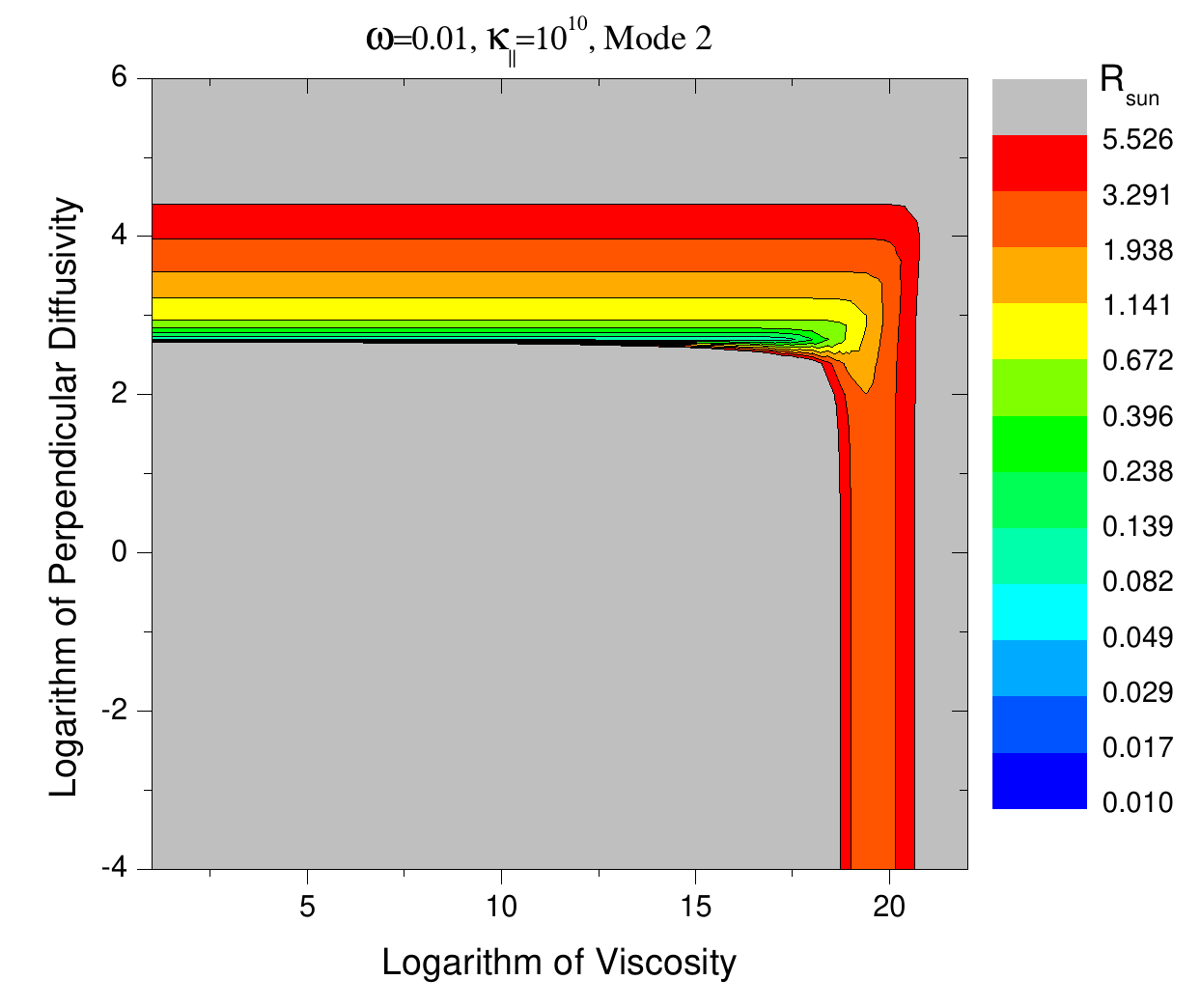}
\includegraphics[angle=0,scale=.49]{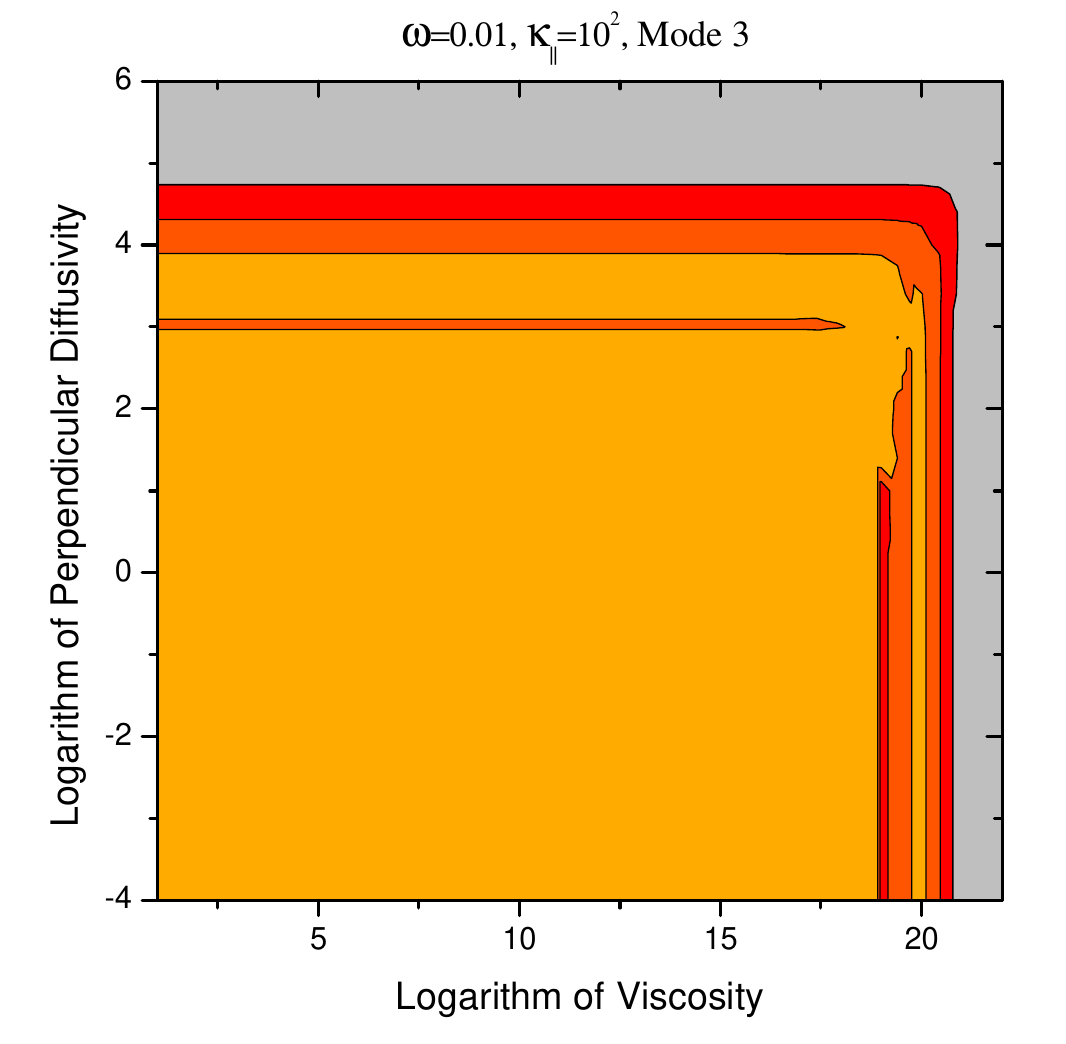}
\includegraphics[angle=0,scale=.49]{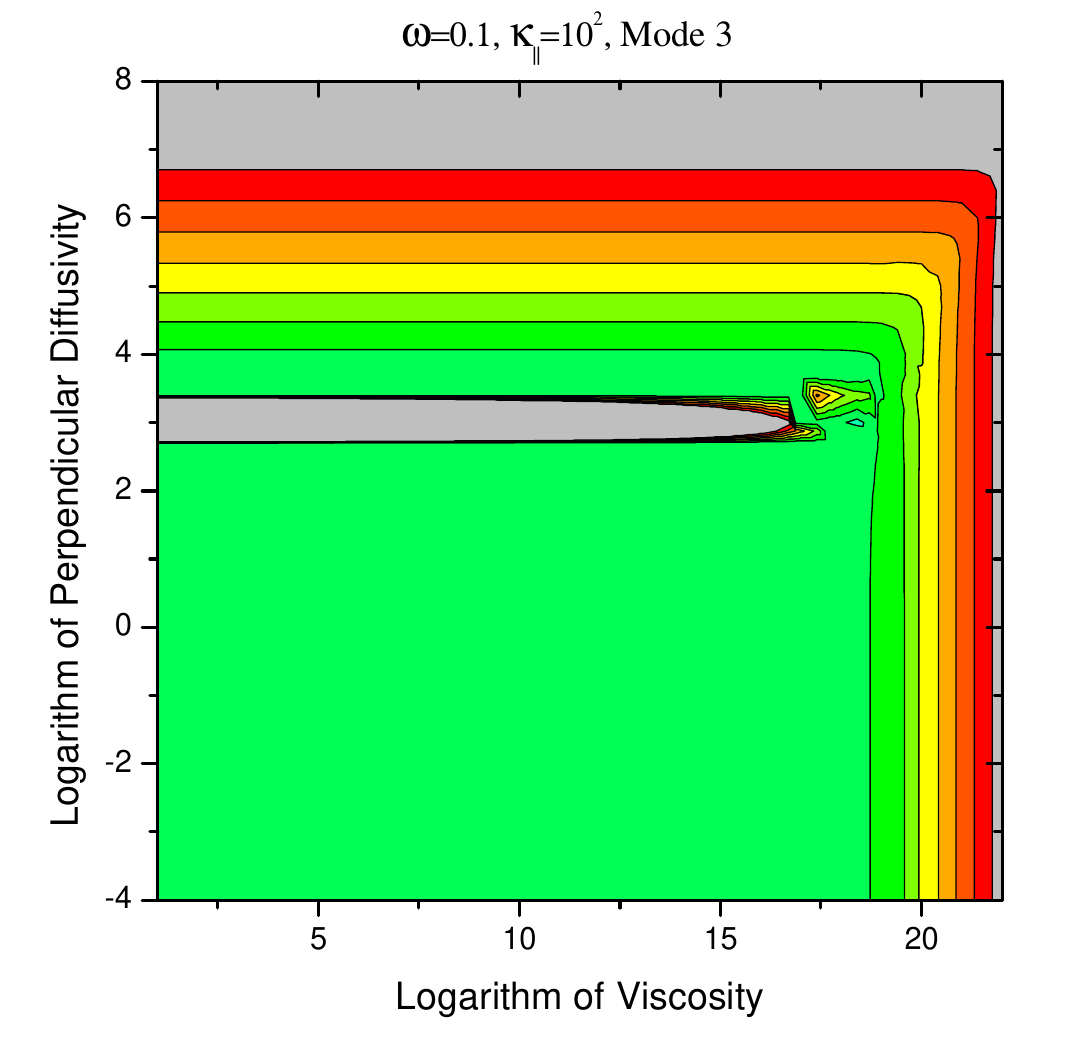}
\includegraphics[angle=0,scale=.5]{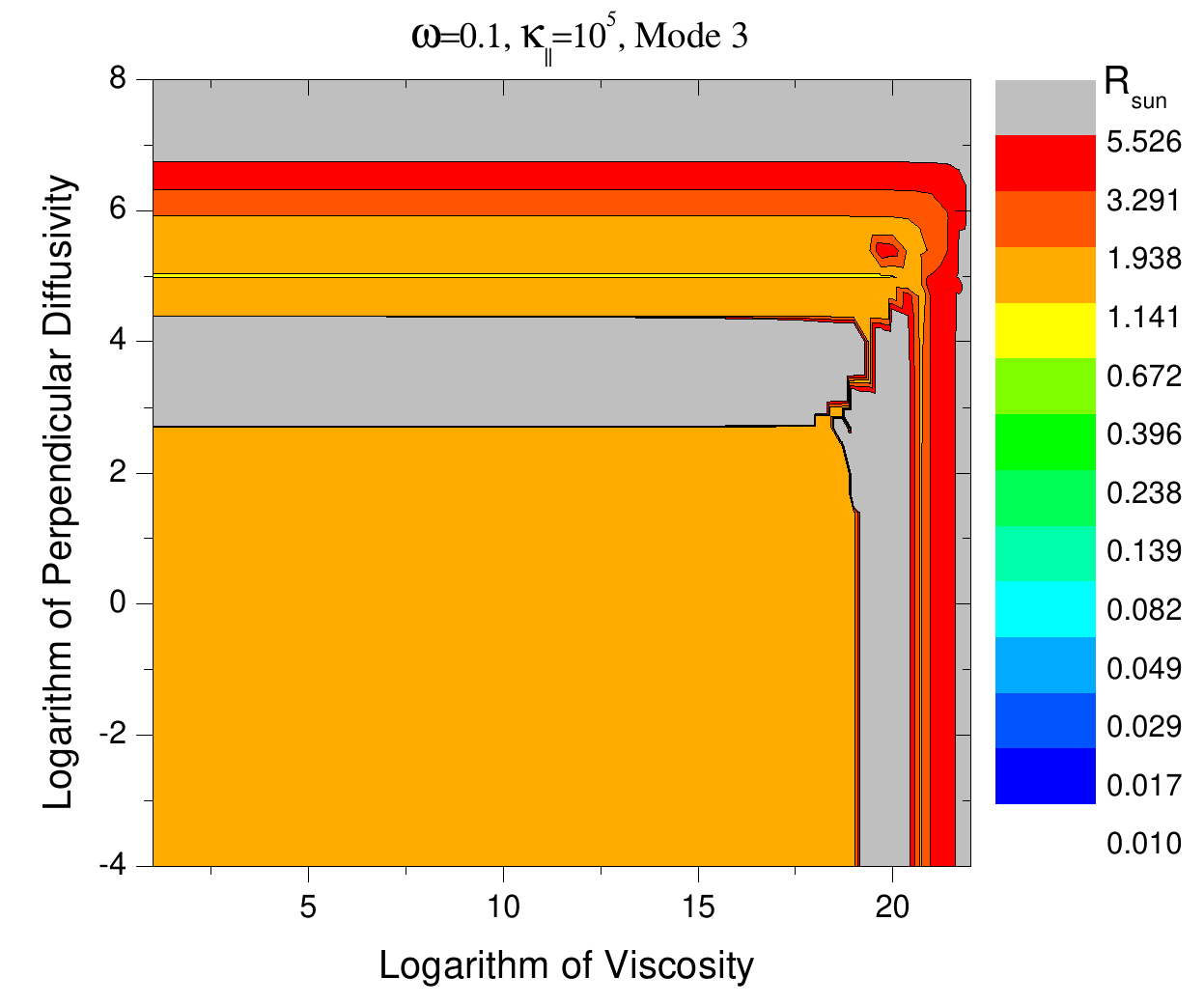}
\caption{The variation of damping length scales with viscosity and perpendicular diffusivity of Mode 1 (top) and Mode 2 (middle) having a frequency $\omega=0.01 \,rad/s$  when $\kappa_{\parallel}$ values are assumed to be $10^2, 10^5$ and $10^{10} \,\,erg\,cm^{-1}s^{-1}K^{-1}$.  Bottom graphics show the variation of damping length scale of Mode 3 having a frequency $\omega=0.01 \,rad/s$ and $\kappa_{\parallel}=10^2 \,\,erg\,cm^{-1}s^{-1}K^{-1} $ and the variation of damping length scale of the same mode (Mode 3) having a frequency  $\omega=0.1 \,rad/s$ and $\kappa_{\parallel}=10^2, 10^5 \,\,erg\,cm^{-1}s^{-1}K^{-1}$ values.}
\end{center}
\end{figure*}

\subsubsection{Solution for ($X_1,Y_1$)}

Let us substitute ($X_1$,$Y_1$) into equation (24) using Taylor series expansion and after a lengthy effort we reach the dispersion relation as Equation (A1) in the Appendix.

Figures 5 and 6  show the solutions of ten degree dispersion relation (Equation A1). The variation of damping length scales of Mode 1 (upper graphics) and Mode 2 (lower graphics) versus frequency and perpendicular diffusivity is shown in Figure 5 wherein viscosity and three parallel diffusivity $\kappa_{\parallel}$  values are assumed to be $10^{19} \,cm^{2}\,s^{-1} $ and $10, 10^3, 10^{10} \,\,erg\,cm^{-1}s^{-1}K^{-1}$, respectively. For Mode 1 in case $\kappa_{\parallel}=10 \,\,erg\,cm^{-1}s^{-1}K^{-1}$, there appears damping length scales only dependent on frequency at the values smaller than $\sim100 \,\,erg\,cm^{-1}s^{-1}K^{-1}$ for perpendicular diffusivity. At values greater than $\sim100 \,\,erg\,cm^{-1}s^{-1}K^{-1}$ for perpendicular diffusivity, the value of perpendicular diffusivity  increases linearly with increasing frequency for all the damping length scales. At extremely high value of $\kappa_{\parallel}$,  there appears a linear increase of perpendicular diffusivity with increasing frequency for all the damping length scales within the range $10^4 - 10^{7.5} \,\,erg\,cm^{-1}s^{-1}K^{-1}$ for perpendicular diffusivity. For other values, wave damping occurs at either very far or very close distances.  

On the other hand, it is seen that the damping length scales of Mode 2 is independent of the value of $\kappa_{\parallel}$. Waves with lower frequencies ($\omega \la 10^{-3}$) get damped at very far distances. Similiarly, there appears damping length scales independent of perpendicular diffusivity and only dependent on frequency at the values smaller than $\sim10^3 \,\,erg\,cm^{-1}s^{-1}K^{-1}$ for perpendicular diffusivity while there appears a linear increase of perpendicular diffusivity with increasing frequency for all damping length scales at the values greater than $\sim10^3 \,\,erg\,cm^{-1}s^{-1}K^{-1}$ for perpendicular diffusivity.  

Mode 3 shows a similar tendency with Mode 2. While Mode 4 gets damped very near to the sun, Mode 5 propagates to very far distances before damping.

If we assume $H_{\parallel}=0$ in the dispersion relation (A1), the solution gives an acoustic wave given by the equation (27).  

When only $H_{\parallel}$  is taken into consideration and other effects are assumed to be negligible in Equation (A1), damping length scale increases with increasing $\kappa_{\parallel}$  for a fixed frequency. At lower frequencies, required damping length scales for heating corona and acceleration of solar wind obtained only for smaller values of $\kappa_{\parallel}$.

When only $H_{\parallel}$  and viscosity are taken into consideration, Mode 1 gets damped at distances $0.01-0.5 R_{\odot}$.  Mode 2 waves with frequencies within the range $10^{-1.75} - 10^{-2.25} \,rad/s $  get damped within the required radial distances. Damping is independent of the value of $\kappa_{\parallel}$.  Mode 3 has required damping length scales when $\kappa_{\parallel}$  values fall within a narrow range, i.e.,  $10-10^5 \,\,erg\,cm^{-1}s^{-1}K^{-1}$. Waves with frequencies lesser than $0.01 \,rad/s $  get damped at very far distances (i.e. further than $5.5 R_{\odot}$).  
  
Figure 6 (top and middle) shows the variation of damping length scales with viscosity and perpendicular diffusivity of Mode 1 and Mode 2 having a frequency $\omega=0.01 \,rad/s$  when $\kappa_{\parallel}$ values are assumed to be $10^2, 10^5$ and $10^{10} \,\,erg\,cm^{-1}s^{-1}K^{-1}$.  Plots in the bottom of Figure 6 show the variation of damping length scale of Mode 3 having a frequency $\omega=0.01 \,rad/s$ and $\kappa_{\parallel}=10^2 \,\,erg\,cm^{-1}s^{-1}K^{-1}$ and the variation of damping length scale of the same mode (Mode 3) having a frequency  $\omega=0.1 \,rad/s$ and $\kappa_{\parallel}=10^2, 10^5 \,\,erg\,cm^{-1}s^{-1}K^{-1}$ values. 

For Mode 1 when $\kappa_{\parallel}=10^2 \,\,erg\,cm^{-1}s^{-1}K^{-1}$, there is a region wherein waves get damped independent of viscosity for values of perpendicular diffusivity between $10^{3.8}$ and $10^{4.8} \,\,erg\,cm^{-1}s^{-1}K^{-1} $. At smaller values of perpendicular diffusivity  and within viscosity value range $\sim10^{13}- 10^{19} \,cm^{2}\,s^{-1} $ there appears the second damping region. First damping region disappears while $\kappa_{\parallel}$ values increase and the second region enlarges towards smaller viscosity values. 

Mode 2 has only the first damping region like the Mode 1. This mode has not been significantly affected by the variation of $\kappa_{\parallel}$. 

When perpendicular diffusivity value is smaller than $\sim10^{4.8} \,\,erg\,cm^{-1}s^{-1}K^{-1}$, Mode 3 with $\omega=0.01 \,rad/s$ gets damped within the radial distance range $1.2-5.5 R_{\odot}$. When $\kappa_{\parallel}$ assumes higher values, wave gets damped at further radial distances. For a wave with a frequency $\omega=0.1  \,rad/s$ at the value of $\kappa_{\parallel}=100 \,\,erg\,cm^{-1}s^{-1}K^{-1}$ damping occurs, on the average, at  $0.3 R_{\odot}$ for smaller values of perpendicular diffusivity. But when $\kappa_{\parallel}=10^5 \,\,erg\,cm^{-1}s^{-1}K^{-1}$, the wave with a same frequency gets damped, on the average, at $1.5 R_{\odot}$.

The energy flux densities of the three modes are calculated as high as $10^{10} \,erg\, cm^{-2}\, s^{-1}$.  The ``inflationary'' values of these energy flux densities result from the group velocities of the waves which are three orders of magnitude higher than the speed of light in vacuum! Therefore, energy flux densities of these modes have no physical meaning since they are calculated by equation (28) containing the group velocity of the waves. 

\citet{jac75} noted that the group velocity often becomes larger than the speed of light in a dispersive medium, but ideas of special relativity does not violated, because approximations caused that the transport of energy occurs with the group velocity are no longer valid.  Besides, \citet{ven10} referring to \citet{lom02} conclude that physical properties of a wave packet should not be described by group velocity in a medium with anomalous dispersion.

\begin{figure*}
\begin{center}
\includegraphics[angle=0,scale=.48]{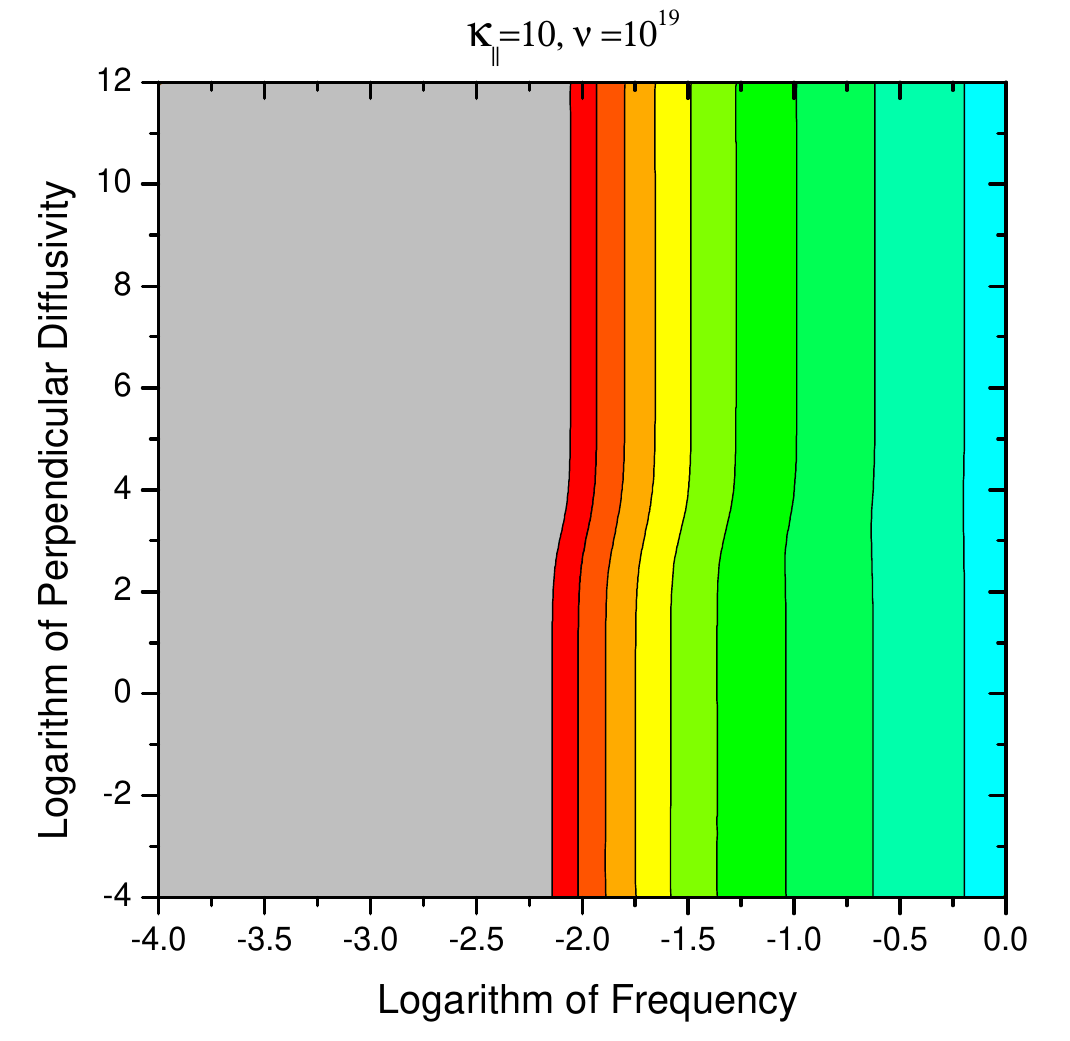}
\includegraphics[angle=0,scale=.48]{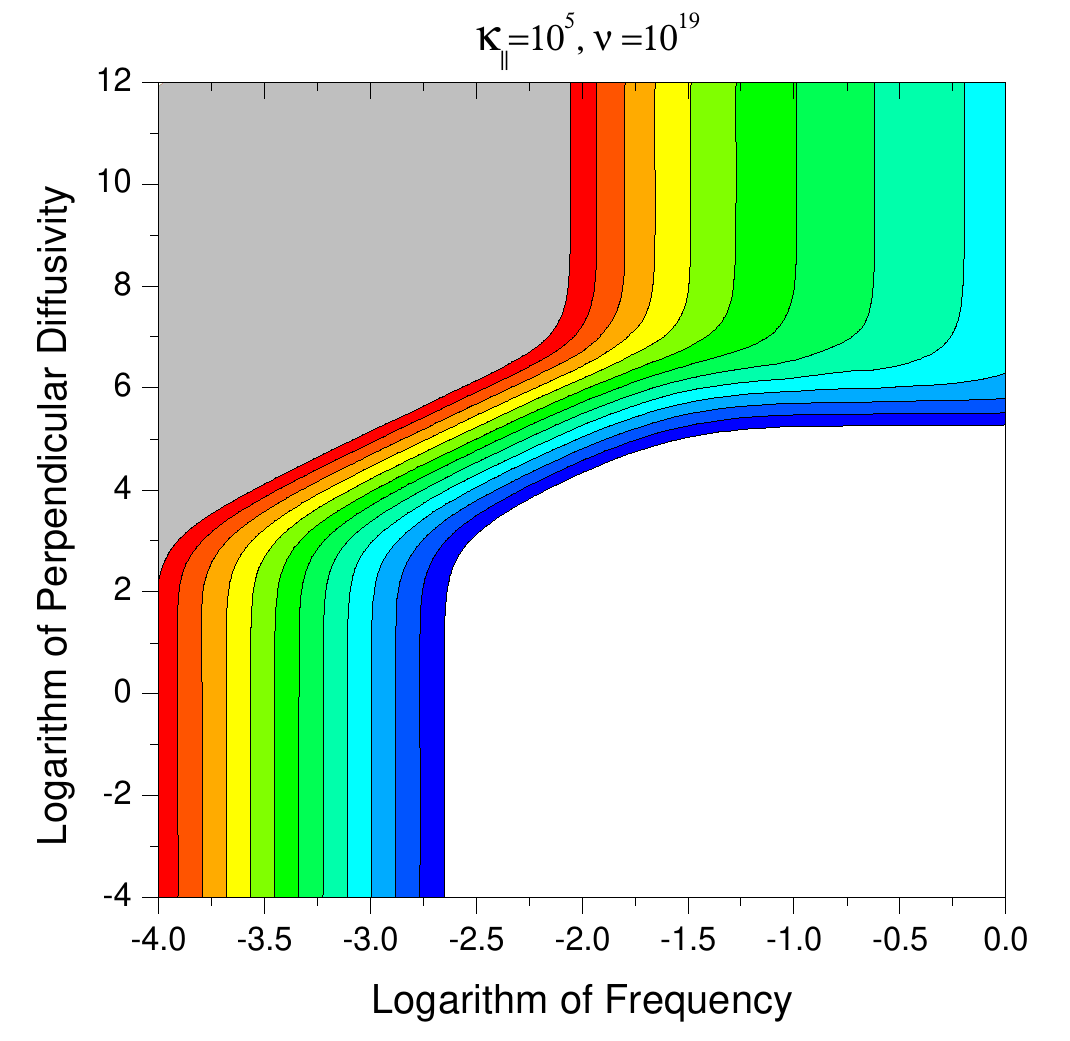}
\includegraphics[angle=0,scale=.53]{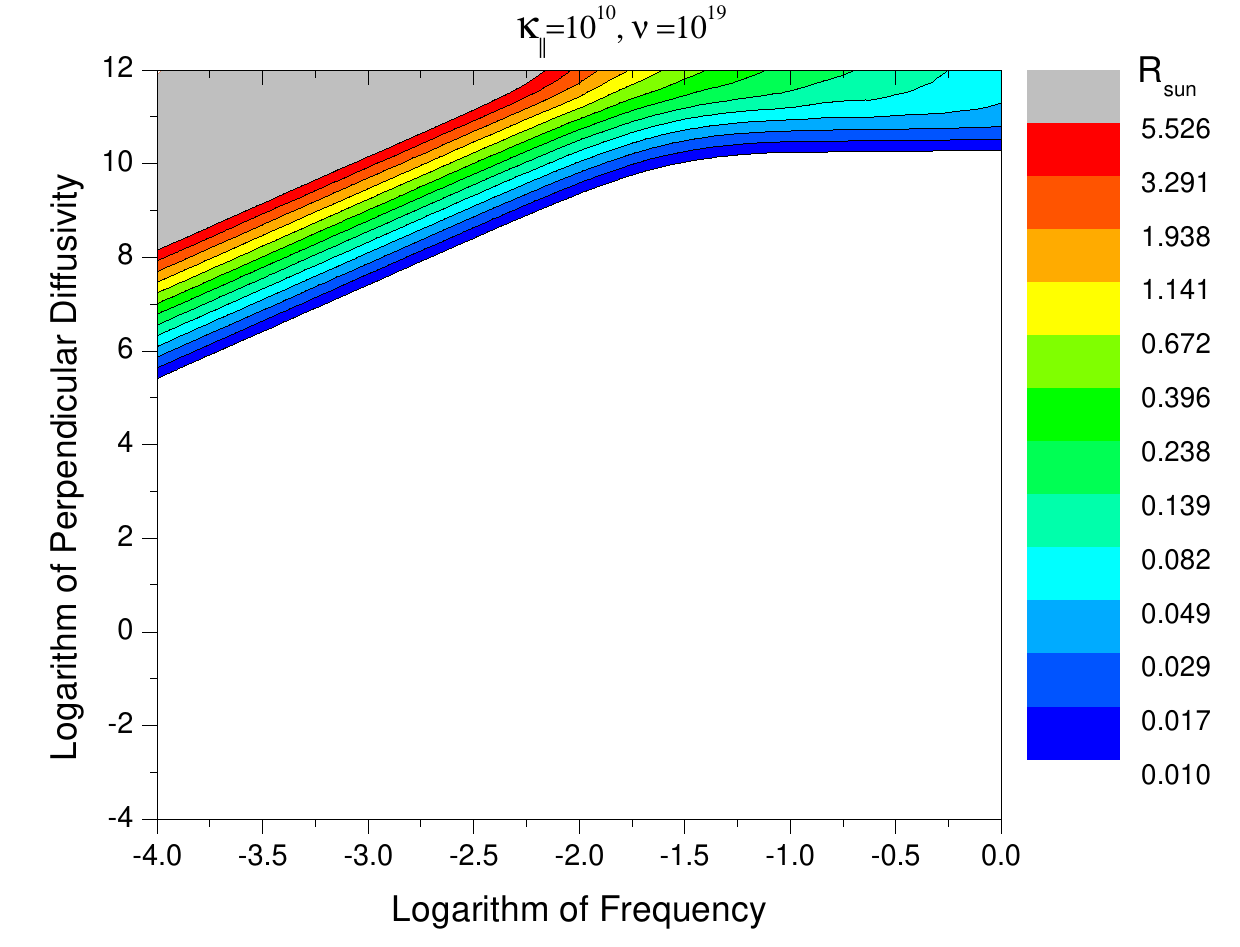}
\caption{The variation of the damping length scale of a magnetosonic wave (Equation 30) with frequency and perpendicular diffusivity. Values of viscosity and $\kappa_{\parallel}$ are assumed as $10^{19} \,cm^{2}\,s^{-1} $ and  $10, 10^5$ and $ 10^{10} \,\,erg\,cm^{-1}s^{-1}K^{-1}$, respectively.}
\end{center}
\end{figure*}

\begin{figure*}
\begin{center}
\includegraphics[angle=0,scale=.48]{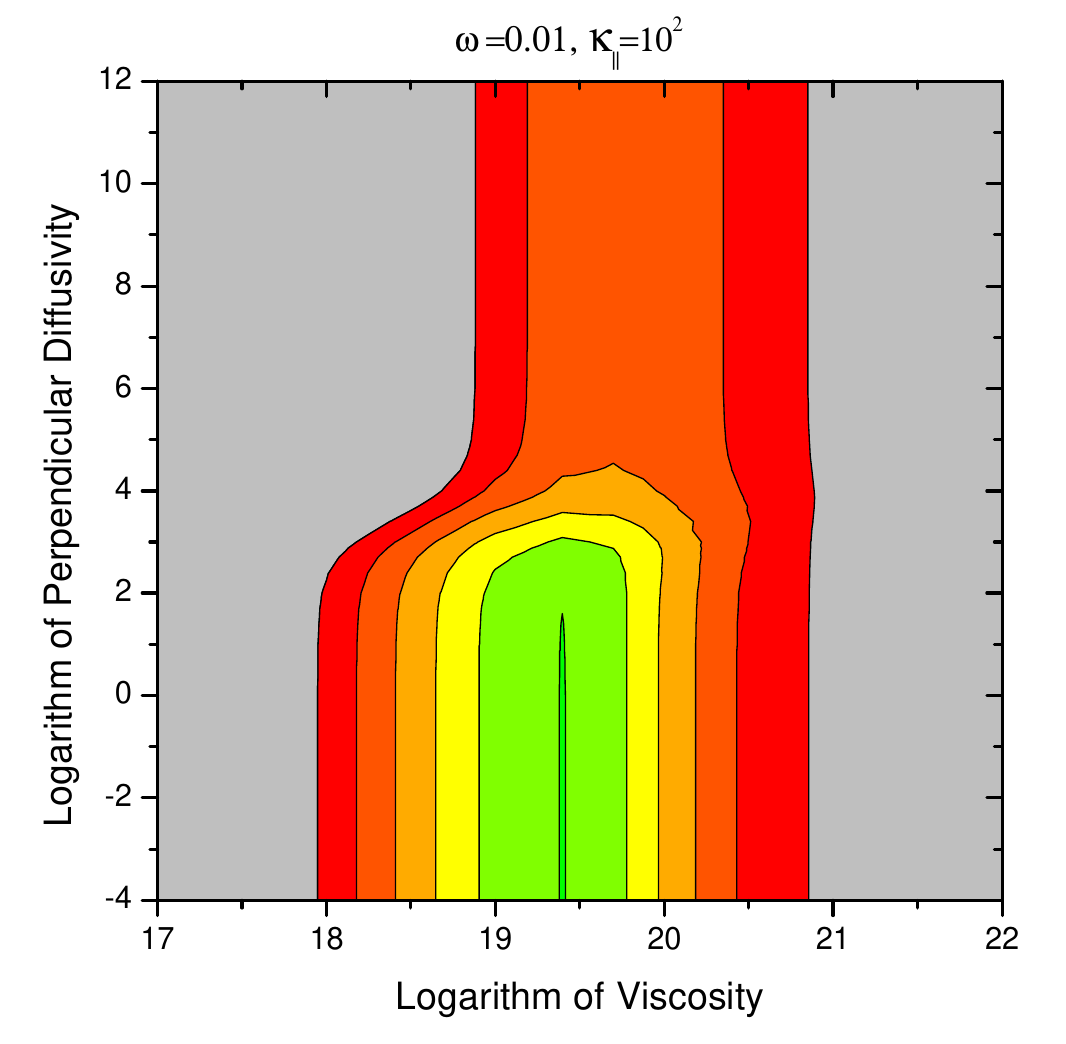}
\includegraphics[angle=0,scale=.53]{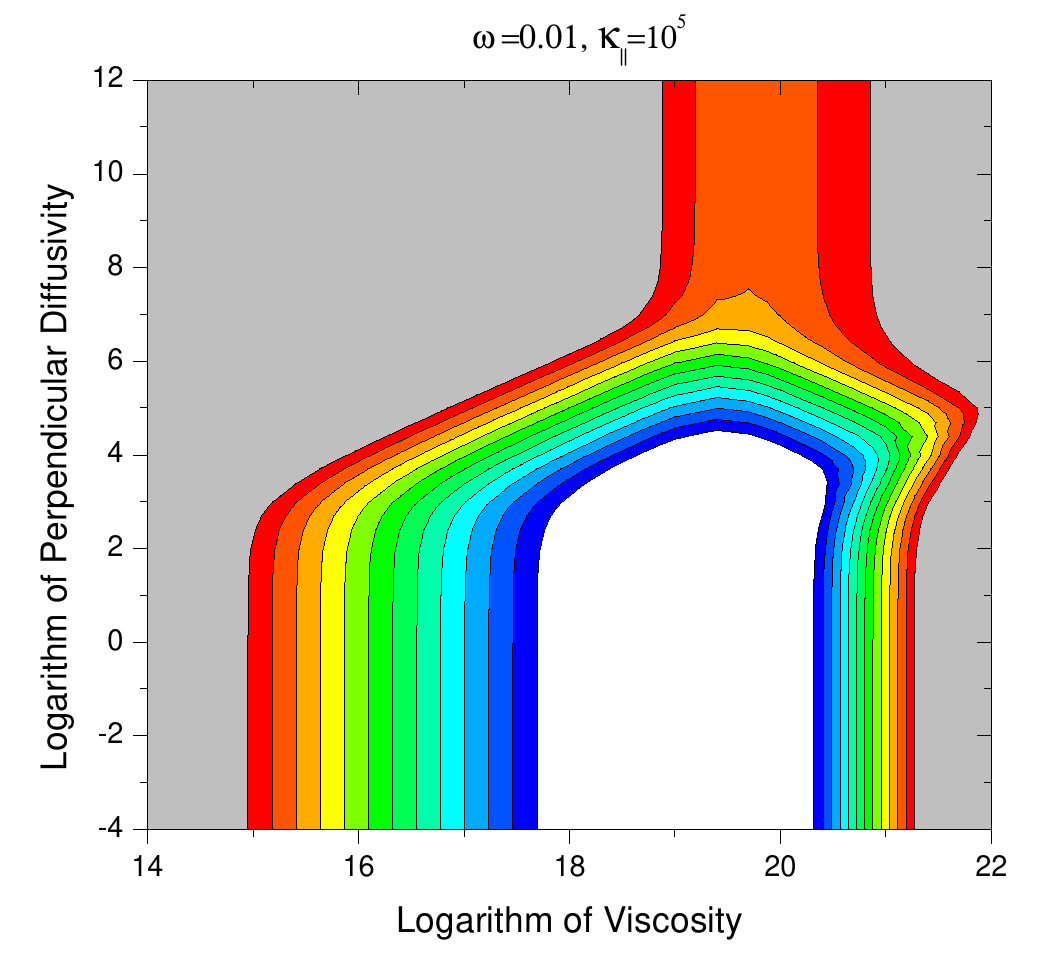}
\includegraphics[angle=0,scale=.5]{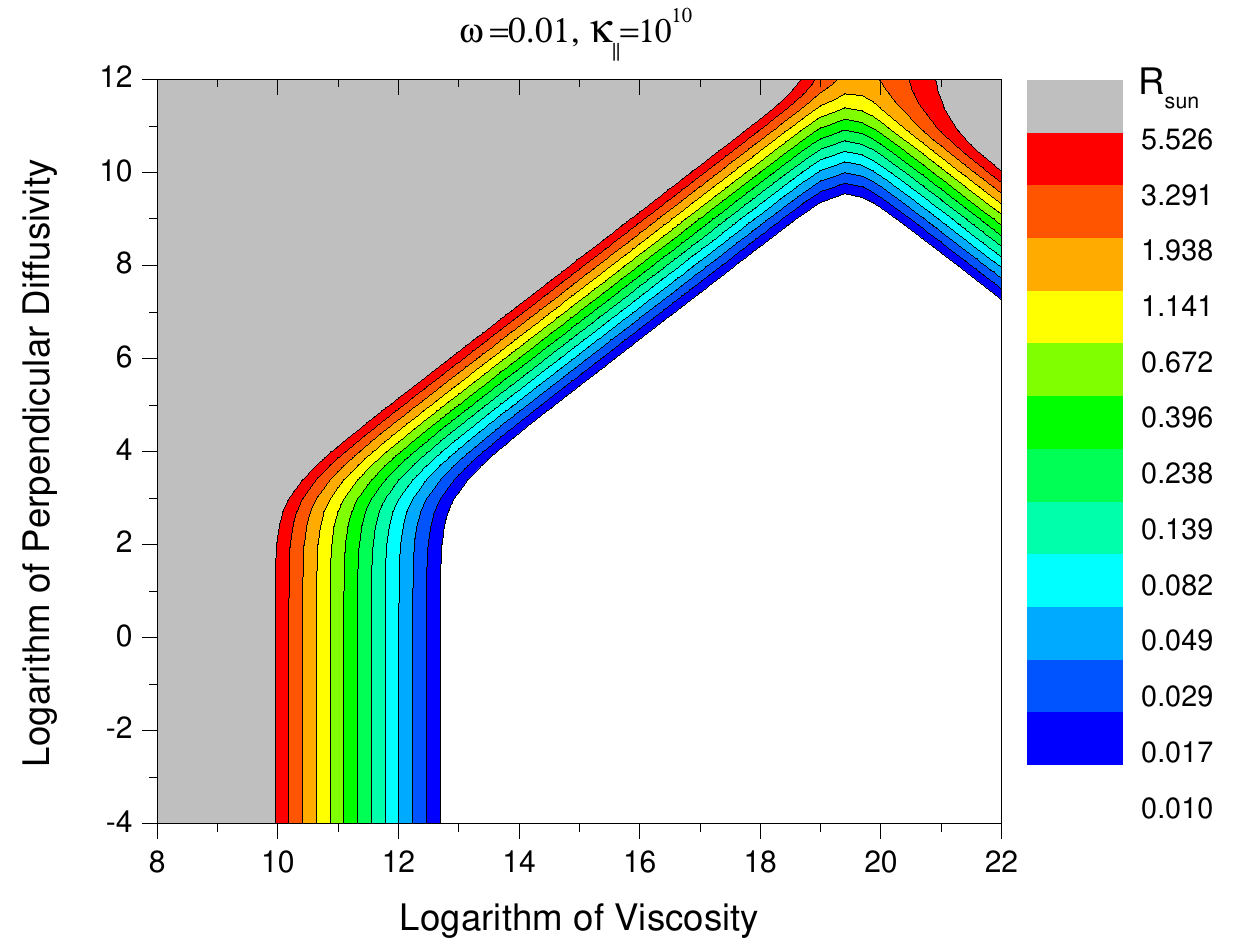}
\caption{The variation of the damping length scale of the wave having a frequency $\omega=0.01 \,rad/s$ with viscosity and perpendicular diffusivity for different $\kappa_{\parallel}=10^2, 10^5, 10^{10}  \,\,erg\,cm^{-1}s^{-1}K^{-1} $ values.}
\end{center}
\end{figure*}

\subsubsection{Solution for ($X_2,Y_2$)}

If we substitute $(X_2,Y_2)$ into Equation (24) we obtain dispersion relation given as


\begin{equation}
\quad k^2 \left(M-{v_A}^2\right)\left({v_A}^2-i\omega\nu \right)+4\omega 0.6 \frac{1}{\rho_0} H_\parallel k -\left(M-{v_A}^2\right) \omega^2=0.
\end{equation}When assumed that $H_{\parallel}=0 $ in this equation,  Alfven wave given with equation (29) is obtained. 
   
Figure 7 shows the variation of the damping length scale of a magnetosonic wave with frequency and perpendicular diffusivity, assuming the value of viscosity as $10^{19} \,cm^{2}\,s^{-1} $ and $\kappa_{\parallel}$ values as $10, 10^5$ and $ 10^{10} \,\,erg\,cm^{-1}s^{-1}K^{-1}$. For $\kappa_{\parallel}=10 \,\,erg\,cm^{-1}s^{-1}K^{-1}$, while the waves with frequencies greater than $ 10^{-2.1}\,rad/s$  get damped at the required distances for the heating of corona and solar wind acceleration, waves with smaller frequencies get damped at greater radial distances. When $\kappa_{\parallel}=10^5 \,\,erg\,cm^{-1}s^{-1}K^{-1} $, while the lesser frequency waves get damped at the required distances for perpendicular diffusivity smaller than $\sim10^5 \,\,erg\,cm^{-1}s^{-1}K^{-1}$,  higher frequency waves get damped at diffusivity values greater than $ 10^5 \,\,erg\,cm^{-1}s^{-1}K^{-1}$. When the $\kappa_{\parallel}=10^{10} \,\,erg\,cm^{-1}s^{-1}K^{-1}$, waves get damped generally at smaller radial distances. There is a small region wherein required damping occurs for greater perpendicular diffusivity values.   
 
In Figure 8, by assuming different $\kappa_{\parallel}$ values (i.e.  $10^2, 10^5, 10^{10}  \,\,erg\,cm^{-1}s^{-1}K^{-1} $) we plotted the variation of the damping length scale of the wave having a frequency $\omega=0.01 \,rad/s$ with viscosity and perpendicular diffusivity. While damping occurs independent of diffusivity when the viscosity has a value within the range $10^{18}- \sim10^{21} \,cm^{2}\,s^{-1} $ for $\kappa_{\parallel}=100  \,\,erg\,cm^{-1}s^{-1}K^{-1} $, in case  $\kappa_{\parallel}$ have greater values, damping of waves occurs within the larger range of viscosity. When $\kappa_{\parallel}= 10^{10}  \,\,erg\,cm^{-1}s^{-1}K^{-1} $ damping of waves occurs independent of diffusivity for small viscosity ($10^{10}- \sim10^{13} \,cm^{2}\,s^{-1} $) and perpendicular diffusivity values within the range  $10^{-4}-10^3  \,\,erg\,cm^{-1}s^{-1}K^{-1} $. For the viscosity values within the range  $10^{10}-10^{19} \,cm^{2}\,s^{-1} $ it is seen that the value of the perpendicular diffusion coefficient increases linearly with viscosity for all the damping length scales. 

Although the parameter ranges giving required damping length scales for these waves have been determined, it is seen that the calculated wave fluxes are inadequate for heating the corona and solar wind acceleration. These waves can not replace the loss energy in NPCH.

\begin{figure*}
\begin{center}
\includegraphics[angle=0,scale=.49]{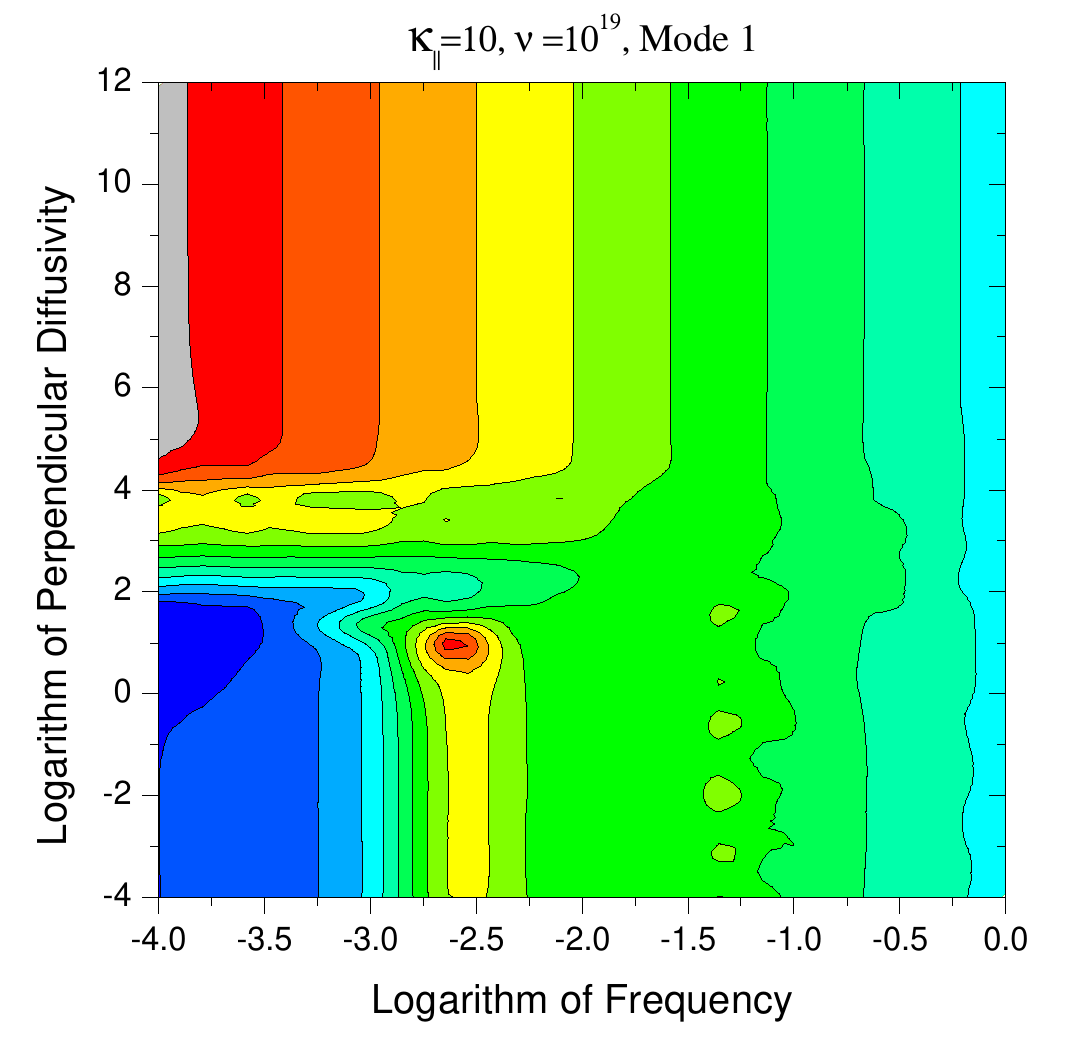}
\includegraphics[angle=0,scale=.49]{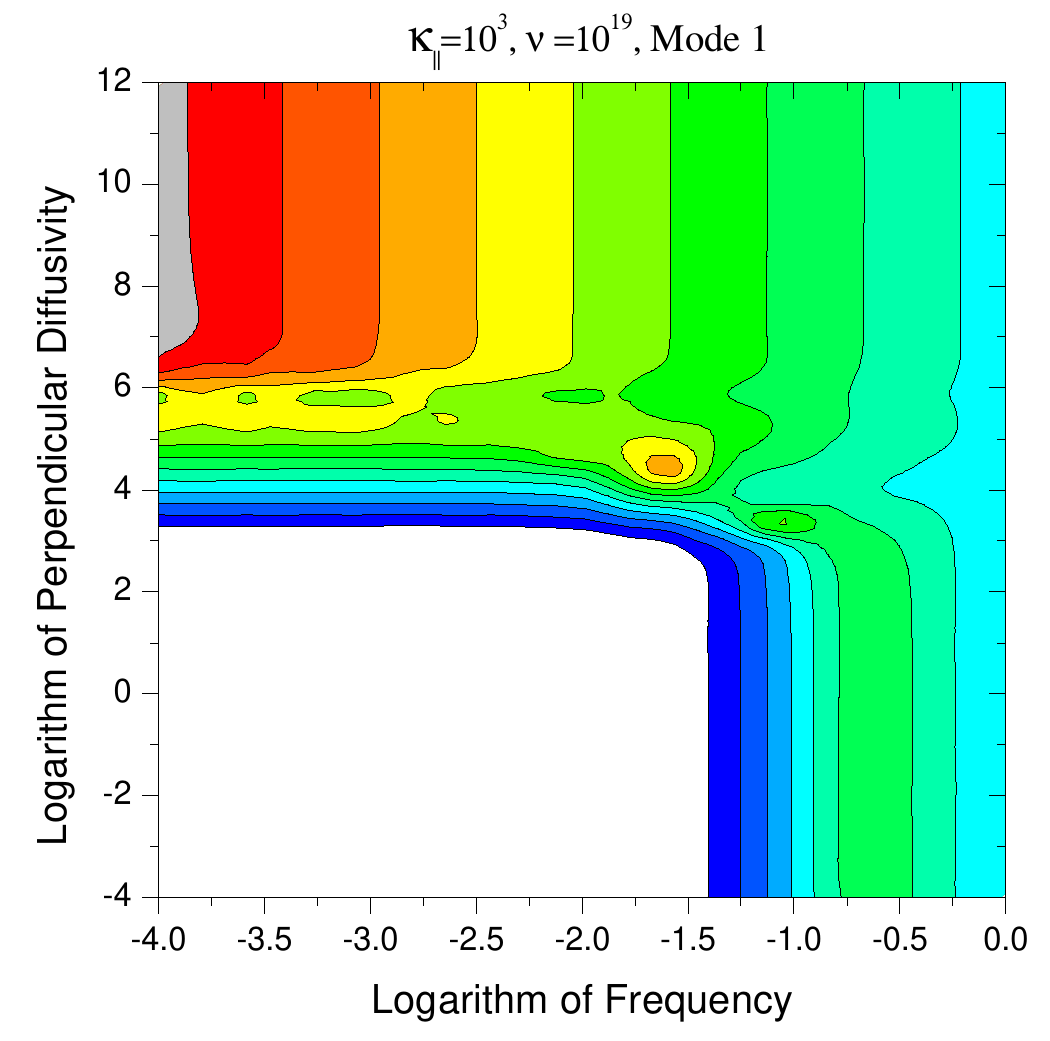}
\includegraphics[angle=0,scale=.53]{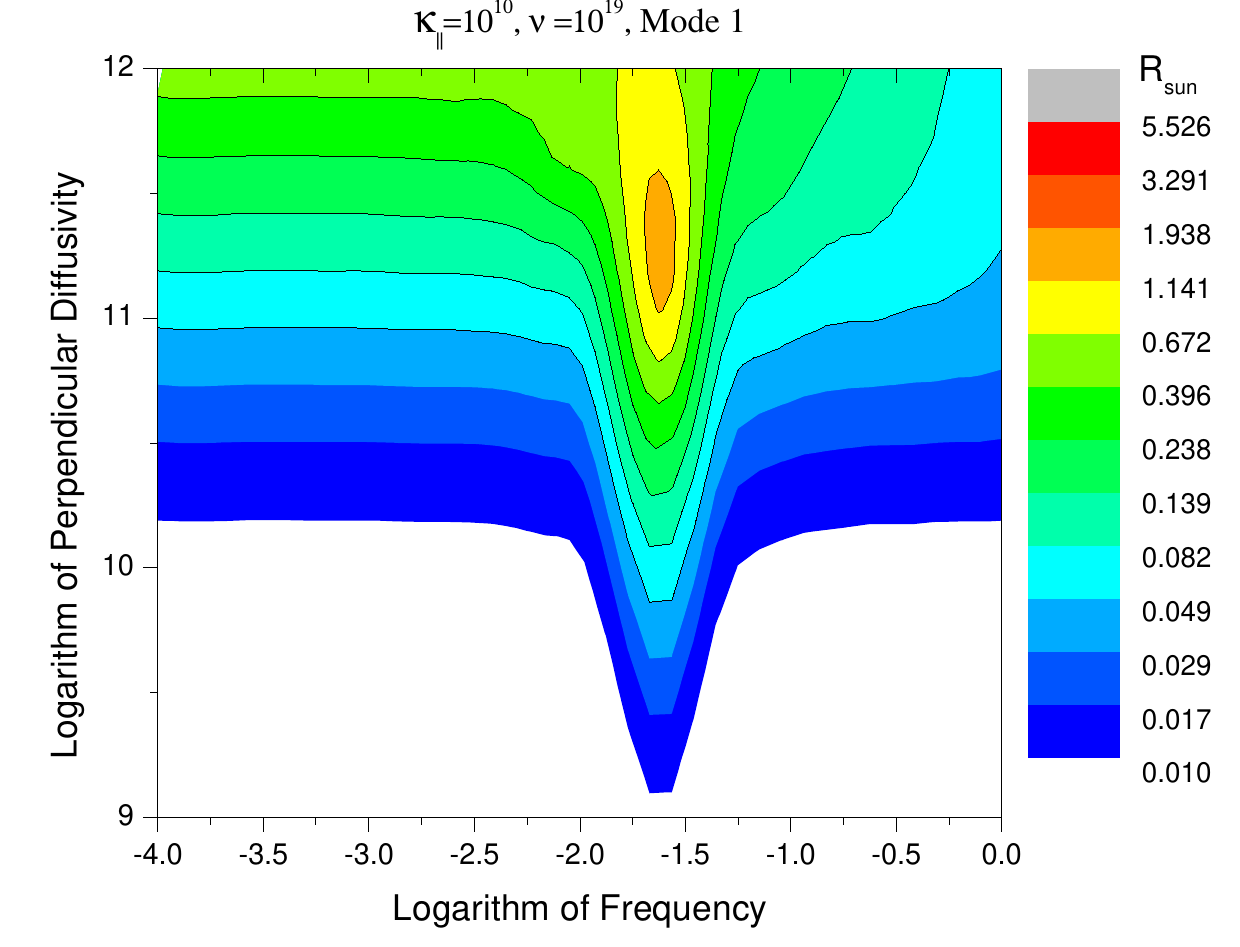}
\includegraphics[angle=0,scale=.49]{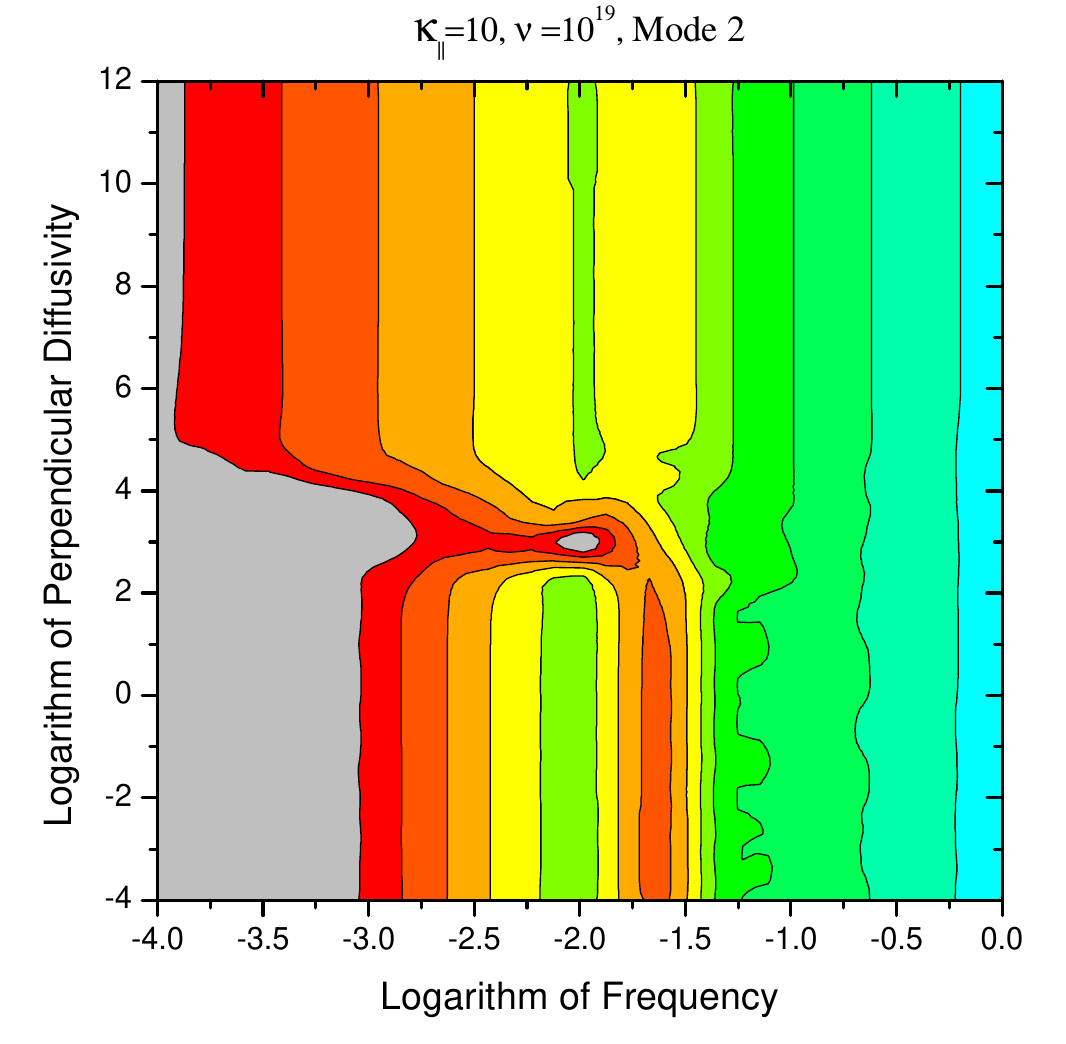}
\includegraphics[angle=0,scale=.49]{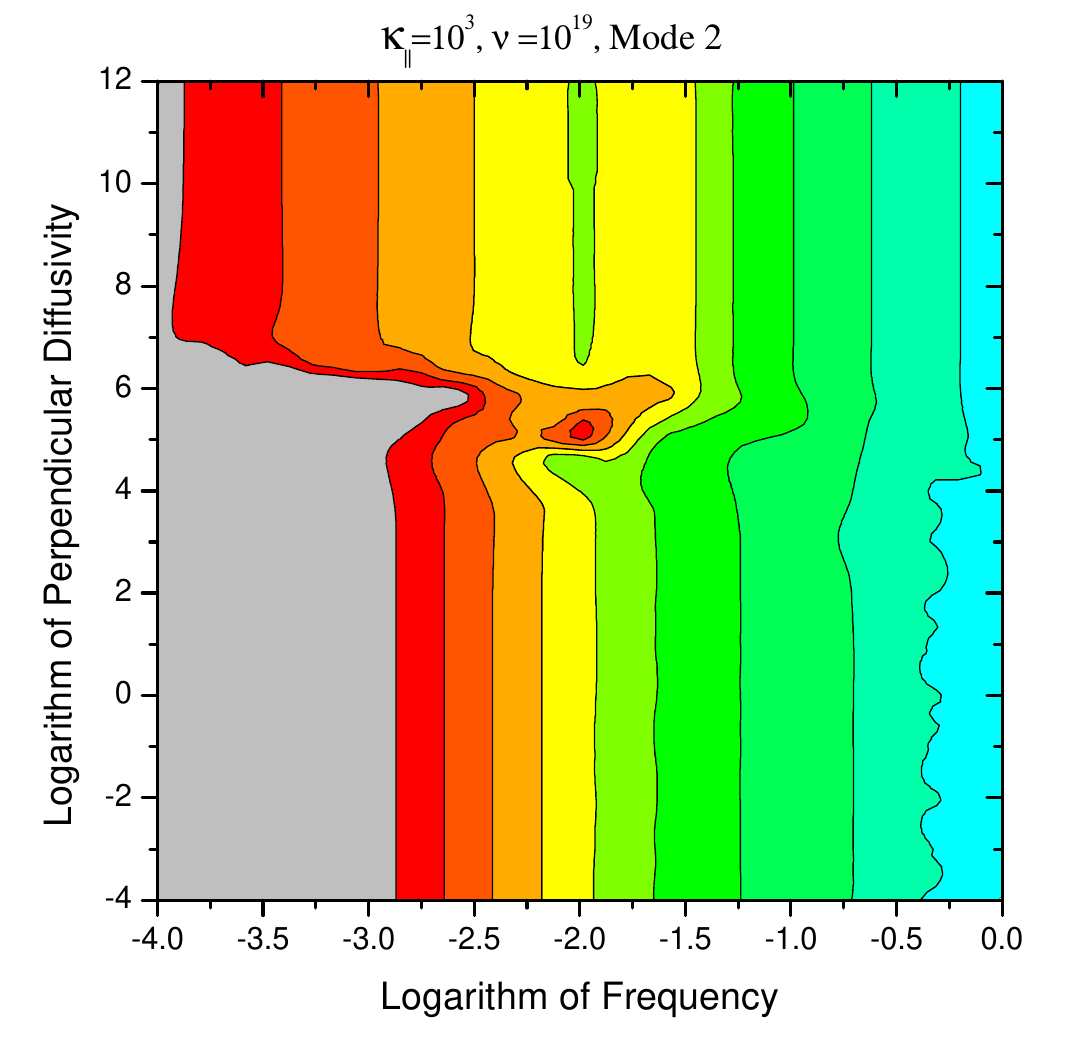}
\includegraphics[angle=0,scale=.53]{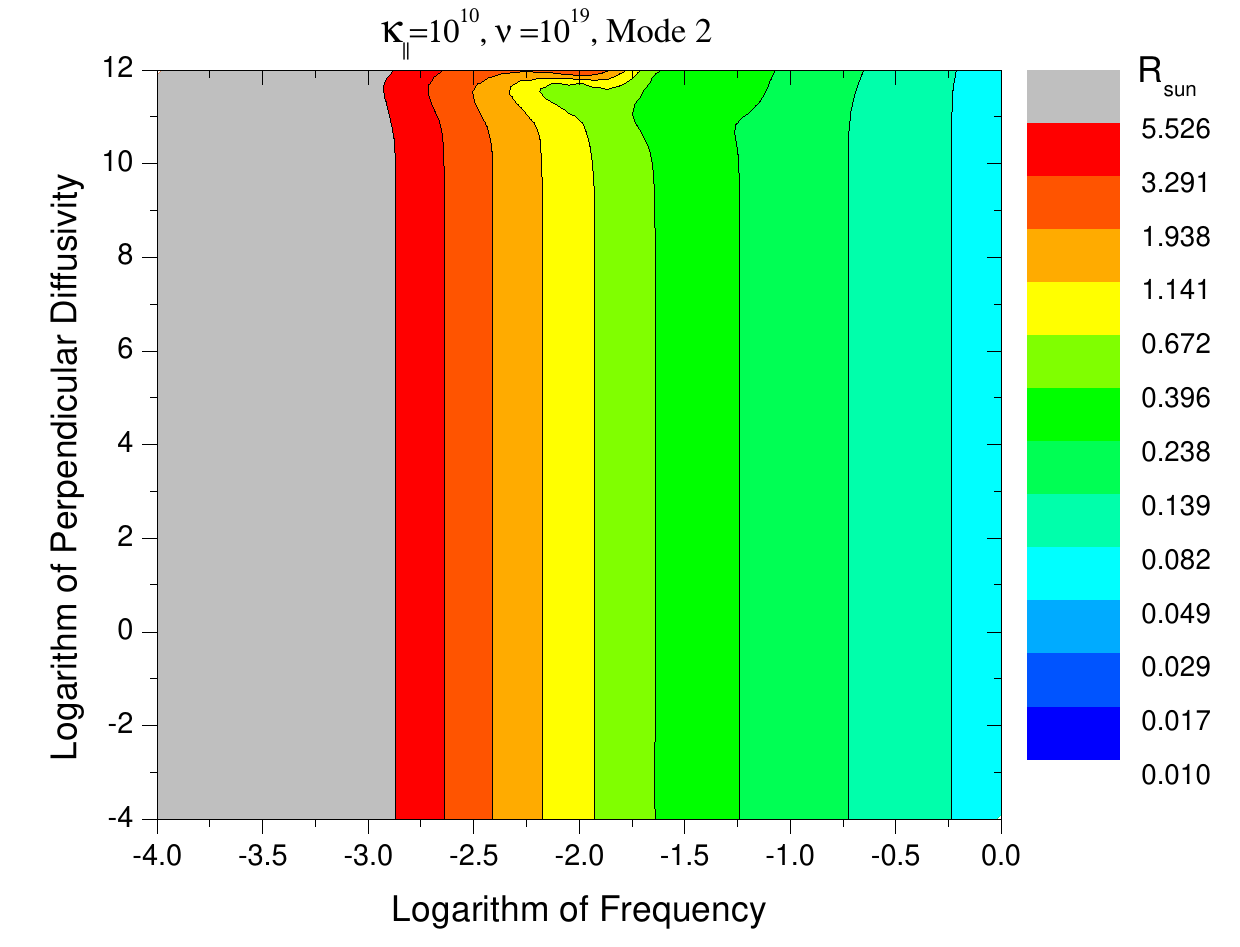}
\includegraphics[angle=0,scale=.49]{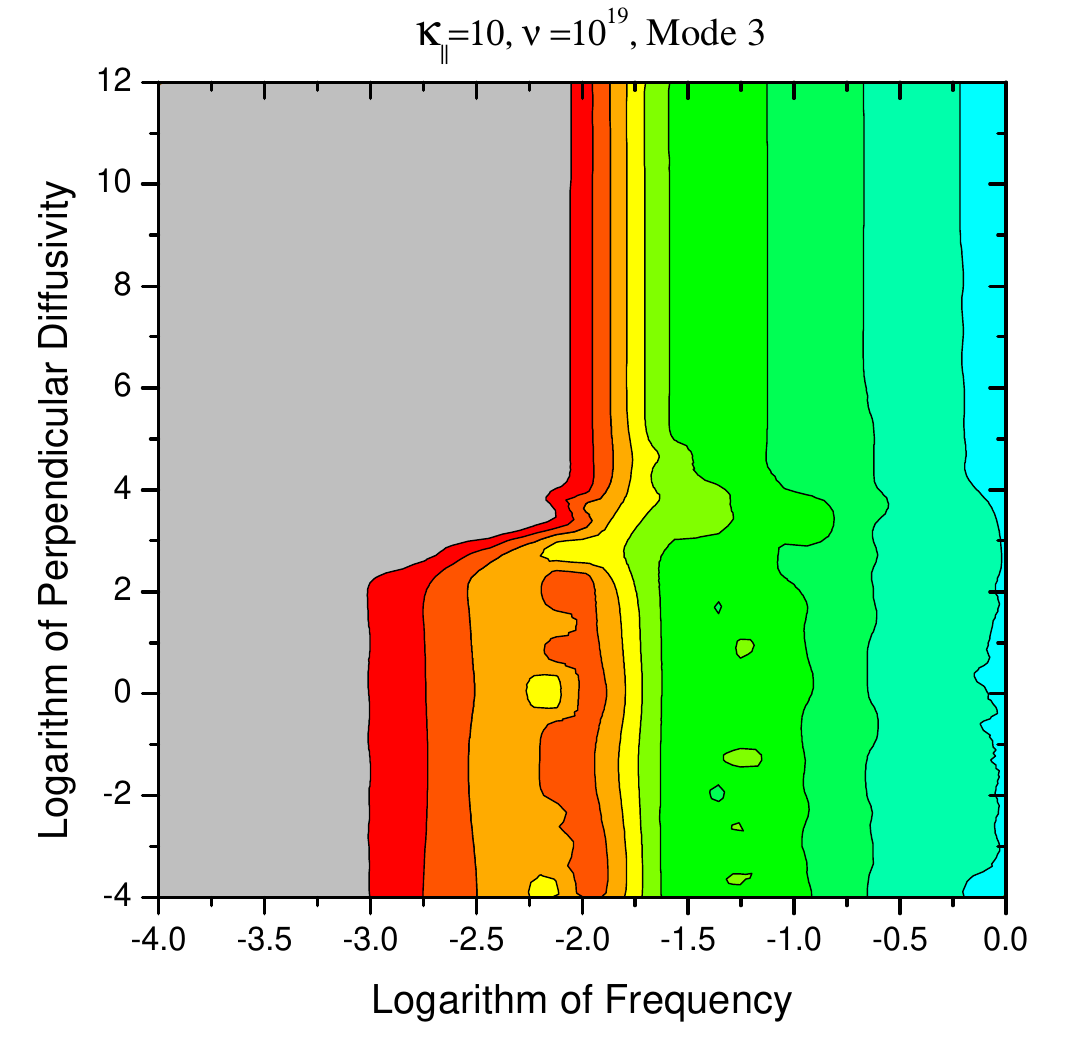}
\includegraphics[angle=0,scale=.49]{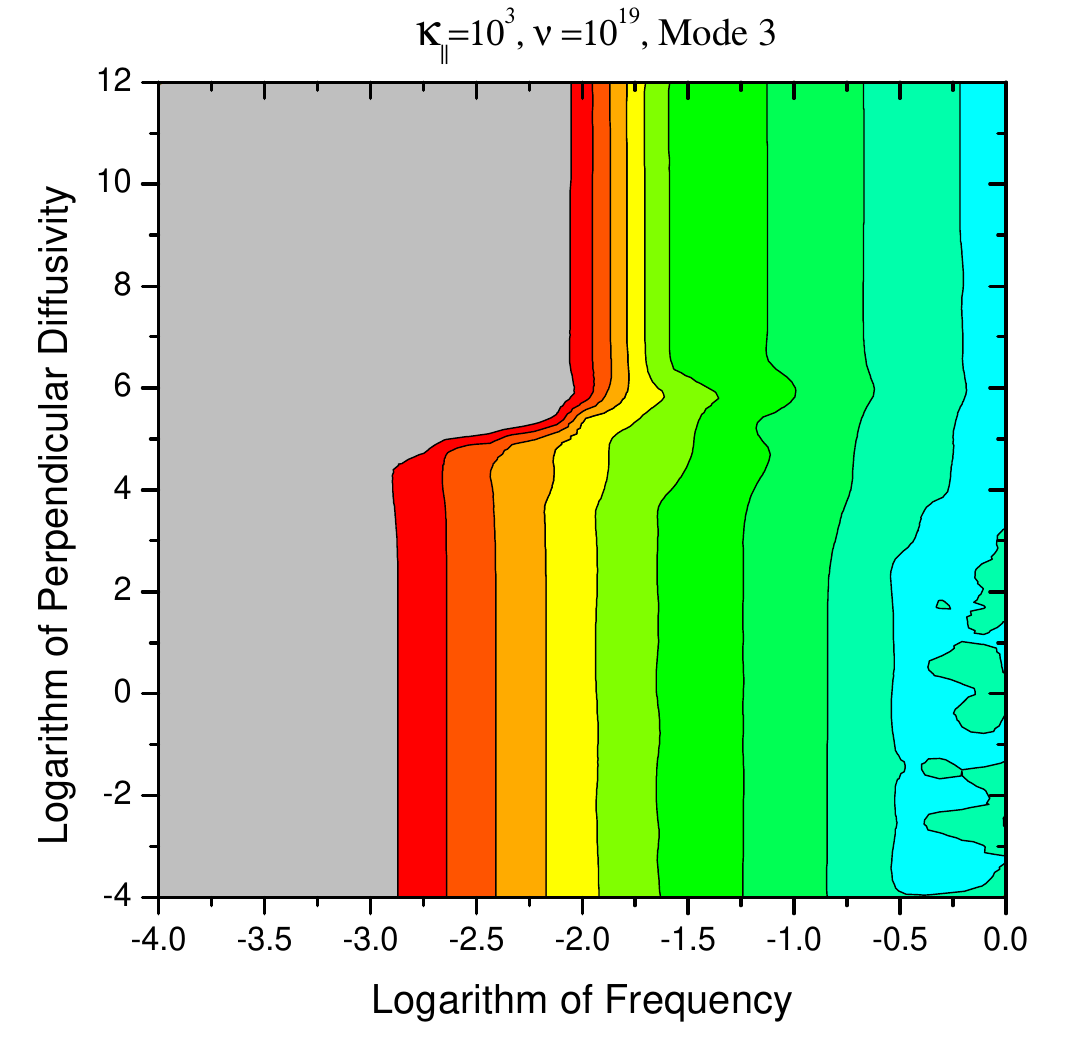}
\includegraphics[angle=0,scale=.5]{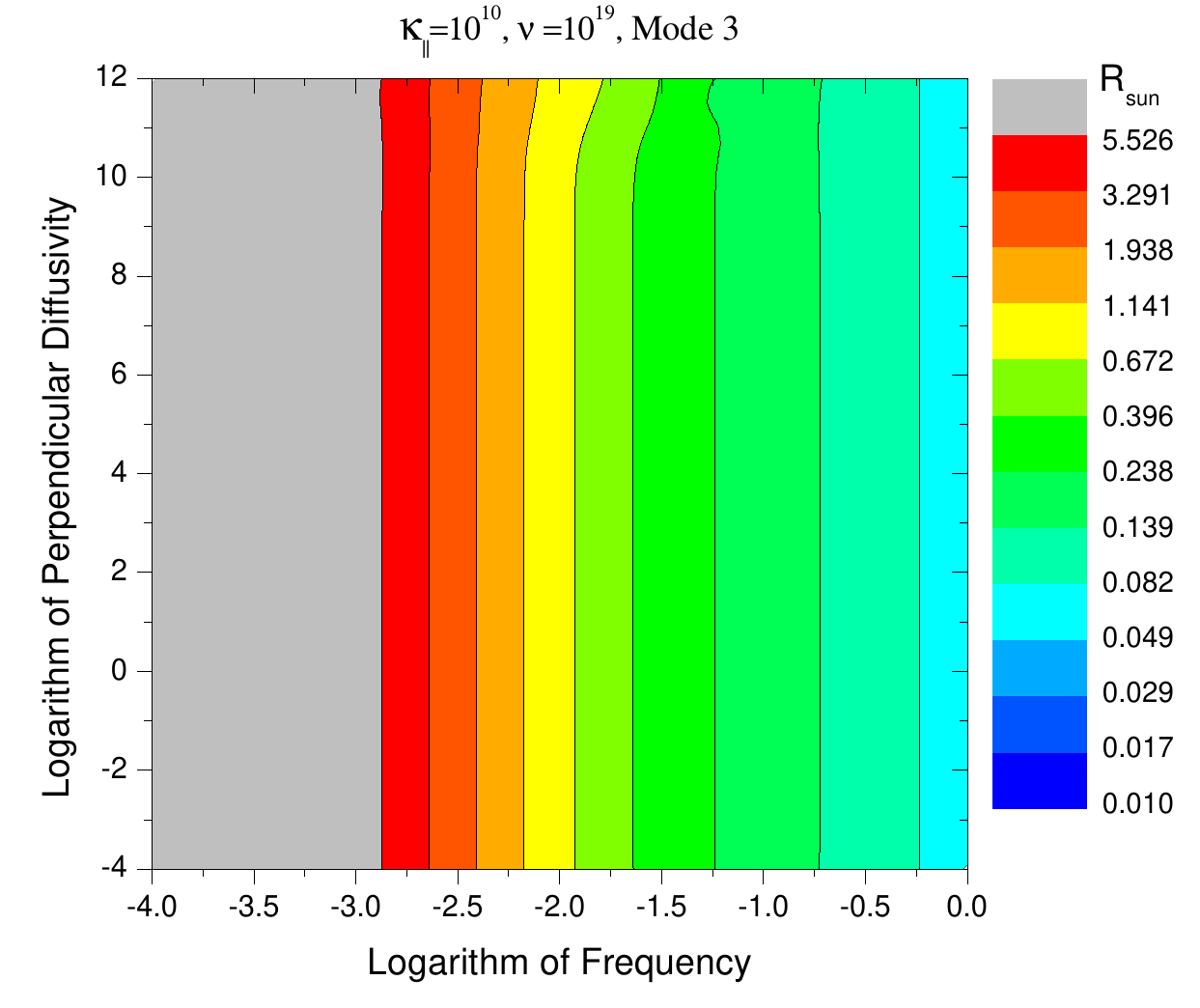}
\caption{The variations of the damping length scales of Mode 1 (top), Mode 2 (middle) and Mode 3 (bottom) with wave frequency and perpendicular diffusivity for $\kappa_{\parallel}=10, 10^3, 10^{10} \,\,erg\,cm^{-1}s^{-1}K^{-1} $ values. Viscosity is assumed as ($10^{19}\,cm^{2}\,s^{-1}$) in all graphics.}
\end{center}
\end{figure*}

\begin{figure*}
\begin{center}
\includegraphics[angle=0,scale=.48]{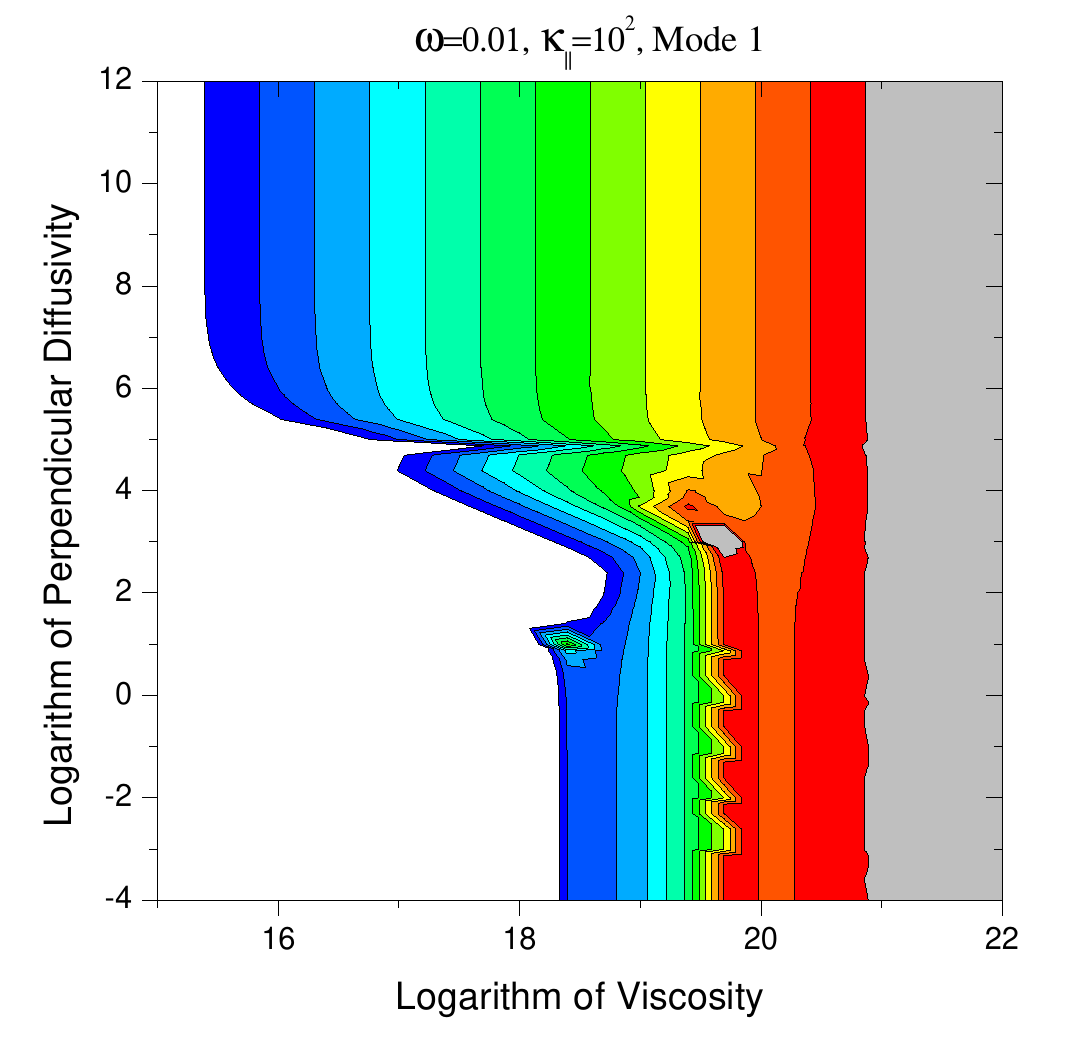}
\includegraphics[angle=0,scale=.52]{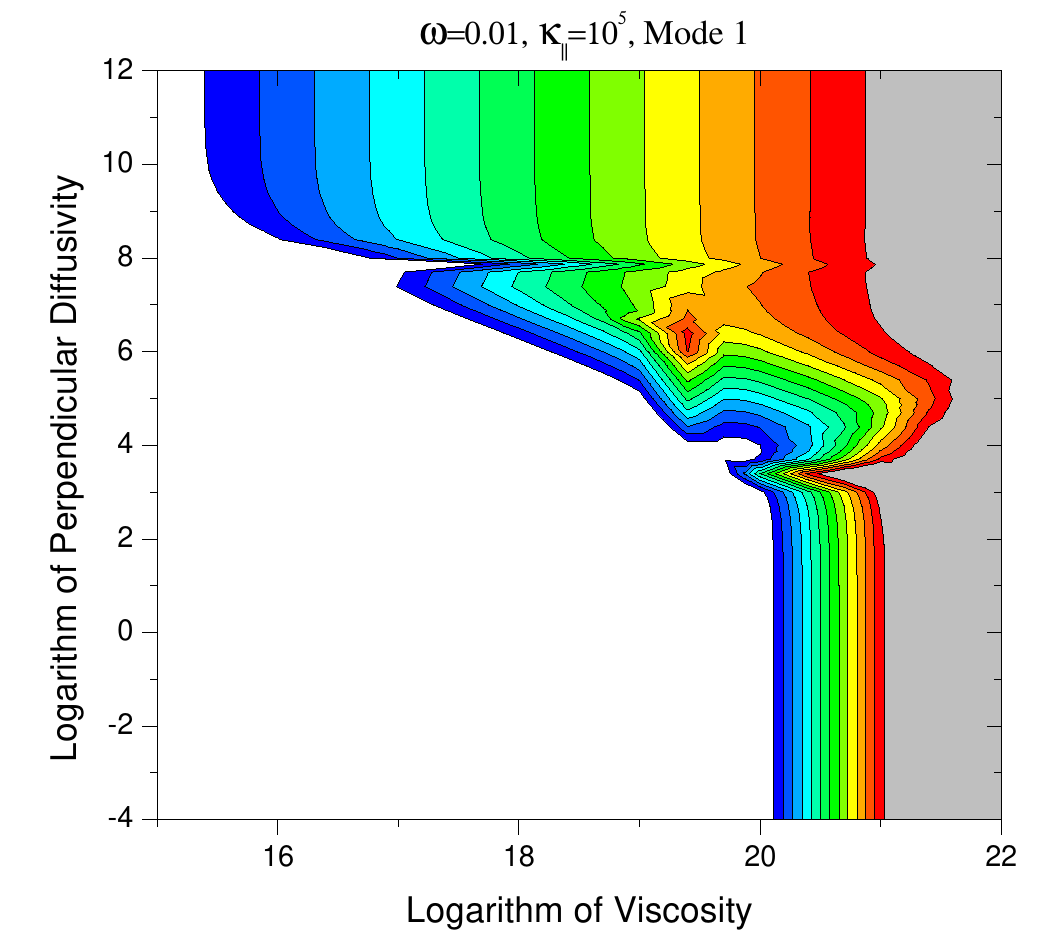}
\includegraphics[angle=0,scale=.48]{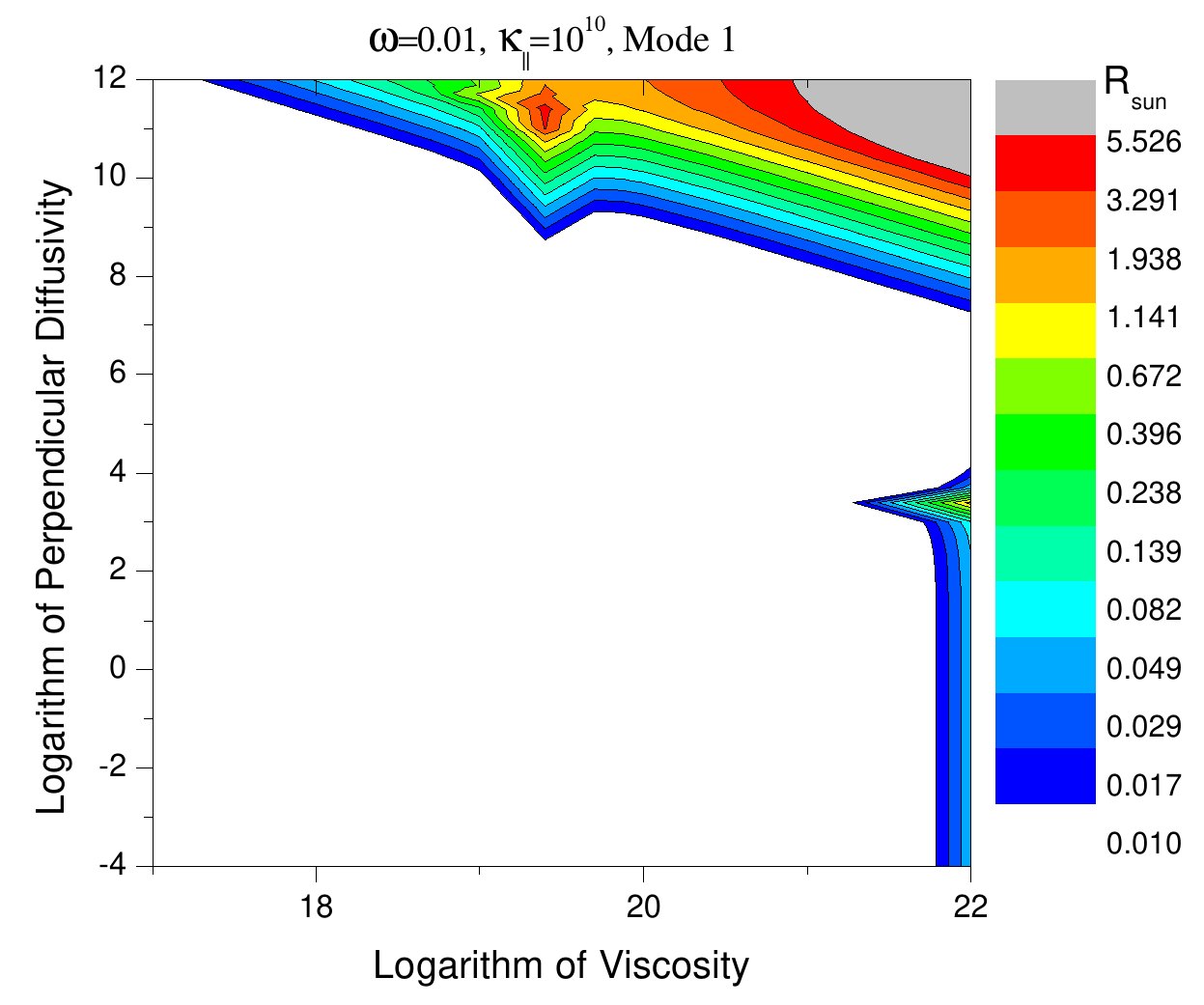}
\includegraphics[angle=0,scale=.49]{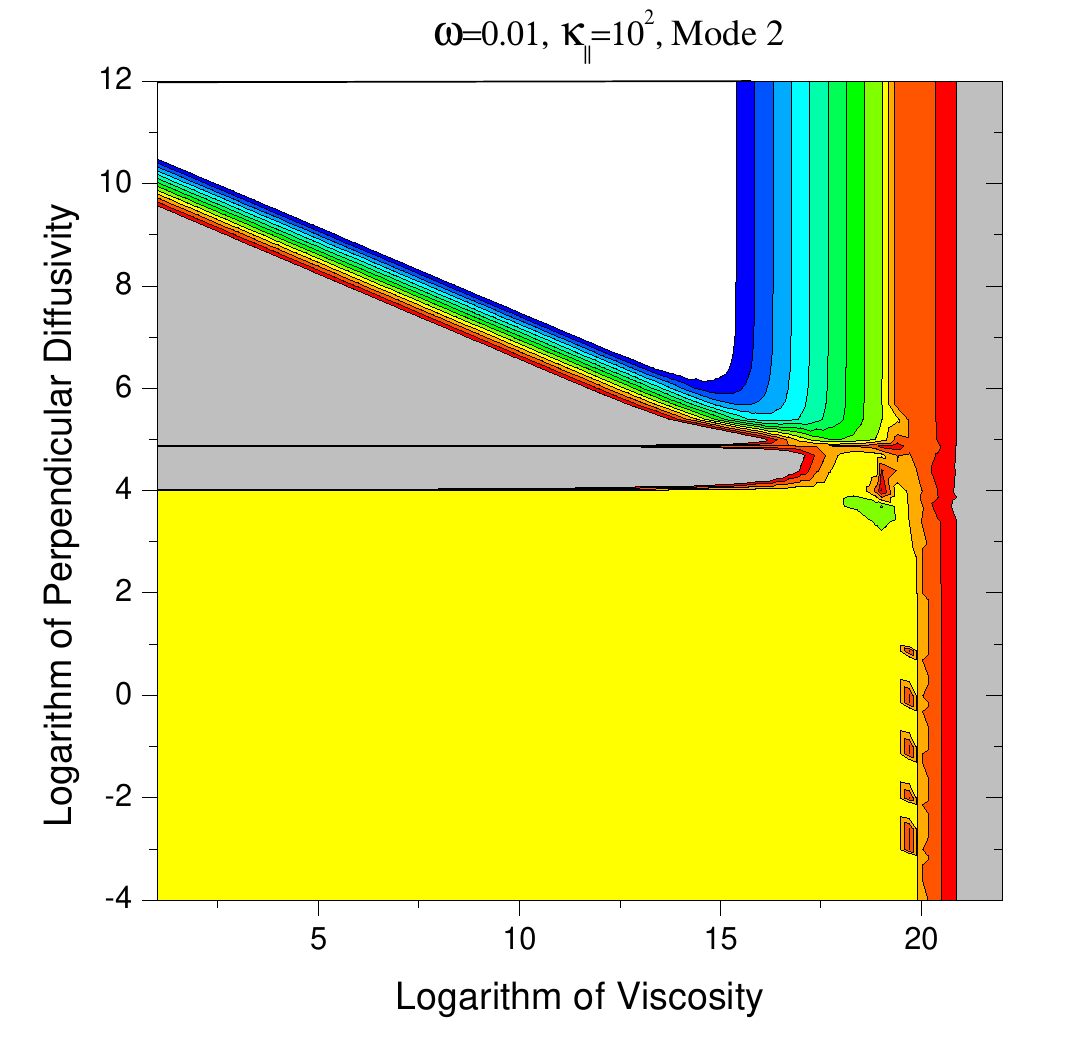}
\includegraphics[angle=0,scale=.52]{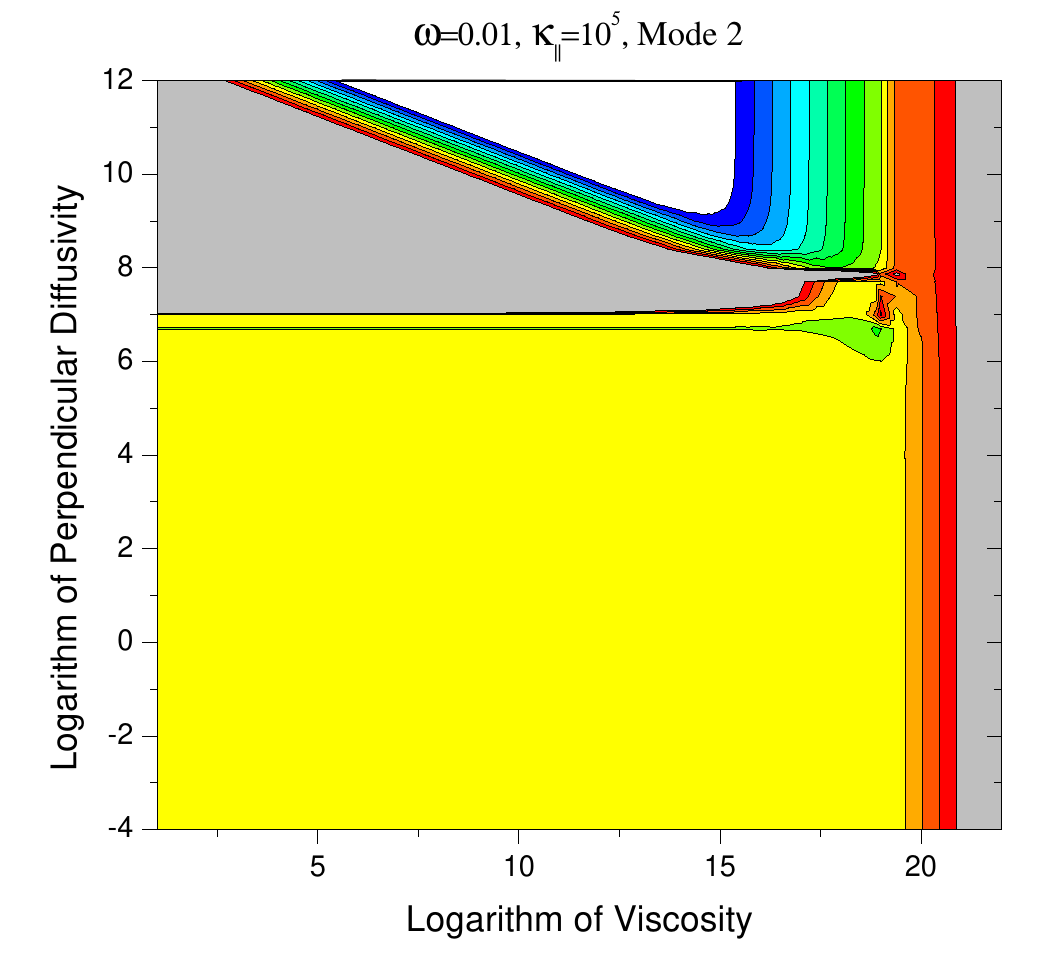}
\includegraphics[angle=0,scale=.52]{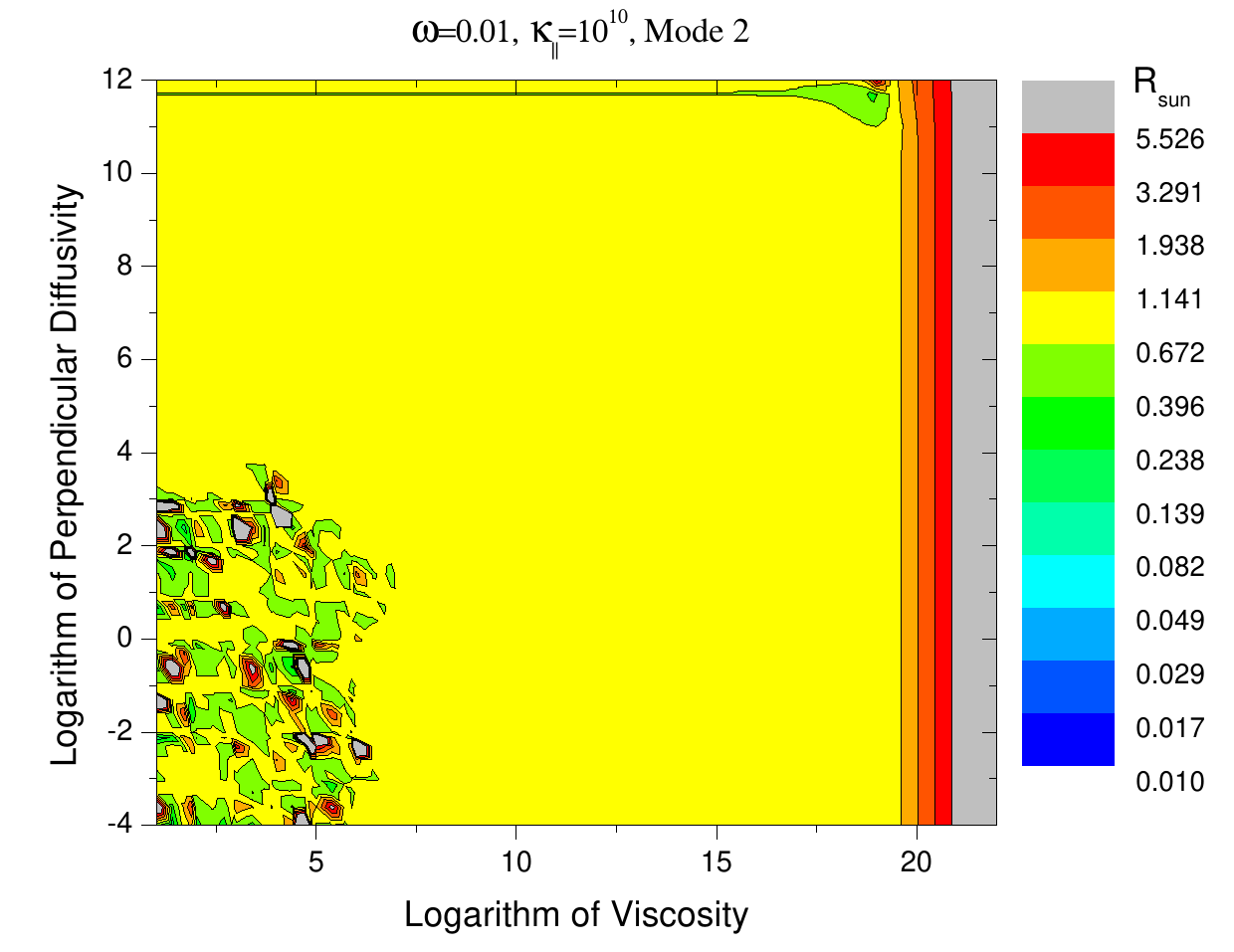}
\includegraphics[angle=0,scale=.49]{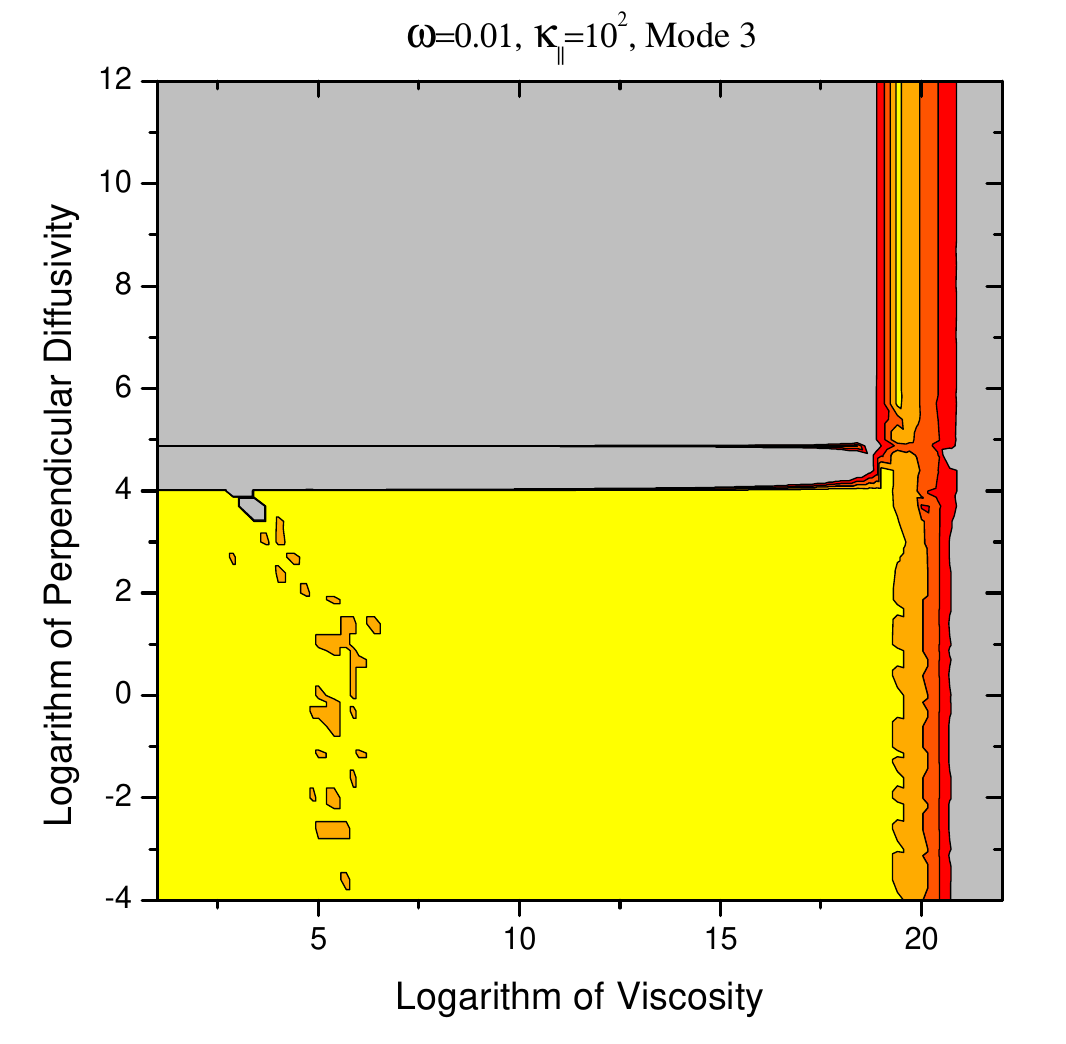}
\includegraphics[angle=0,scale=.49]{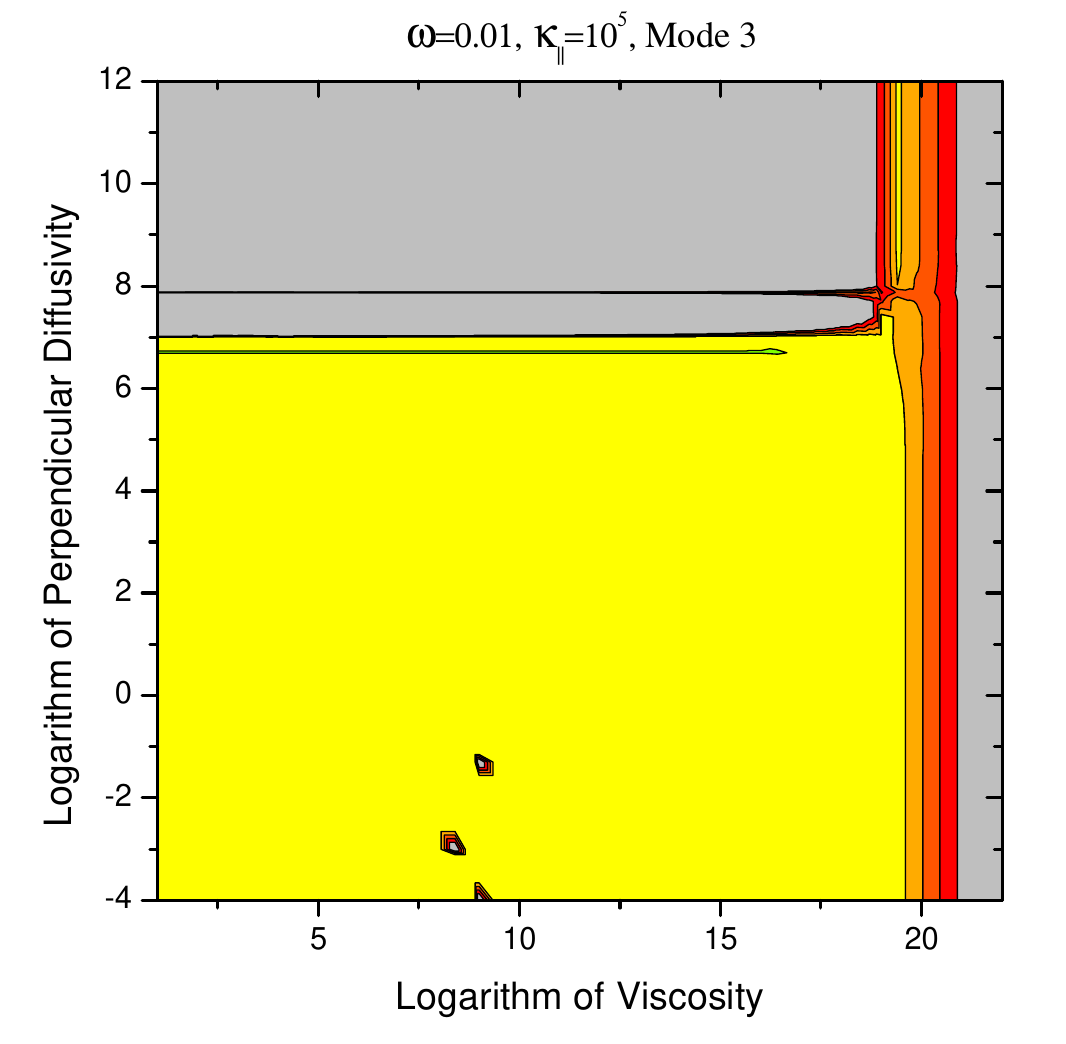}
\includegraphics[angle=0,scale=.53]{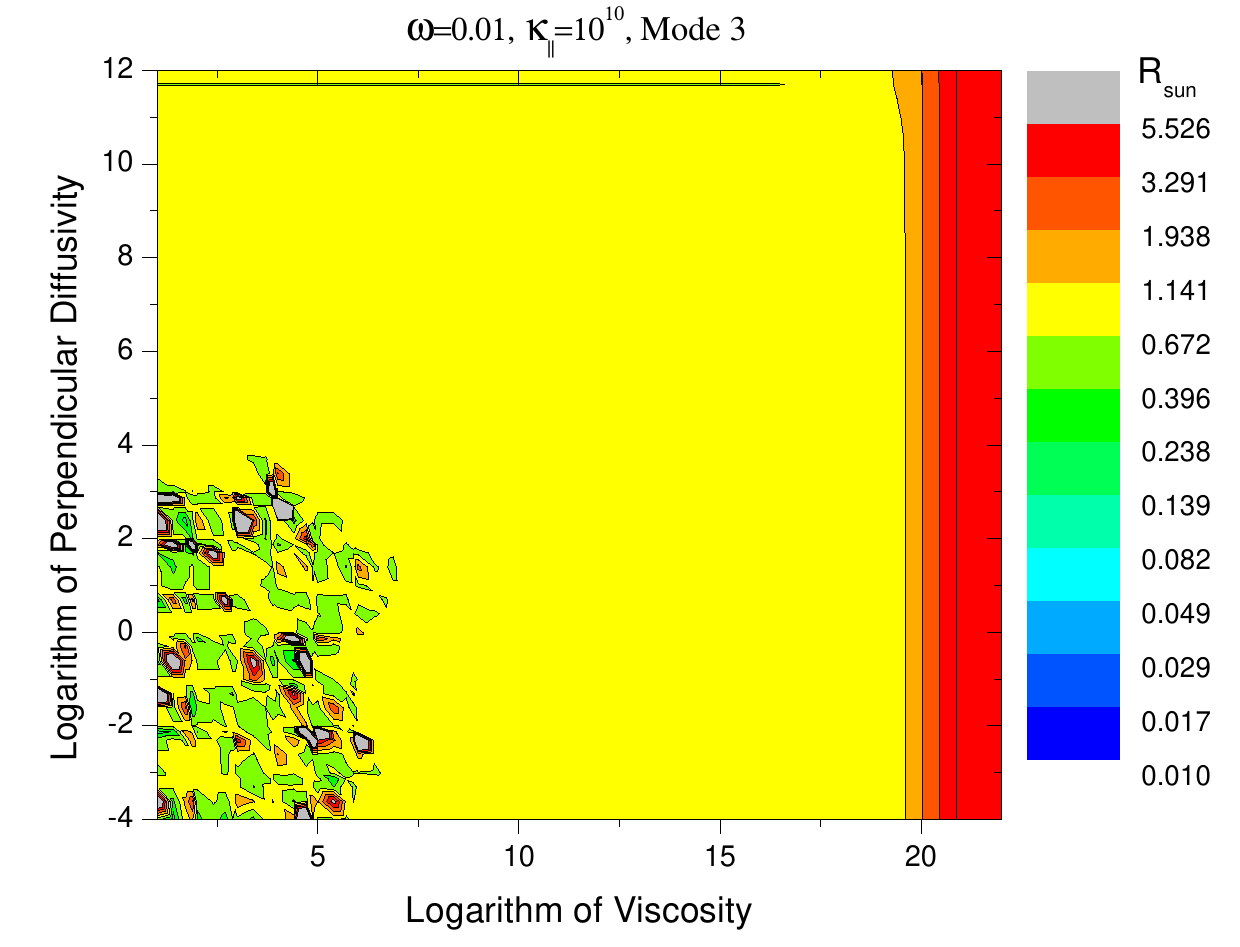}
\caption{The variation of  damping length scales of Mode 1 (top), Mode 2 (middle) and Mode 3 (bottom) having frequency $\omega=0.01 \,rad/s$ with viscosity and perpendicular diffusivity for $\kappa_{\parallel}=10^2, 10^5$ and $10^{10}\,\,erg\,cm^{-1}s^{-1}K^{-1} $.}
\end{center}
\end{figure*}

\subsubsection{Solution for ($X_3,Y_3$)}

Let us substitute ($X_3$,$Y_3$) given into equation (24) using Taylor series expansion  and after a lengthy effort we reach the dispersion relation as Equation (A2) in the Appendix.

The solution of the dispersion relation (A2) yields five modes. Solutions of three modes are shown  in Figures 9 and 10. Other two mode have very long damping length scales, therefore we didn't give their graphs. 

In Figure 9, assuming that viscosity has a constant value ($10^{19}\,cm^{2}\,s^{-1} $), variations of the damping length scales of Mode 1 (top), Mode 2 (middle) and Mode 3 (bottom) with wave frequency and perpendicular diffusivity are shown for $\kappa_{\parallel}=10, 10^3, 10^{10} \,\,erg\,cm^{-1}s^{-1}K^{-1} $ values.   

In case $\kappa_{\parallel}=10 \,\,erg\,cm^{-1}s^{-1}K^{-1} $, the damping length scale of Mode 1 have required values for heating corona and solar wind acceleration. When $\kappa_{\parallel}=10^3 \,\,erg\,cm^{-1}s^{-1}K^{-1} $, in order the waves with frequencies smaller than $\omega=10^{-1.5}\,rad/s$ to be damped within required distances, perpendicular diffusivity should have values greater than $10^3$. For $\kappa_{\parallel}=10^{10} \,\,erg\,cm^{-1}s^{-1}K^{-1} $, the region of the damping length scale with $R\ga 1.2 R_{\odot}$ range disappears. Waves get damped very close to the sun.

Damping length scale of Mode 2 is more or less independent of perpendicular diffusivity. However, as $\kappa_{\parallel}$ increases, waves with lower frequencies (i.e. $\omega\la 10^{-2.9}\,rad/s$) get damped at further radial distances instead of required ones for perpendicular diffusivity value $>10^5 \,\,erg\,cm^{-1}s^{-1}K^{-1} $.

Mode 3 shows similar variation to Mode 2. But, in any case, waves with frequencies smaller than $\omega\la 10^{-3}\,rad/s$ get damped at very far radial distances.

In Figure 10, assuming $\kappa_{\parallel}=10^2, 10^5, 10^{10}\,\,erg\,cm^{-1}s^{-1}K^{-1} $, the variation of  damping length scales of Mode 1 (top), Mode 2 (middle) and Mode 3 (bottom) having frequency $\omega=0.01 \,rad/s$ with viscosity and perpendicular diffusivity are shown.

When viscosity values are taken within the range $10^{15.5}-10^{21} \,cm^{2}\,s^{-1}$ and the value of perpendicular diffusivity is greater than $10^6 \,\,erg\,cm^{-1}s^{-1}K^{-1} $, damping of Mode 1  occurs at the required distances. For this wave, larger viscosity values are required in order the wave damping occuring at required distances when smaller values of perpendicular diffusivity are assumed.  When $\kappa_{\parallel}$  reaches a value $10^{10} \,\,erg\,cm^{-1}s^{-1}K^{-1} $, waves get damped at the required distances only when higher perpendicular diffusivity values are assumed.  

For Mode 2 in cases of $\kappa_{\parallel}=10^2$ and $10^5 \,\,erg\,cm^{-1}s^{-1}K^{-1} $, when the perpendicular diffusivity is less than $10^4 \,\,erg\,cm^{-1}s^{-1}K^{-1}$, there is a region around $1 \,R_{\odot}$  wherein waves get damped. In this region, damping is independent of viscosity values. For greater perpendicular diffusivity values ($> 10^5 \,\,erg\,cm^{-1}s^{-1}K^{-1} $) a secondary damping region emerges. This region disappears as the value of $\kappa_{\parallel}$  increases and the first damping region takes over.  

Mode 3 displays a similar structure as the Mode 2 but does not show a secondary damping region.

\begin{figure*}
\begin{center}
\includegraphics[angle=0,scale=.48]{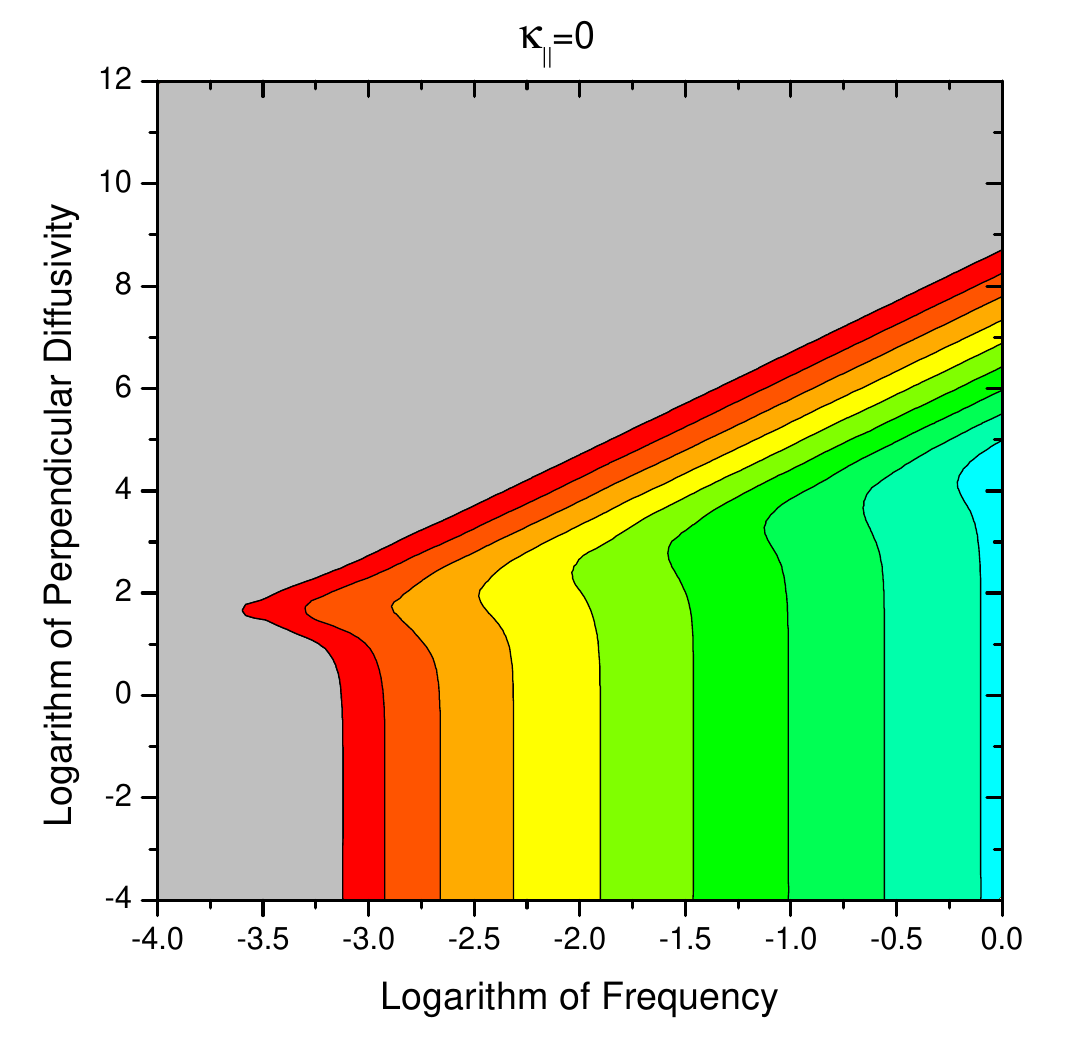}
\includegraphics[angle=0,scale=.48]{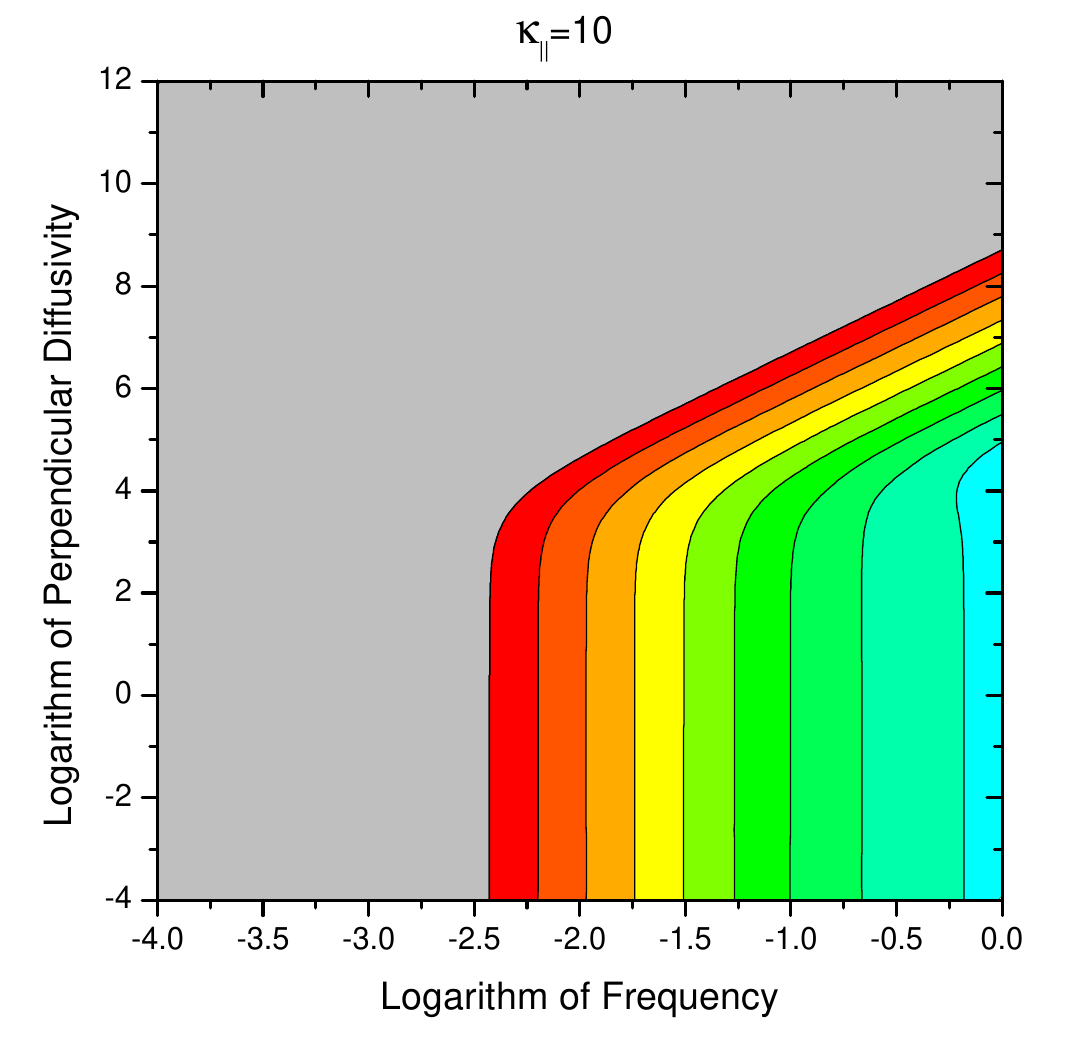}
\includegraphics[angle=0,scale=.52]{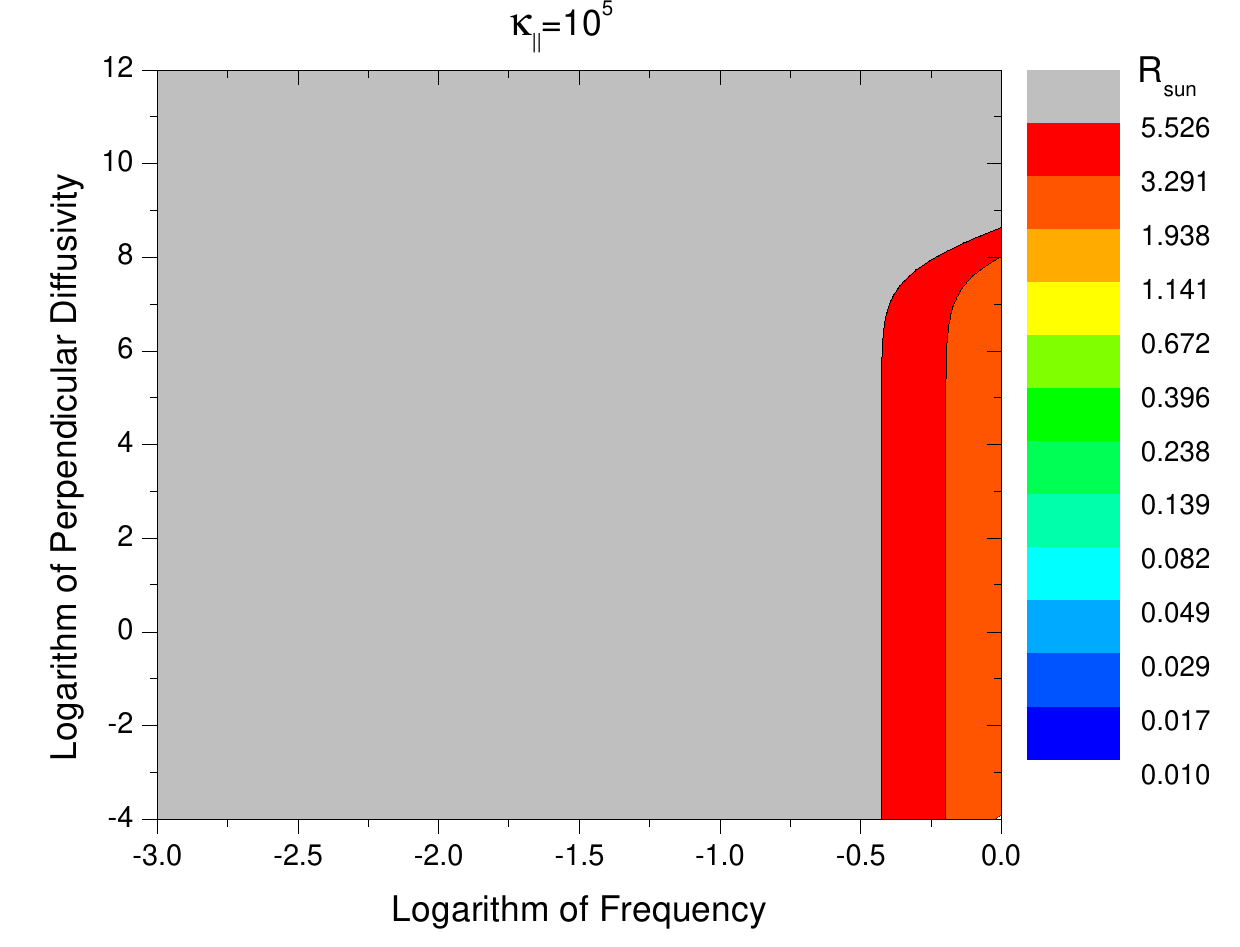}
\includegraphics[angle=0,scale=.48]{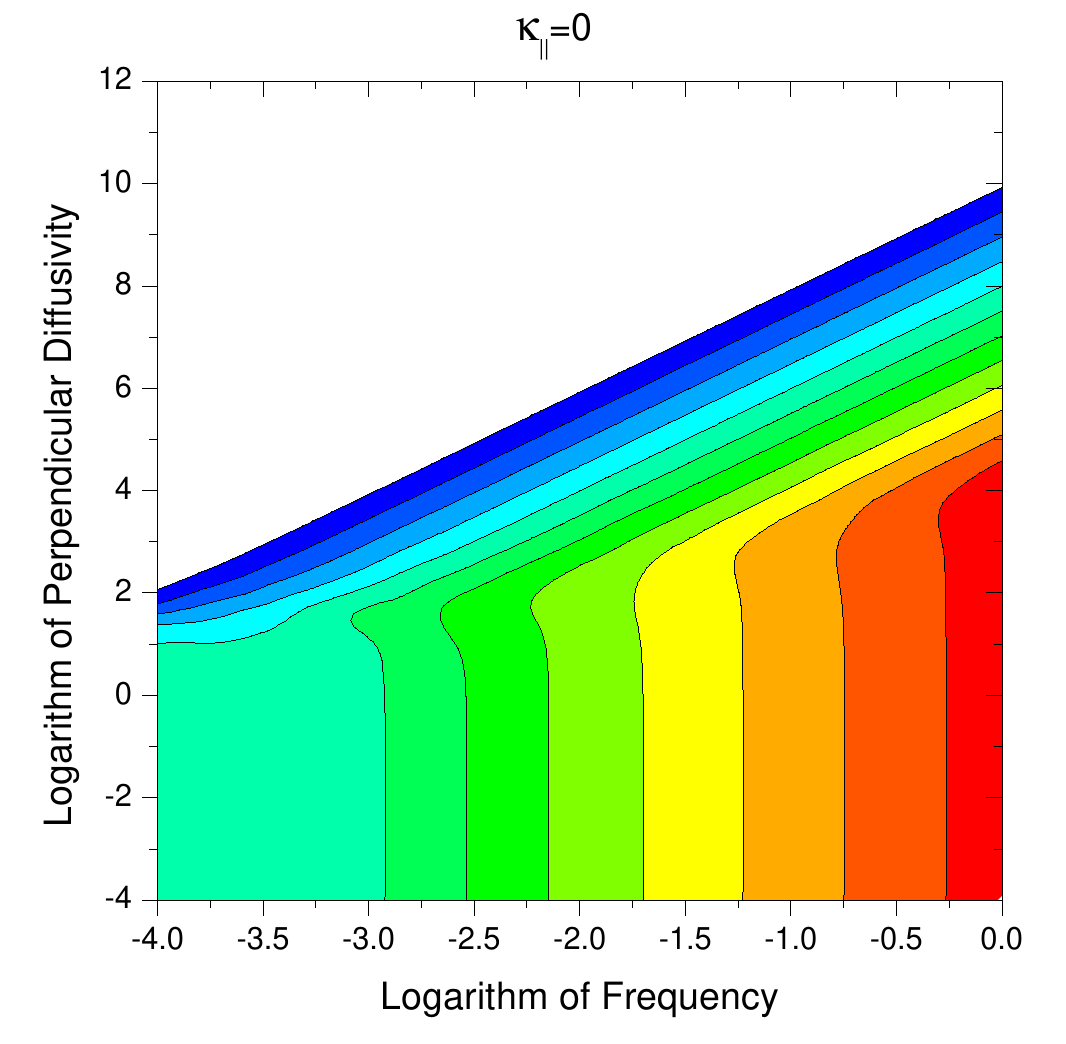}
\includegraphics[angle=0,scale=.48]{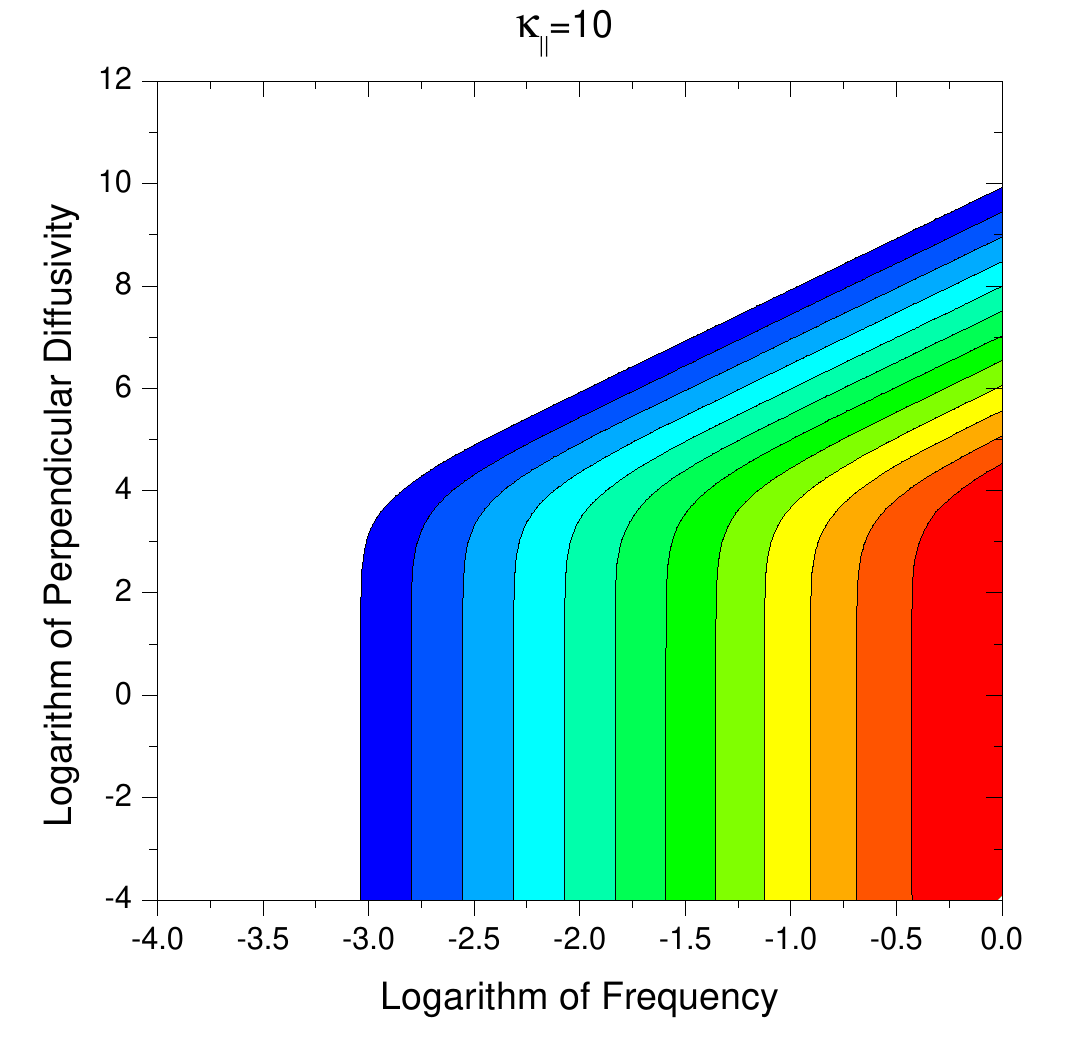}
\includegraphics[angle=0,scale=.48]{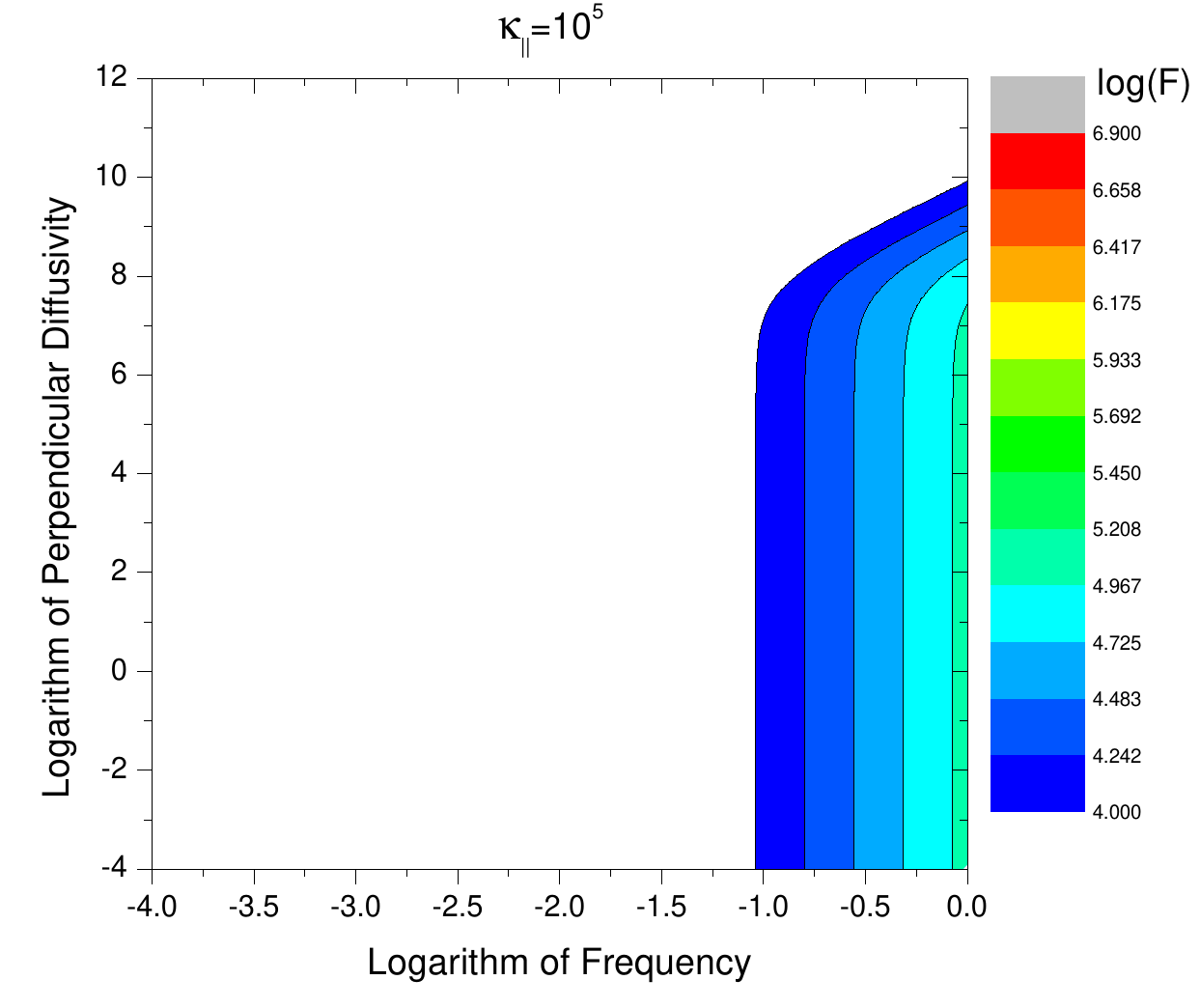}
\caption{The variation of  damping length scales (top) and energy flux densities (bottom) of fast mode which is given by Equation (32)   with frequency and perpendicular diffusivity for $\nu=10^{19} \,cm^{2}\,s^{-1}$ and $\kappa_{\parallel}=0, 10$ and $10^{5} \,\,erg\,cm^{-1}s^{-1}K^{-1} $.}
\end{center}
\end{figure*}

\subsection{Linear Approximation}

In order to compare the results of quasi-linear and linear approximations in this sub-section, we neglect $(\mathbf{v}_1 \cdot \nabla)\mathbf{v}_1$ nonlinear term  in the momentum equation (16) and then we obtain dispersion relation  as given:


\begin{equation}
\left( \omega ^2 - {v_A}^2 k^2+i \omega \nu k^2 \right)^3 \left\lbrace
\begin{array}{c}
 k^4 \left[ \left(  M-i \omega \nu \right)^2 - \left(\frac{0.6}{\rho_0}\right)^2  \left( \frac{H_{\|}} {\xi} \right)^2  \right] 
 \\ - k^2 \left[ 2 \omega ^2 \left( M- i \omega \nu  \right)   \right] +\omega ^4   
 \end{array}
 \right\rbrace =0.
\end{equation}

The dispersion relation has two roots: one root gives damping Alfven waves. We have given the solutions of this wave already in subsection 3.1.2. The other quartic root corresponding to the modified acoustic waves represents thermal pressure, viscosity and heat conduction acting in concert. This is in quadratic roots in $k^2$ and is related to the fast and slow modes; one root  propagates with a higher phase velocity.

The fast (corresponding to the plus sign) and slow (corresponding to the minus sign)  modes  have a dispersion relation as below:

\begin{equation}
\omega^2 - k^2 \left( M -i \omega \nu \mp \left(\frac{0.6}{\rho_0}\right) \left( \frac{H_{\|}} {\xi} \right)  \right) =0.
\end{equation}

For $\kappa_{\parallel}\ga 10^5 \,\,erg\,cm^{-1}s^{-1}K^{-1} $ these two waves have a damping length scale about ${3\times10^{13}\,cm}$ $\backsim 431\, R_{\odot}$, whatever the value of viscosity is. It means that these waves propagate without dissipation, in other words, they don't heat NPCH plasma. For $\kappa_{\parallel}<10^5 \,\,erg\,cm^{-1}s^{-1}K^{-1} $, a wave with a given frequency damps at the required distances. The energy flux density of these waves varies between $10^{4.5}$ and $10^{5.5} \,erg\, cm^{-2}\, s^{-1}$(see Fig. 11).  

If take $H_{\|}=0$ into consideration, the resulting dispersion relation becomes:

\begin{equation}
\left(  \omega^2 - {v_A}^2 k^2 + i \omega \nu k^2\right)^3 \left( \omega^2 - k^2 \left( M -i \omega \nu \right)   \right)^2  =0.
\end{equation}

The second factor in this relation is similar with equation (27)  in subsection 3.1.1. except $2\xi^2$ term within the factor of $k^2$. This factor corresponds with modified acoustic waves, but they haven't fast or slow wave branch. A wave with given frequency has a  given damping length scale. Besides, waves with smaller damping length scales have greater energy flux densities(see Fig. 11).

Thus one can easily see that the dispersion relation obtained in the linear approximation doesn't give modified acoustic mode which is damped for $\kappa_{\parallel}\ga 10^5 \,\,erg\,cm^{-1}s^{-1}K^{-1} $. Although the solution of dispersion relation is obtained after a very lengthy effort and is complicated mathematically in the quasi-linear approximation, it gives damping waves. In the case of $H_{\|}=0$,  we have also damping solutions for two approximations. The damping length scale of acoustic wave with a given frequency is short enough and energy flux density is high enough to heat \ion{O}{VI} ions in NPCH plasma  in the linear approximation.  But, $2\xi^2$ term comes from neglecting term $(\mathbf{v}_1 \cdot \nabla)\mathbf{v}_1$ in the linear approximation produces the fast and slow wave branch. These waves damp at required damping length scales
and have about $>10^5\, erg\,cm^{-2}\,s^{-1}$  energy flux densities. Thus these waves may be responsible for perpendicular heating of \ion{O}{VI} ions and thus contribute to the heating of the coronal base in NPCH, besides they may be an extra accelerating agents for fast solar wind in NPCH.

\section{SUMMARY AND DISCUSSION}

In this paper we analyze the damping of acoustic, Alfven and magnetosonic waves taking into consideration viscosity, parallel and perpendicular heat conduction terms. The thermal conductivity and viscosity have anisotropic
character due to the presence of magnetic field in NPCH. The damping length scales are estimated for waves with angular frequency range of $10^{-4} $ to $1 \,rad/s$ which corresponds to observed period interval $6 \,s > P> 1048 \, min$ (see \citet{ban11}). Using equation (8), the parallel thermal diffusivity for \ion{O}{VI} ions for our model is estimated as $\sim 10^9-10^{10} \,\,erg\,cm^{-1}s^{-1}K^{-1}$. In NPCH the ratio $\kappa_{\perp}/\kappa_{\parallel}$ is unknown. For solar corona, \citet{van84} takes it as $10^{-10}$, therefore the perpendicular thermal diffusivity is usually neglected. But it may be argued that for solar coronal conditions $\kappa_{\perp}\ll\kappa_{\parallel}$. \citet{van84} argue that even the ratio $\kappa_{\perp}/\kappa_{\parallel}$ assumes the value $10^{-14}$ it should be taken into consideration because it removes unphysical singularity in the case of $\kappa_{\perp}=0$ for thermal instability. For this reason we took into consideration the perpendicular thermal diffusivity  as a free parameter which varies between $10^{-4}$ and $10^{12} \,\,erg\,cm^{-1}s^{-1}K^{-1}$. Similiarly viscosity coefficient for solar corona is quite uncertain. It is taken within a range of $1$ to $10^{22} \,cm^{2}\,s^{-1}$. Thus  we tried to determine the necessary values of these parameters by considering the damping length scales which would cause the solar wind acceleration and heating corona. 

The main conclusions that can be drawn from our study are as follows:

Assuming the temperature gradient does not depend on $R$ in the interval $R=1.7-2.24$ (i.e., $\nabla_{\parallel} T=0$) and thus  $H_{\parallel}=0$ at the coronal base of NPCH, we obtained two dispersion relations in relation to modified acoustic waves and Alfven waves.

At the values of $\kappa_{\perp}$  smaller than 100, damping of acoustic waves at the required distances sensitively depends on the viscosity  value (Figure 2). The viscosity should have values higher than $10^{15} \,cm^{2}\,s^{-1}$. 

Waves with angular frequency $\omega= 0.01 \,\,rad/s$ and period $10.5 \, minute$  propagating in plumes damp for all the values of viscosity and perpendicular diffusivity while waves with same frequency  propagating in interplume lanes damp just for a given viscosity range ($\sim10^{16.7}$ - $10^{20.7} \,cm^{2}\,s^{-1}$). But in  both of lanes, the energy flux densities of these waves vary between $ 10^{4.7}$ and $10^7 \,erg \,cm^{-2}\,s^{-1}$ (Figure 3).

For damping of the Alfven wave with a given frequency at the required distances, viscosity should have a value within a given range. On the other hand, the energy flux density of Alfven waves is in the range $\sim10^6-10^{8.6} \,erg\, cm^{-2}\, s^{-1}$ (Figure 4).

It is seen that $1/k_i$ is directly proportional to $R$ and wave energy flux densities continuously decrease with increasing $R$ from Table 1. This decay of energy implies transfer of energy between \ion{O}{VI} ions and waves in NPCH. Some knowledge about energy fluxes and damping length scales of waves are given in literature: i.e., Acoustic waves are not taken into consideration in the coronal plasma heating problem because they have too low energy fluxes and short dissipation length scales \citep{ath78,mei81}. The total Alfven wave energy flux required to heat the quiet corona is $3\times10^5\,erg\, cm^{-2}\, s^{-1}$ \citep{wit77}. Most of the heat supply to the corona should be within $1-2$ $R_\odot$ and also in order to produce fast solar wind, a little more energy flux density ($1-2\times10^5\,erg\, cm^{-2}\, s^{-1}$) should be deposited at the sound point (a few solar radii) \citep{wit88}; Both the slow and the fast solar wind reveal the existence of an extra accelerating agent beyond $2.3R_\odot$ \citep{str12}.

The energy flux densities  of acoustic and Alfven waves reported here are  larger than what is required to heat corona. Besides, different waves damp at different heliocentric distances.  After all, our results suggest energy deposition by acoustic and Alfven waves in the NPCH, thus providing a significant source for coronal heating and  mainly preferentially heating of \ion{O}{VI} ions and observed wind acceleration.

When we retain the parallel heat conduction term, the first solution ($X_1$,$Y_1$) gives ten degree dispersion relation (Equation A1). Up to the value of $\kappa_{\perp}=10^2-10^3$ damping length scales of Mode 1 and Mode 2 are independent of the value of perpendicular diffusivity. At larger values of perpendicular diffusivity, the value of perpendicular diffusivity  increases linearly with increasing frequency for all the damping length scales.

On the other hand, it is seen that the damping length scale of Mode 2 is independent of the value of $\kappa_{\parallel}$. Mode 3 shows a similar tendency with Mode 2. While Mode 4 gets damped very near to the sun, Mode 5 propagates to very far distances before damping.

When only $H_{\parallel}$  and viscosity are taken into consideration in Equation (A1), Mode 1 gets damped at distances $0.01-0.5 R_{\odot}$.  Mode 2 waves with frequencies within the range $10^{-1.75} - 10^{-2.25} \,rad/s $  get damped within the required radial distances. Damping is independent of the value of $\kappa_{\parallel}$.  Mode 3 has a required damping length scale when $\kappa_{\parallel}$  values fall within a narrow range, i.e.,  $10-10^5 \,\,erg\,cm^{-1}s^{-1}K^{-1}$. Waves with frequencies lesser than $0.01 \,rad/s $  get damped at very far distances (i.e. further than $5.5 R_{\odot}$).  As seen in Figure 5, to obtain required damping length scales in all frequency values we should consider perpendicular diffusivity effect even if it has a negligible value.
 
Although the second solution ($X_2$,$Y_2$) gives the parameter ranges giving required damping length scales for waves, it is seen that the calculated wave fluxes are very small. So, these waves can not replace the lost energy in NPCH.

The third solution ($X_3$,$Y_3$) also gives tenth order dispersion relation (A2). Three modes give required damping length scales for all frequencies at a given parallel diffusivity value. Damping length scale of these modes is more or less independent of perpendicular diffusivity (Figure 9). The other two modes have very long damping length scales.

The energy flux densities of the waves which are derived from the solutions of ($X_1$,$Y_1$) and ($X_3$,$Y_3$) turned out to be as high as $10^{10} \,erg\, cm^{-2}\, s^{-1}$. Since the group velocities turned out to be superluminal, energy flux densities are unphysically huge. We can not estimate energy flux densities of these waves using Equation (28) because physical properties of a wave packet should not be described by group velocity in a medium with anomalous dispersion (\citet{jac75}, \citet{ven10}).

But for modified acoustic wave and Alfven wave we found that driven waves in the NPCH have enough energy deposition which gives rise not only to substantial heating and acceleration of the \ion{O}{VI} ions but also coronal plasma heating and an extra accelerating agent for fast solar wind in NPCH.

In this paper we used isotropic viscosity.  Anisotropic viscosity may significantly change the dynamics of MHD waves. Therefore, this study should be extended using anisotropic viscosity to obtain more realistic results.


\section*{Acknowledgments}

We thank to  Donald B. Melrose and Luca Teriaca for useful suggestions. We also thank to the referee for constructive comments.

 
%

\medskip

\bibliographystyle{mnras}	

\onecolumn



\appendix
\section{Long Dispersion Relations}

The dispersion relation which is obtained for (X1,Y1) is given as

\begin{eqnarray}
&& k^{10} \left[-M\left(M-{v_A}^2 \right) M \left(M-i\omega\nu \right)^5   \right]+k^9 \left[ -2\omega 0.6 \frac{1}{\rho_0} H_\parallel  \left(M- i\omega\nu \right)^4 \left( 2M-{v_A}^2 \right) -4 \omega M 0.6 \frac{1}{\rho_0} H_\parallel \left(M-i\omega\nu \right)^3  \left(M-{v_A}^2 \right) \right] \nonumber \\
&& +k^8 \left[ \begin{array}{l}
 \omega^2 M \left( M-i\omega\nu \right)^4 \left[ 5\left(M-{v_A}^2 \right) +4\xi^2 \right] -4 \omega^2\left(M-i\omega\nu \right)^3  \left[\left(0.6 \frac{1}{\rho_0} H_\parallel \right)^2 +\xi^2 M\left({v_A}^2- i\omega\nu \right) \right] \nonumber \\
 -12 \omega^2 \left(0.6 \frac{1}{\rho_0} H_\parallel \right)^2 M\left(M-i\omega\nu \right)^2 -4 \omega^2 \left(M- i\omega\nu \right) \left(0.6 \frac{1}{\rho_0} H_\parallel \right)^2 M \left(M-{v_A}^2 \right)
\end{array}
\right] \nonumber \\
&& +k^7\left[\begin{array}{l}
 8\omega^3\left( M- i\omega\nu \right)^3 0.6 \frac{1}{\rho_0} H_\parallel \left(2M-{v_A}^2 +\xi^2 \right) + \omega^3 (M-i\nu\omega)^2 0.6 \frac{1}{\rho_0} H_\parallel \left[12M \left(M -{v_A}^2\right) +8M\xi^2 -8\xi^2 \left({v_A}^2-i\omega\nu \right)\right] \nonumber \\
 -8\omega^3 \left(0.6 \frac{1}{\rho_0} H_\parallel\right)^3 \left(M-i\nu\omega \right)-8\omega^3\left(0.6 \frac{1}{\rho_0} H_\parallel \right)^3 \left(M+{v_A}^2 \right) \nonumber \\ 
\end{array}
\right] \nonumber \\
&& +k^6 \left[ \begin{array}{l}
-M\omega^4 \left(M- i\omega\nu \right)^3 \left[10\left( M-{v_A}^2 \right)+12 \xi^2 \right]  +12 \omega^4 \left(M-i\omega\nu \right)^2 \left[\left(0.6 \frac{1}{\rho_0} H_\parallel\right)^2 +\xi^2 M\left({v_A}^2 - i\omega\nu \right) \right] \nonumber \\
+\omega^4 \left(M- i\omega\nu \right)\left(0.6 \frac{1}{\rho_0} H_\parallel\right)^2 \left(24M+16\xi^2\right)+4\omega^4 M\left(0.6\frac{1}{\rho_0} H_\parallel \right)^2\left(M-{v_A}^2\right) \nonumber \\
\end{array}
\right] \nonumber \\
&& +k^5 \left[ \begin{array}{l}
-\omega^5 \left(M-i\omega\nu \right)^2  0.6 \frac{1}{\rho_0} H_\parallel \left[12\left(2M-{v_A}^2\right)+16\xi^2\right] \nonumber \\
-\omega^5 \left(M- i\omega\nu \right)0.6 \frac{1}{\rho_0} H_\parallel \left[12M\left(M-{v_A}^2 \right)+16M\xi^2-16\xi^2 \left({v_A}^2 -i\omega\nu \right) \right] +8\omega^5 \left(0.6 \frac{1}{\rho_0} H_\parallel \right)^3 \nonumber \\
\end{array}
\right] \nonumber \\
&& +k^4 \left[ M\omega^6\left( M-i\omega\nu \right)^2 \left[10\left(M-{v_A}^2 \right)+12\xi^2\right]-12 \left(M-i\omega\nu \right)\omega^6 \left[\left(0.6 \frac{1}{\rho_0} H_\parallel \right)^2+\xi^2 M \left({v_A}^2-i\omega\nu \right)\right]-\omega^6 \left(12M+16\xi^2 \right)\left(0.6 \frac{1}{\rho_0} H_\parallel \right)^2 \right] \nonumber \\
&& +k^3 \left[ 8\omega^7\left( M-i\omega\nu\right)0.6 \frac{1}{\rho_0} H_\parallel \left( 2M-{v_A}^2+\xi^2 \right)+4\omega^7 0.6 \frac{1}{\rho_0} H_\parallel \left[M\left(M-{v_A}^2\right)+2M\xi^2-2\xi^2\left({v_A}^2-i\omega\nu \right)\right] \right] \nonumber  \\
&& +k^2 \left[ -\left(M-i\omega\nu\right)M\omega^8\left[5\left(M-{v_A}^2 \right)+4\xi^2\right]+4\omega^8 \left[\left(0.6 \frac{1}{\rho_0} H_\parallel \right)^2+\xi^2 M \left({v_A}^2-i\omega\nu \right)\right] \right] \nonumber \\
&& +k \left[ -2\omega^9 0.6 \frac{1}{\rho_0} H_\parallel \left(2M-{v_A}^2 \right)\right]  +\left[ \omega^{10}\left( M-{v_A}^2 \right) M\right] =0
\end{eqnarray}

The dispersion relation which is obtained for (X3,Y3) is given as

\begin{eqnarray}
&& k^{10} \left[ \begin{array}{l}
-\xi^2 i\omega\nu \left(M-i\omega\nu\right)^3\left({v_A}^2-i\omega\nu\right)-\left(0.6\frac{1}{\rho_0} H_\parallel\right)^2 \left(M-i\omega\nu\right)^2 \left(2{v_A}^2-i\omega\nu\right) \\
-i\omega\nu \left(M-i\omega\nu\right) \left(M+{v_A}^2\right) \left(0.6\frac{1}{\rho_0} H_\parallel\right)^2+2 {v_A}^2 M\left(0.6 \frac{1}{\rho_0} H_\parallel\right)^2\left(M-i\omega\nu\right)
\end{array}
\right] \nonumber \\
&& +k^9 \left[ 2\omega\xi^2 0.6\frac{1}{\rho_0} H_\parallel \left(M-i\omega\nu\right)^2\left({v_A}^2-2i\omega\nu\right)+2\omega\left( 0.6\frac{1}{\rho_0} H_\parallel \right)^3 \left({v_A}^2+i\omega\nu \right) \right] \nonumber \\
&& +k^8 \left[ \begin{array}{l}
-\omega^2\left(0.6\frac{1}{\rho_0} H_\parallel \right)^2 \left(M+{v_A}^2\right) \left(M-2i\omega\nu\right) +4\xi^2\omega^2\left(M-i\omega\nu\right)\left(0.6 \frac{1}{\rho_0} H_\parallel \right)^2 +2 \left(0.6 \frac{1}{\rho_0} H_\parallel \right)^2 \omega^2\left( M-i\omega\nu \right) \left(2{v_A}^2-i\omega\nu \right) \\
-\xi^2\omega^2 \left( M-i\omega\nu \right)^3 \left({v_A}^2-2i\omega\nu \right)+3i\nu\xi^2 \left(M-i\omega\nu\right)^2 \omega^3\left({v_A}^2-i\omega\nu\right)-2\omega^2 {v_A}^2 M \left(0.6 \frac{1}{\rho_0} H_\parallel\right)^2 +\left(0.6 \frac{1}{\rho_0} H_\parallel\right)^2 \omega^2 \left(M-i\omega\nu\right)^2 \\
\end{array}
\right] \nonumber \\
&& +k^7 \left[ 2\left(0.6\frac{1}{\rho_0} H_\parallel\right)^3 \omega^3-4\omega^3\xi^2 0.6\frac{1}{\rho_0} H_\parallel \left(M-i\omega\nu \right)^2 -4\omega^3\xi^2 0.6\frac{1}{\rho_0} H_\parallel \left({v_A}^2-2i\omega\nu \right) \left(M-i\omega\nu\right) \right] \nonumber \\
&& +k^6 \left[ \begin{array}{l}
\omega^4\left(0.6 \frac{1}{\rho_0} H_\parallel \right)^2 \left(-{v_A}^2-M+3i\omega\nu\right)-4\xi^2\omega^4 \left(0.6 \frac{1}{\rho_0} H_\parallel \right)^2+\omega^4\xi^2 \left(M-i\omega\nu \right)^3 +3\xi^2\left(M-i\omega\nu\right)^2\omega^4 \left({v_A}^2-2i\omega\nu\right) \\
-3i\nu\xi^2 \omega^5\left(M-i\omega\nu \right)\left({v_A}^2-i\omega\nu\right)
\end{array}
\right] \nonumber \\
&& +k^5 \left[2\omega^5\xi^2  0.6 \frac{1}{\rho_0} H_\parallel \left(4M+{v_A}^2-6i\omega\nu\right) \right] \nonumber \\
&& +k^4 \left[ \omega^6\left(0.6\frac{1}{\rho_0} H_\parallel \right)^2 -3\xi^2 \left(M-i\omega\nu \right)^2 \omega^6 -3\xi^2 \left(M-i\omega\nu \right) \omega^6 \left( {v_A}^2 -2i\omega\nu \right) +\xi^2 i\nu\omega^7 \left( {v_A}^2 -i\omega\nu \right) \right] \nonumber \\
&& +k^3 \left[-4\xi^2 0.6 \frac{1}{\rho_0} H_\parallel \omega^7\right]+k^2 \left[ \omega^8\xi^2\left(3M-5i\omega\nu+{v_A}^2\right)\right]-\omega^{10}\xi^2=0
\end{eqnarray}


\bsp	
\label{lastpage}
\end{document}